\documentstyle[12pt,twoside]{report}

 \def\be{\begin{equation}} \def\ee{\end{equation}}
 \def\bdm{\begin{displaymath}} \def\edm{\end{displaymath}}
 \def\ba{\begin{array}} \def\ea{\end{array}}
 \def\bea{\begin{eqnarray}} \def\eea{\end{eqnarray}}
 \def\beas{\begin{eqnarray*}} \def\eeas{\end{eqnarray*}}
 \def\disp{\displaystyle}
 
 \newcommand{\mysection}[1]{\section{#1}\setcounter{equation}{0}}
 \newcommand{\dfrac}[2]{{\displaystyle \frac{#1}{#2}}}
 \newcommand{\tfrac}[2]{{\textstyle \frac{#1}{#2}}}

      \def\cut{\hfill\break}
      
      \def\arabeq{\arabic{equation}}
 \newcounter{remno}
 \def\remarks{\setcounter{remno}{0}}
 \newcommand{\remark}[1]{\addtocounter{remno}{1} \subsubsection*{Remark
           \arabic{remno}#1}}

\def\ct{canonical transformation}
      \def\gr{general relativity}
      
      \def\qg{quantum gravity}
      
      \def\st{spacetime}
      \def\wrt{with\ respect\ to}
      \def\ie{i.e.}
      \def\eg{e.g.}
      \def\iff{if\ and\ only\ if}
      \def\ps{phase space}

      \def\bra#1{\langle\, #1\, |}
      \def\d{\partial}
      \def\Div#1{{\rm Div}_{#1}}                
      \def\half{{\textstyle{1\over2}}}
      
      \def\IP#1#2{\langle\, #1\, |\, #2\, \rangle}
      \def\ket#1{|\, #1\, \rangle}
      \def\Lie#1{{\cal L}_{\tenrm #1}}
      \def\lint{\int\nolimits}
      
      \def\ovr{\overline}
      \def\pb#1{\rlap{\lower1ex\hbox{$\leftarrow$}}#1{}}
      \def\real{{\rm I\!R}}
      \def\rd{{\rm d}}
      
      \def\tw{\widetilde}
      \def\und{\underline}
      \def\wh{\widehat}        
      \newcommand{\fr}[2]{\frac{#1}{#2}}
      \newcommand{\sfr}[2]{{\textstyle \frac{#1}{#2} }}

 \def\Gbar{\hbox{$\ovr{\Gamma}$}}
 \def\Ghat{\hbox{$\hat{\Gamma}$}}

 \def\cC{\hbox{${\cal C}$}}                    
 \def\cV{\hbox{${\cal V}$}}                    
 \def\cVp{\hbox{${\cal V}_{phy}$}}             
 \def\cA{\hbox{$\cal A$}}                      
 \def\cS{\hbox{$\cal S$}}                 
 \def\cAp{\hbox{${\cal A}_{phy}$}}             
 \def\cAs{\hbox{${\cal A}^{(\star)}$}}         
 \def\cAps{\hbox{${\cal A}_{phy}^{(\star)}$}}  
 \def\hca{\hbox{${\cal A}_{red}$}}               
 \def\hcas{\hbox{${\cal A}_{red}^{(\star)}$}}  

 \def\newbook{Ashtekar A 1991 Lectures on {\it Non-perturbative canonical
  gravity} Notes prepared in collaboration with Tate R S (World Scientific
  Singapore)}
 \def\UP{University Press}
  \def\OUP{Oxford University Press}
 \def\SV{Springer-\negthinspace Verlag}
 \def\pc{{\it private communication}}

 \def\CMP{{\it  Comm.\ Math.\ Phys.}}
 \def\CQG{{\it  Class. Quantum Grav.}}
 \def\GRG{{\it  Gen.\ Rel. \&\ Grav.}}
 \def\JMP{{\it  J.\ Math.\ Phys.}}

 \def\PR{{\it   Phys.\ Rev.}}
 
 \def\PRL{{\it  \PR\ Lett.}}
 \def\RPP{{\it  Rep. Prog. Phys.}}
 \def\SUpp{ Syracuse University pre-print}
 
 \def\auth{$\und{\hphantom{aaaaaa}}$}

\begin{document}

\pagenumbering{roman}
\begin{flushright}
SU-GP-92/8-1\\
gr-qc/9304043
\end{flushright}

\begin{center}
An Algebraic Approach to the Quantization of Constrained Systems: \\
Finite Dimensional Examples\\
(Ph.D. dissertation, Syracuse University, August 1992)
\end{center}
\medskip
\begin{center}
Ranjeet S. Tate \\
Department of Physics,
University of California - Santa Barbara,
CA 93106-9530\\
e-mail: rstate@cosmic.physics.ucsb.edu
\end{center}
\medskip

\begin{center}
$\und{\cal ABSTRACT}$
\end{center}

General relativity has two features in particular, which make it difficult to
apply to it existing schemes for the quantization of constrained systems.
First, there is {\em no background structure} in the theory, which could be
used, e.g., to regularize constraint operators, to identify a ``time'' or to
define an inner product on physical states. Second, in the Ashtekar formulation
of general relativity, which is a promising avenue to quantum gravity, the
natural variables for quantization are {\it not} canonical; and, classically,
there are algebraic identities between them. Existing schemes are usually not
concerned with such identities. Thus, from the point of view of canonical
quantum gravity, it has become imperative to find a framework for quantization
which provides a {\em general} prescription to find the physical inner product,
and is flexible enough to accommodate non-canonical variables.

In this dissertation I present an algebraic formulation of the Dirac approach
to the quantization of constrained systems. The Dirac quantization program is
augmented by a general principle to find the inner product on physical states.
Essentially, the Hermiticity conditions on physical operators determine this
inner product. I also clarify the role in quantum theory of possible algebraic
identities between the elementary variables.

I use this approach to quantize various finite dimensional systems. Some of
these models test the new aspects of the algebraic framework. Others bear
qualitative similarities to \gr, and may give some insight into the pitfalls
lurking in \qg. The previous quantizations of one such model had many
surprising features. When this model is quantized using the algebraic program,
there is no longer any unexpected behaviour. I also construct the complete
quantum theory for a previously unsolved relativistic cosmology. All these
models indicate that the algebraic formulation provides powerful new tools for
quantization.

In (spatially compact) general relativity, the Hamiltonian is constrained to
vanish. I present various approaches one can take to obtain an interpretation
of the quantum theory of such ``dynamically constrained'' systems. I apply some
of these ideas to the Bianchi I cosmology, and analyze the issue of the initial
singularity in quantum theory.

\newpage
\setcounter{page}{3}
\tableofcontents

\chapter*{Preface}
\addcontentsline{toc}{chapter}{Preface}
\pagestyle{myheadings}
\markboth{{\sf Preface}}{{\sf Preface}}
Further work has been done on parts of this thesis. A preprint (on
which chapter 2 and some of the examples are based) is in preparation
(A.~Ashtekar and R.~S.~Tate) on the algebraic approach to the
quantization of constrained systems. The approach to deparametrization
applied to the quantized Bianchi type I model has been improved upon,
and the analysis of the singularity has been extended to a class of
``solvable'' Bianchi models, by A.~Ashtekar, R.~S.~Tate and C.~Uggla,
see gr-qc/9302027.
\begin{center}
$\und{\cal ACKNOWLEDGEMENTS}$
\end{center}

First and foremost I would like to thank my thesis advisor, Prof.\ Abhay
Ashtekar. He has been a very good advisor, in terms of the guidance and
encouragement he has provided, both scientific and personal. Discussions with
him are always exciting, and our collaboration has been very enjoyable. His
detailed comments on this thesis --based on his uncanny ability to place
himself in the shoes of the reader-- have resulted in a much more readable
thesis than the one I first presented to him.

Members of the relativity group at Syracuse University --faculty, post-docs,
visitors and graduate students-- foster an atmosphere which is both
academically stimulating and personally friendly. All of them have always been
willing to discuss things with me, even if the ideas were not of direct
interest to them. Many ideas in my graduate work have arisen from discussions
with various members of the group. I have enjoyed discussions with Bernd
Br\"ugmann, Josh Goldberg, Jorge Pullin, Joseph Samuel, Lee Smolin, Charles
Torre and Claes Uggla; all of whom have also provided help on technical
matters.

I thank Karel Kucha\v r and Carlo Rovelli for their patient explanations of
their ideas.

Jorma Louko read preliminary drafts of my thesis very carefully; his
criticisms and thought-provoking questions have clarified many conceptual and
technical issues.

Most graduate students help each other out by reading first drafts, helping
prepare talks, providing ideas or listening to them --and then helping to work
them out. As my office-mates, (for four years and one year
respectively) Joe Romano and Gabriela Gonzalez have always gone out of their
way to provide such help.

I thank the Syracuse University Outing Club for providing wonderful
alternatives to academics. The idea for at least one paper occurred to me on a
club trip.

I thank HarperCollins Publishers for their permission to use Achilles and the
Tortoise as characters in a dialogue.

I have been supported by funds provided by Syracuse University, a Syracuse
University Graduate Fellowship and a research assistantship under NSF grants
PHY86-12424 and PHY90-16733.

\chapter{INTRODUCTION}
\pagenumbering{arabic}
\pagestyle{myheadings}
\markboth{{\sf Chapter 1}}{{\sf Introduction}}

      \def\comp{{\rm C}\llap{\vrule height7.1pt width1pt depth-.4pt\phantom t}}

\mysection{Motivation}

For many physically interesting systems, not all points of the \ps\ are
accessible. The allowed classical states of the system are confined to lie on a
sub-manifold of the \ps, called the constraint surface. Now, all empirical
evidence supports the widely held view that the fundamental behaviour of all
systems must be quantum mechanical in nature, and not classical. As such, we
are concerned with the {\it canonical quantization} of such {\it constrained
systems}.

A pre-eminent constrained system --one that has withstood all assaults to
quantize it through at least 4 decades \cite{qg:rev}-- is general relativity
\cite{wald1}. Of the difficulties related to quantizing gravity, a significant
part arises from the fact that, in one way or the other, current approaches to
the quantization of constrained systems have not been quite general enough to
apply to \gr. Almost all current approaches \cite{gq1,gtq1,feyn1} rely on the
presence of  specific structures, on either configuration space or \ps; and a
generic constrained system may or may not possess these structures. Certainly,
it is not known whether these approaches can be taken to quantize \gr.

Before I describe the features peculiar to \gr, let us briefly recall gauge
theories, to provide points of reference.  Like \gr, gauge theories are field
theories with constraints on the \ps, (As fundamental theories of matter, they
are of course important in their own right.)

Gauge field theories are defined on a background \st, usually taken to be
Minkowski \st. The elementary configuration variable is a Lie algebra valued
connection on spatial slices of the \st\ (Cauchy surfaces), and the canonical
momentum is a triad which takes values in (the dual to) the Lie algebra. The
constraint surface can be specified by the vanishing of the Gauss constraint
functional. Dirac has pointed out \cite{pamd1} --in the general context of
constrained systems-- that due to the ambiguity in specifying the constraint
function, one is forced to consider that canonical transformations generated by
the constraint connect  physically equivalent classical states. As we know, the
Gauss constraint generates gauge transformations --i.e.\ rotations in the
internal symmetry group-- on the canonical variables. Dynamics is specified by
a Hamiltonian, which is constructed however,  not just from the canonical
variables, but also from the non-dynamical spatial metric. All progress in the
quantization of gauge theories has made essential use of this background \st\
metric. The metric is used, e.g., to regularize operators (via point splitting
or other means), to construct an inner product on physical quantum states, and
even to define perturbation theory.

General relativity is markedly different from gauge theories. It is a theory of
the structure of \st\ itself: in the geometrodynamical formulation the
canonical variables \cite{adm1} are the 3-dimensional spatial metric and its
conjugate momentum, the extrinsic curvature of the spatial hypersurface. There
are four constraints on the geometrodynamical \ps: the three independent vector
constraints --which generate spatial diffeomorphisms of the canonical fields--
and the scalar constraint. In the \st\ one can construct from any solution, the
scalar constraint generates a time-like diffeomorphism of the canonical
variables. Hence, in the spatially compact cases at least, the scalar
constraint is the generator of dynamical evolution. This constraint is the
primary source of difficulty, in both classical and quantum theory.

The most significant difference between \gr\ and gauge theories is of course
the complete absence in \gr\ of any background fields: be they metrics or
connections. As one can imagine, this absence of background structure has
prevented the construction of universally accepted and well-defined
regularizations of the constraints in quantum theory. Eventually, the absence
of background structure will also make itself felt when one has quantum
physical states and wishes to construct an inner product on them, {\it unless}
one has a prescription --to find the physical inner product-- which does {\it
not} depend on the presence of background fields.
Of course, the lack of a non-dynamical background will also cause difficulties
in the construction of a physical interpretation of \qg.

Recall that \gr\ is the theory of space{\it time} itself, and hence there
cannot be an external time parameter. In the Hamiltonian formulation, time is
sometimes considered as ``hidden'' in the canonical variables themselves.
However, there is no {\it preferred} internal time. This difficulty arises from
the fact that, up to the possible addition of a surface term, the Hamiltonian
of \gr\ {\it is} the scalar constraint. Thus, on the one hand, the scalar
constraint generates time evolution, where, classically, time can be identified
as an affine parameter along the orbits of the \ct\ generated by the scalar
constraint. On the other hand, since it is after all a constraint, the \ct s it
generates must be viewed as gauge; and since there is considerable ambiguity in
defining the constraint function, there is at least that much ambiguity in
identifying a time.

One is not used to dealing with time as a dynamical entity. In ordinary quantum
mechanics and in field theories, the time parameter is available to us
externally, from the background \st. Furthermore, time  plays no role in
constructing the kinematical quantum theory --i.e.\ in solving the constraints
and finding the inner product on the space of physical states. From the point
of view of canonical \qg, an important issue then is whether or not ``time''
has first to be singled out from the other variables before one can proceed
with the construction of the kinematical theory.

The difficulties associated with the canonical quantization of \gr\ which I
have discussed up to this point have been independent of any particular choice
of canonical variables. Now, I will concentrate on a particular formulation of
\gr, which appears to be the most promising for \qg.

In 1986 Ashtekar \cite{aa:new1} introduced new variables for \gr, and
formulated it as a theory of the dynamics of a certain connection.
In terms of these variables, a fair amount of progress has been made towards
a quantum theory of gravity%
\footnote{For recent reviews of the new variables approach to \qg, see
 \cite{newbook1,carlo:review1,lee:spain1}.}.
{}From the point of view of canonical quantization, there are some
difficulties endemic to this connection dynamical formulation of \gr.
\begin{itemize}

\item The new canonical variables are a complex $SU(2)$ connection $A$ and a
real triad $E$. For connections which describe real \gr, the complex conjugate
connection is given by the ``reality condition'': $\bar{A}=2\Gamma(E)-A$, where
$\Gamma(E)$ is the spin connection of the triad. Immediately, a seeming
difficulty presents itself: Consider e.g., the connection representation, in
which states are (holomorphic) functions of the connection; $\hat{A}$, the
operator corresponding to the connection, acts via multiplication and the
canonically conjugate momentum $\hat{E}$ is represented by a {\it holomorphic}
derivative operator, symbolically $(\d/\d A)$. Now, the reality conditions on
the classical variables imply Hermiticity conditions on the corresponding
quantum operators. Thus, for real \qg, $\hat{E}$ should be a Hermitian
operator. How though, can a {\it holomorphic derivative operator} be {\it
Hermitian}? Thus, quite apart from the difficulties associated with the
constraints of \gr, within the contexts of the existing frameworks for
quantization it is not immediately clear that a description in terms of such
hybrid variables will lead to any consistent quantum theory.

\item In the new variables, much of the progress towards a quantum theory of
gravity has come about because the reformulation of \gr\ as a
connection-dynamical theory has allowed one to import and adapt techniques from
Yang-Mills theory. Of particular interest for this discussion is the loop
representation for a theory of a connection: quantum states are functionals of
loops in the spatial manifold, and the action of various operators is to
``break'' and ``join'' the loops in the argument of the states. In some ways
the loop representation is like the Fock representation for Maxwell theory:
states are not functions on some configuration space, neither are they
functions on the \ps\ which satisfy some polarization condition. Of course,
operators on this space do correspond to functions on the \ps. However, the
most natural, regularized operators on the loop states are not the canonically
conjugate pair of $(A,E)$, but certain loop variables (see e.g\
\cite{newbook1,carlo:review1,lee:spain1}),  built out of the holonomies of the
connection around loops in the spatial manifold (the Wilson loops), with triads
inserted in appropriate ways. These loop variables are not canonical
coordinates on the \ps, though in a (reasonably) well-understood sense they do
code the entire symplectic structure of the gravitational \ps. To {\it some}
extent one is familiar with dealing with non-canonical coordinates in quantum
theory. However, with the loop variables there is another difficulty: They are
an {\it overcomplete} set of functions on the \ps\ --i.e., there are ``too
many'' of them-- and they satisfy certain non-trivial algebraic identities. How
are these algebraic identities to be incorporated in quantum theory?

\item The new variables approach is a connection-dynamical  reformulation of
\gr. There is the Gauss constraint of Yang-Mills  theory, which generates
transformations in the internal gauge group --$\comp SU(2)$ in this case. In
addition, there are the usual diffeomorphism and scalar constraints of \gr. A
salient feature of the new variables is that the scalar constraint is
expressible in a very simple form. In terms of the loop variables, a
well-defined, regularized operator corresponding to the scalar constraint has
been constructed, and a large number of solutions to all the quantum
constraints has been found [9-12]. An inner product on physical states is no
longer of only conceptual interest in \gr; due to the existence of the recent
solutions, it has now become imperative to establish a criterion governing the
completeness of a set of solutions to the constraint equations, and a criterion
to select a physical inner product on these states.
\end{itemize}

Now, let us return to the problem of the canonical quantization of constrained
systems in general. As we have seen, one of the most important potential
applications of a quantization scheme will be to \qg. Hence, we must take an
approach which is general enough to include relativity in its framework.

A prominent approach to canonical quantization, which has proven successful in
the quantization of many well known physical systems, is Dirac quantization, or
the operator constraint method \cite{pamd1}. Here,  one has to first choose a
representation of quantum operators on some complex vector space, then solve
the quantum constraints by finding states which are annihilated by the
constraint operators and restrict attention to these {\it physical states}.
Physically interesting observables correspond to  operators that leave the
space of physical states invariant. However, before one can calculate physical
probabilities, expectation values or transition amplitudes one has to find an
inner product and thus impose a Hilbert space structure on the physical quantum
states. In this respect the original procedure outlined by Dirac is incomplete:
no {\rm general} principle is outlined to find the physical inner product.

Why has this problem not surfaced before, and how, within the framework of
Dirac quantization, has one been able to successfully quantize a vast variety
of physically interesting constrained systems? Though no general principle has
been used, most theories have been quantized on a case by case basis, with very
essential use being made of the symmetries of the background \st\ metric. For
example, in Schr\"odinger particle mechanics one naturally uses the volume
element of the background spatial metric to define an inner product on the
states. In Minkowskian quantum field theories, the inner-product --or,
equivalently, the vacuum state-- is selected by making an appeal to Poincar\'e
invariance (equivalently, Lorentz invariance on Fourier space): the vacuum is
the unique Poincar\'e invariant state and the vacuum expectation values of all
operators provide us with the inner product and the Hilbert space (e.g.,
through the Gelfand-Naimark-Segal construction \cite{gns1}). When quantizing a
field theory on a curved, stationary \st, on the other hand, one can use the
time-like Killing vector field of the background \st\ metric to carry out a
positive/negative frequency decomposition of the fields, and hence construct an
inner product \cite{aa:qft1}. Even in linearized gravity, one uses the
structure
of the background Minkowski \st\ to construct an inner product on the solutions
to the linearized constraints.

None of the above techniques (mostly field theoretic) can be used to find an
inner product for \gr. In all the above cases, the availability of the {\it
background} \st\ metric and the simplicity of its structure (flatness or
stationarity) is essential to find the measure, and has been explicitly used to
do so. In full \gr, of course one does not have this luxury. The metric is now
no longer a kinematical, background field but a {\it dynamical} entity. Except
in reduced and simplified theories, one has to consider arbitrary
positive-definite 3-metrics.

Consider the usual metric representation, in which quantum states are functions
of 3-metrics. In this case, for the inner product on quantum states, one is
concerned with the metric structure not of \st\ but of the superspace, i.e. the
space of all 3-metrics. Now, the scalar constraint consists of two parts: the
kinetic part, which is quadratic in momenta; and the potential term, which
depends only on the configuration variables. (In the connection-dynamical
formulation, there is no potential term.) The coefficient of the kinetic term
in the scalar constraint defines a {\it supermetric} on the space of 3-metrics.
One might be tempted to use the volume element of the supermetric to construct
an inner product on quantum states, integrating over the configuration space,
as one does in ordinary quantum mechanics. The first difficulty is field
theoretic: there are an infinite number of degrees of freedom, and such
infinite dimensional integrals are not, in general, well-defined. Next,
consider spatially homogeneous cosmological models, in which the components of
the metric are constants on the spatial manifold; and, since the resulting
configuration space is finite-dimensional, one does not have to worry about the
field degrees of freedom. Even in these simplified models of \gr,  since the
superspace metric is of  {\it Lorentzian signature}, its volume element cannot
yield the correct inner product on physical states. By analogy with the
relativistic particle, one might then consider the following alternative:
foliate superspace with submanifolds which are space-like w.r.t.\ the
supermetric; and use the volume element of the  supermetric induced on these
spatial slices (which is then positive definite) to define an inner product.
For the spatially homogeneous cases, this {\it  might} work, though there is no
guarantee that it will. However, as a strategy for the full theory, this will
run into the field theoretic difficulties we discussed before.

In the absence of a nondynamical background \st\ metric, can one use the {\em
symmetries} of the supermetric instead, to construct an inner product or select
a unique ground state? The analog of Poincar\'e invariance in field theories is
diffeomorphism invariance in \gr. Hence, one might imagine using the spatial
diffeomorphism group to select the inner product on physical states: by
requiring the (unique) vacuum state to be spatially diffeomorphism invariant.
However, we cannot select the vacuum expectation value in this manner: now the
vector constraint requires that {\it every} physical state be spatially
diffeomorphism invariant! Are there other symmetries of the supermetric that
one could use? In the geometrodynamical formulation, the supermetric has no
symmetries  which also appropriately scale the potential \cite{kk:jmp1}.
Specifically, the supermetric admits no conformal Killing vector field, along
the integral curves of which the potential is rescaled by the same conformal
factor. While conformal rescaling is a symmetry of the supermetric, it is not a
physical operator, and thus it has limited usefulness in singling out a ground
state%
\footnote{For certain Bianchi cosmologies, the scalar constraint does admit
 causal symmetries which can be used to construct a physical inner product
 \cite{atu:I1}.}.

Apart from the lack of a general prescription for the inner product, there is
another reason that the procedure outlined by Dirac is unsuitable for \gr: it
implicitly relies on {\it canonical} coordinates, and there is no discussion of
the role, in quantum theory, of algebraic identities between non-canonical
coordinates.

Due to the difficulties, which I have discussed,  associated with quantizing
\gr, and the features peculiar to the new variables, the Dirac quantization
program has to be supplemented. We need a {\it general principle} to  select an
inner product on a complete set of physical states. This principle should not
rely on background kinematical structure, as there is none in \gr.  The new
framework must be general enough to include non-canonical choices of elementary
variables and must establish the role of any algebraic identities that the
classical variables may satisfy.  In addition, the framework should be able
to deal with  representation spaces which are not directly constructed on the
\ps.

In order to find or state such a principle, we need to work with a specific
technique to implement Dirac quantization. Several quantization schemes exist,
like geometric quantization \cite{gq1}, group theoretic quantization
\cite{gtq1} or Feynmann path integrals \cite{feyn1} which have been successful
in many cases. Path integrals for example have been tremendously useful in
perturbative approaches in field theory. However, as I mentioned earlier, these
techniques are too specific, they rely on the presence of structure on the \ps\
which may or may not be present in \gr. For example, both geometric
quantization and path integrals rely too heavily on configuration space (or at
least a polarization on \ps) to construct representations. While these
techniques have worked very well for Yang-Mills theory, they are too
restrictive for many constrained systems, and in particular for \gr. Our
interest here is to construct a general {\it framework} applicable at least in
principle to all or most constrained systems; and encompassing some of the
above techniques. The algebraic formulation of Dirac quantization \cite[see
Appendix 5]{aa:rg1}, provides a good starting framework which we can extend and
complete.

In this dissertation I will present the algebraic approach to the quantization
of constrained systems, which is broad enough to apply to \gr.

\mysection{Chronology}

A general framework is now available which establishes criteria and provides
guidelines for the canonical quantization of constrained systems. At least as
early as the Osgood Hill conference (Boston University 1987), Ashtekar
\cite[pp. 407-8]{aa:osgood1} presented the essential steps. Based on an
algebraic approach to Dirac quantization, a criterion for the choice of the
physical inner product was established, which at the same time elucidated the
role in quantum theory of the reality conditions on the new canonical variables
for \gr.

The general strategy outlined by Ashtekar --and the one I will present here--
is to choose an inner product on the space of physical states such that with
respect to it, physical operators corresponding to any two classical
observables that are complex conjugates of each other become Hermitian adjoint
operators of one another in the chosen representation. Though on the face of it
this strategy appears obvious, recall that this is {\it  not} one of the
strategies commonly employed, as the preceding discussion shows. Typically, an
inner product is found by other means, and then one checks that various
operators are Hermitian with respect to it. Here, one {\it uses} the reality
conditions on the physical operator algebra to {\it find} the inner product.

In addition, based on ideas from the quantum mechanics of a particle on a
manifold (see e.g.\ \cite{aa:cmp1} and \cite[Appendix 2]{aa:rg1}) Ashtekar
realized that the classical algebraic identities between functions on \ps\ have
to be satisfied by the corresponding operators in quantum theory, in order to
regain the correct physical sector.

The algebraic approach has now been adopted by leading researchers in the new
variables, and the essential features have been discussed briefly in recent
reviews (see \cite{carlo:review1,lee:spain1}), and in somewhat greater detail
in various lecture notes \cite{newbook1}. The criterion to select the physical
inner product has been applied successfully in the many field theories (e.g.,
Maxwell theory, 2+1 gravity and linearized gravity) which have been quantized
using the loop representation as well as the connection representation. For
full \qg\ in the new variables, the loop representation has led to a number of
major advances. Of the various techniques for quantization which are currently
in use, the algebraic approach is the only one which is complete and general
enough to encompass the loop representation.

I personally became involved with the algebraic approach when I tried
quantizing a certain model (the coupled oscillators model solved in chapter 5).
Since then, I have studied various other finite dimensional test models. Based
on the difficulties and questions that arose in the quantization of these
models, Ashtekar and I have completed the construction of a reasonably coherent
algebraic framework for the canonical quantization of constrained systems
\cite{aa:rst1}.

The focus of this framework is the construction of the kinematical quantum
theory, i.e.\ a unitary representation on physical states of a complete set of
physical operators. Within the quantization program itself there is no
discussion of either dynamics or physical interpretation. A further extension
of the algebraic framework is required to extract dynamical information and
physical interpretation from the quantum theory. Illuminating discussions,
first with Rovelli \cite{cr:pc1}, and soon after with Ashtekar \cite{aa:pc1},
clarified many of these issues for me. Some of the essential steps in a
particular approach to physical interpretation are present in Rovelli's work
\cite{cr:II1}. While my ideas --on the dynamical and physical interpretations
of canonical quantum theories of constrained systems-- are preliminary, and
possibly ill-formed, I will discuss these issues too.

\mysection{Outline}

In chapter 2 I will present the algebraic approach to quantum mechanics, paying
close attention to the two new features, namely the role of algebraic relations
in quantum theory, and the criteria for selecting an inner product on a
complete set of physical states. Initially the approach will be somewhat
heuristic, but in a later section I will also discuss various technical
details.

An alternative approach to the canonical quantization of constrained systems is
the reduced space approach (see e.g. \cite{aa:rg1,jr:rst1}). In this approach,
the constraints are first solved classically, and then the reduced system is
quantized. It is not always clear that the resulting theory is equivalent to
the Dirac quantum theory; in fact there are interesting examples where they
differ. In order to compare the results of these two approaches, I will
reformulate the reduced space approach in the algebraic framework in chapter
2.

In chapters 3-5, I will construct the kinematical quantum theories for
some finite dimensional systems. While many of the models share a number of
features in common with \gr, some models are included specifically to
illustrate the new features of the quantization program.
\begin{itemize}

\item In chapter 3, I will first consider the quantization of a (free) particle
moving on a circle. This is the simplest model in which non-trivial algebraic
identities arise between the elementary operators, and so provides a good
illustration of the absolute necessity of correctly taking these identities
into account in the quantum theory. The next example, in section 3.2, is that
of the 1-dimensional oscillator in a ``hybrid'' set of variables (one variable
is real; and the canonically conjugate variable to this one is complex). This
model illustrates the use of the Hermiticity conditions to find an inner
product. Also, the hybrid variables are the simplest analogs of the new
variables, and this example indicates that it is consistent to do quantum
theory with such hybrid variables. In section 3.3 I will quantize the Bianchi
II cosmology, which is a spatially homogeneous reduction of \gr. We will see
that even though the supermetric in the obvious polarization does not possess
enough symmetries of the required kind, a complete set of physical operators
exists, and can be used to construct an explicit physical inner product.

\item The next model, introduced by Ashtekar and Horowitz \cite{aa:gh1}, is
somewhat more complicated, as one can see from some of the previous attempts to
quantize it. The previous quantizations either displayed various unexpected
features \cite{aa:gh1,db:ag1} or were incomplete in some critical way
\cite{db:ag1}; in other quantum theories the classically unexpected behaviour
was ruled out by fiat \cite{mjg:ag1}. As we will see in chapter 4, algebraic
quantization leads finally to a quantum theory which is complete and consistent
in every way. The final quantum theory displays no strange features.

\item In chapter 5, I will consider a model in which two oscillators are
coupled to each other via a first class constraint. This is a model for a
homogeneous, isotropic relativistic cosmology, and in addition shares
qualitative features with full \gr. From the point of view of the quantization
program, this model illustrates again the role of algebraic identities on
physical operators, the importance of the Hermiticity conditions, and the
importance of discrete symmetries in pinning down a unique inner product. In
the quantum theory of this model, there is an unavoidable violation of
classical intuition.
\end{itemize}

There are many classical systems in which the dynamics is generated by a
Hamiltonian which is constrained to vanish; \gr\ in the spatially compact case,
and in particular the homogeneous cosmologies, are prime examples. The
non-relativistic parametrized particle is another such model, and in chapter 6
I will use it to illustrate the need for a framework to extract a physical
interpretation from the quantum theory. I will present two such frameworks: one
quantum mechanical, following Ashtekar \cite{aa:pc1}, Kucha\v r \cite{kk:pc1}
and Rovelli \cite{cr:pc1,cr:II1}; and another ``classical'' framework, after
Rovelli \cite{cr:pc1}. I will then apply this framework to the issue of the
initial singularity in quantum cosmology.

In chapter 7, I will discuss --in the form of a dialogue-- the lessons we have
learned about the quantization program from the various models to which it has
been applied, and some of the implications for \qg.

In the main portion of the dissertation I deal only with first class
constraints, since as is widely known, second class constraints have to be
solved classically, before the quantum theory is constructed. In Appendix A,
however, I will present a way to ``quantize and solve'' second class
constraints, using a certain ``holomorphic $\delta$-function''. The
applications of this are not immediately obvious, but some possibilities do
exist.

I will assume that the reader is familiar with the following topics: {\it i)}
symplectic formulation of classical constrained systems
\cite{gq1,newbook1,rst:cl1} {\it ii)} construction of the abstract Poisson
algebra of functions on \ps\ and the algebraic approach to quantum mechanics
\cite[(see appendices 2,4)]{aa:cmp1,aa:rg1}, and {\it iii)} elementary
representation theory of (Lie) algebras. While some knowledge of these topics
is helpful to understand details of the program, I hope that the various models
will enable even readers unfamiliar with the above topics to grasp the ideas
and concepts behind the approach. Sections and subsections marked with a
`(\dag)' may be skipped on a first reading, without much loss of continuity.

\newpage\mbox{}
\chapter{ALGEBRAIC QUANTIZATION OF CONSTRAINED SYSTEMS}
\pagestyle{myheadings}
\markboth{{\sf Chapter 2}}{{\sf Algebraic Quantization}}

\newcommand{\skhatA}{\skew4\hat{A}}
\newcommand{\skhatf}{\skew3\hat{f}}

\mysection{Introduction}

In this chapter I will present an algebraic approach to the quantization of
constrained systems which is based on the representation of an algebra of
functions on the \ps. In the section 2 I will present the details of the
program, following \cite[(see chapter 10)]{aa:osgood2,newbook2,aa:rst2}. I will
concentrate on the definitions that are used in the context of algebraic
quantization \cite{aa:rg2,aa:cmp2}, and heuristically establish the criteria to
be satisfied by the choices one has to make at various stages. In section 3, I
will summarize the quantization program, condensing it into an itemized
step-by-step outline. I will then make some technical remarks about various
steps (section 4), and postpone a general discussion about the program until
the concluding chapter, {\em after} the illustrative examples and physical
models have been solved.

Primarily, the approach is an extension and generalization of the {\em Dirac
quantization program} (or the operator constraint method), in the sense that
the operator equations corresponding to the constraints are solved in quantum
theory \cite{pamd2}. An alternate approach to the quantization of constrained
systems is to first solve the constraints classically, and then quantize the
resulting theory free of constraints. In general, the two approaches are
inequivalent (see e.g.\ \cite{aa:gh2,jr:rst2,rloll2}), and then it is of
interest to compare the resulting quantum theories. Hence, in section 5 I will
reformulate {\em reduced space quantization} \cite{aa:rg2,jr:rst2} in the
algebraic framework, and discuss possible sources of difference from the
operator constraint approach.

\mysection{Algebraic quantization}

Consider a classical system described by a {\em \ps} $\Gamma$, a real
(finite-dimensional) symplectic manifold. In many physical systems, not all
points of $\Gamma$ are accessible to the system, only those points which lie on
a {\it constraint surface} $\bar\Gamma$ --described by the vanishing of some
(constraint) functions on $\Gamma$-- are allowed classical states for the
system. Recall that in the terminology of Dirac (\cite{pamd2}, see also
\cite{rst:cl2}), the set of constraints is {\em first class} \iff\ their mutual
Poisson brackets vanish on the constraint surface. Classically, this ensures
that the ``gauge'' orbits, of the canonical transformations generated by the
constraints, lie in the constraint surface. As Dirac concluded after a careful
analysis \cite{pamd2}, points on $\bar\Gamma$ that lie on the same gauge orbit
describe equivalent physical configurations of the classical system. i.e. as
``gauge'' equivalent. Thus the ``true'' degrees of freedom of the classical
system can be represented by the space of orbits of the canonical
transformations generated by the constraints. This space, which has a natural
symplectic structure induced on it, is the {\it reduced \ps}, $\hat\Gamma$.

We are interested in a general approach to the quantization of such first class
constrained systems%
\footnote{Dirac has pointed out \cite{pamd2} that second class constraints,
 i.e.\ those  constraints not of first class, cannot be solved in quantum
theory
 (see,  however, Appendix A) and have to be eliminated classically. Therefore,
I
 will  henceforth assume that such a reduction has been carried out and we are
 left with a first class constrained system.}.
The steps in the canonical quantization of a (first-class) constrained system
are as follows:

\subsubsection*{1: Set of elementary functions on \ps}

In order to eventually construct the quantum theory, we want to start with an
optimum description of the classical system. Let us therefore consider the
vector space \cS\ generated by a suitable subset of the space of (complex
valued) functions on the \ps\ $\Gamma$. This space has to satisfy three
conditions:
\begin{enumerate}
\item In order to capture all the classical physics, this set of elementary
functions should be large enough so that any sufficiently regular function on
the phase space can be obtained as (possibly a suitable limit of) a sum of
products of elements $F^{(i)}$ in \cS. This is ensured by requiring that at
each point of the \ps\ $\Gamma$, \cS\ should contain a subset of functions
which are coordinate functions in an open neighbourhood of the point.
Technically, {\em \cS\ is (locally) {\em complete} if and only if the gradients
of the functions $F$ in \cS\ span the cotangent space of $\Gamma$ at each
point}; that is, we require that \be\label{qp:complete} {\rm Rank}\left(d_\mu
F^{(i)}\right)=2n \ee where $d_\mu F^{(i)}$ are the components in some chart of
the elementary variables generating \cS\ and the \ps\ is $2n$ dimensional.
\item In order to construct the (Poisson bracket) algebra of functions on \ps,
we require that \cS\ should be {\it closed under Poisson brackets}, i.e. for
all functions $F,G$ in \cS, their Poisson bracket $\{ F,G\}$ should also be one
of the functions in \cS.
\item Finally, \cS\ should be {\it closed under complex conjugation}, i.e. for
all $F$ in \cS, the complex conjugate $\bar{F}$ should be a function in \cS.
\end{enumerate}
Each function in \cS\ represents an elementary classical
variable which is to have an unambiguous quantum analog.

In general, \cS\ may be {\em overcomplete}, i.e., the number of generators of
\cS\ (not including the constant function, the gradient of which vanishes) may
be larger than $2n$, the dimension of \ps. In this case, the set of generators
of \cS\ will satisfy some {\it algebraic identities}. Are there any situations
where one is forced to consider such genuinely overcomplete sets? Clearly, if
the \ps\ is $\real^{2n}$, then the global Cartesian coordinates (and the
constant function) are a suitable choice for \cS. However, it is when the \ps\
is a nontrivial manifold that the point about overcompleteness becomes
important. There exist {\em no} global coordinates, and unfortunately, one does
not know how to do quantum mechanics ``in coordinate patches''. One might
attempt to construct a quantum theory on a specific coordinate patch; however,
since for e.g.\ one may not know the ``correct'' boundary conditions, in
general the quantum theory will not be well defined. Furthermore, there is no
way to ``patch together'' the quantum theories on overlapping coordinate
patches. Therefore, an {\it overcomplete} set of globally defined functions
becomes necessary. One can then think of the identities as arising from the
overlap regions between two open neighbourhoods with different sets of
coordinates.

Consider for example the case when the \ps\ is the unit 2-sphere in Euclidean
space (say centered at the origin). (This \ps\ arises in the classical
description of internal spin \cite{ms:thesis2,aa:ms2}.) There are no global
coordinates on $S^2$: the pullbacks to $S^2$ of the Cartesian coordinates
$(x,y)$ fail to be coordinates on $S^2$ at the equator $z=0$; similarly, (the
pullbacks of) $(\theta,\phi)$ are bad coordinates at the poles. However, at
each point of $S^2$ one can find a subset of $(x,y,z)$ which does coordinatize
an open neighbourhood. For example, as we just saw, in the neighbourhood of the
$z=0$ equator the set $(x,y)$ is a bad set of coordinates. However, at all
points of the $z=0$ equator (except where $y=0$), $(x,z)$ is a good set of
coordinates. Thus, the set $(x,y,z)$ is complete on $S^2$, as we can also
verify directly using (\ref{qp:complete}). But now, since $\Gamma$ is a
2-manifold, the set of {\em three} functions $(x,y,z)$ is {\it over}complete:
we know that on $S^2$ one cannot specify arbitrary values for the functions
$(1,x,y,z)$, they have to satisfy the algebraic identity $x^2+y^2+z^2-1=0$. The
vanishing of the function $x^2+y^2+z^2-1$ specifies the embedding of $S^2$ into
$\real^3$.

Thus we see that in some cases at least, the algebraic identities between
elements of \cS\ can be derived from the vanishing of the functions which
specify the embedding of the \ps\ $\Gamma$ into some $\real^{>2n}$. In the
general finite dimensional case, simple counting will usually indicate the
number of such algebraic relations that should be identified.

\subsubsection*{2: Commutator algebra of functions on \ps}

{}From the Poisson bracket relations on \cS, we want to construct the abstract
commutator algebra of quantum operators. Following \eg\ \cite[(appendices
4,5)]{aa:rg2} and \cite{aa:cmp2}, associate with each element $F$ in \cS\ an
abstract operator $\hat{F}$. Construct the free associative algebra generated
by these elementary quantum operators. Impose on this algebra the {\it
canonical commutation relations} (CCRs): for all $F$ and $G$ in the set \cS,
impose
 \be\label{qp:ccr}
  [\hat{F},\hat{G}]:=\hat{F}\cdot\hat{G}-\hat{G}\cdot\hat{F}
          =i\hbar\wh{\{F,G\}},
 \ee
where $\wh{\{F,G\}}$ denotes the operator corresponding to the function
$\{F,G\}$. If \cS\ is overcomplete, then as we just saw there are algebraic
relations between its elements. Thus, also impose the {\it anti-commutation
relations} (ACRs) which capture the algebraic identities between the elementary
classical variables: e.g., if all three functions $F,G$ and $H:=F\cdot G$
belong to \cS, then in the algebra impose the condition
 \be\label{qp:acr}
  \hat{F}\cdot\hat{G}+\hat{G}\cdot\hat{F}=2\hat{H}.
 \ee
Denote the resulting algebra by \cA.

\subsubsection*{3: $\star$-relations on the algebra}

On \cA, introduce an involution operation%
\footnote{An involution is a map from an algebra to itself, satisfying
 the following properties. If $\skhatA,\hat{B}$ are in the algebra and
$\lambda$
 is a complex number: {\it i)} $(\skhatA + \lambda\hat{B})^\star =
\skhatA^\star
 +\bar{\lambda}\hat{B}^\star$; {\it ii)} $(\skhatA\hat{B})^\star =
\hat{B}^\star
 \skhatA^\star$; and {\it iii)} $(\skhatA^\star)^\star=\skhatA$.}
denoted by $\star$, by requiring that if two elementary classical variables $F$
and $G$ are related by $\bar{F} = G$ (where $\bar{F}$ denotes the complex
conjugate of $F$), then $\hat{F}{}^\star\equiv\hat{G}$ in \cA. The properties
of the involution operation then define $\star$ on the rest of the algebra.
Denote the resulting $\star$-algebra by \cAs. The $\star$-relation on \cAs\
reflects just the complex conjugation relations between the functions in \cS.

\subsubsection*{4: Linear representation of \cA}

Ignore for now the $\star$ relation we just constructed on \cA.

In order to construct the quantum theory, we want a representation of the
abstract algebra \cA\ via linear operators on some complex vector space \cV. As
in standard quantum mechanics, the canonical commutation relations
(\ref{qp:ccr}) should be faithfully represented. In addition, the
representation should be such that the anticommutation relations (\ref{qp:acr})
should also be satisfied. Note that for now the $\star$-relations on \cAs\ are
to be ignored.

\subsubsection*{5: Physical states}

Up to this point, the constraints of the classical theory have been ignored. We
now want to solve the constraints and isolate the {\it physical states} of the
quantum theory. Thus we have to obtain explicit operators $\hat{C}$ on \cV,
representing the quantum constraints. In the chosen representation, if e.g. the
constraints are quadratic or higher order in momenta, one has to suitably
factor order them: while the functions in the set \cS\ have unambiguous quantum
analogs, operators corresponding to products of the elementary ones will in
general {\it not} have unambiguous quantum analogs. The space of physical
states \cVp\ is the kernel of the constraint operators, i.e.\ the set of states
annihilated by the constraints. Thus $\ket\psi$ is a physical state \iff\
 \be \label{qp:qcon}
  \hat{C}\ket\psi=0.
 \ee
This is the quantum constraint equation which has to be {\it solved} to yield
the physical states.

\subsubsection*{6: Physical operator algebra}

Recall that in classical theory, the allowed physical states of the system
(solutions to the constraint) are points that lie in the constraint surface
$\bar\Gamma$. As Dirac noted, physically interesting functions are ones which
generate \ct s which leave invariant $\bar\Gamma$, the space of classical
physical states. These functions are the classical Dirac observables.
Similarly, in the quantum theory, we are interested only in those operators
which leave invariant \cVp, the space of quantum physical states. Not all
operators in \cA\ are of this type. An operator $\skhatA$ in \cA\ will leave
\cVp\ invariant if and only if $\skhatA$ commutes weakly with the constraints,
i.e.
 \be\label{qp:physop}
  [\skhatA, \hat{C}_I] = \sum_J \skhatf_I^J \hat{C}_J.
 \ee
Note that $\skhatf^I_J$ are operator valued ``structure functions'', i.e.\ they
may not all be proportional to the identity operator. Furthermore, it is
important that various factor-ordering choices be made such that on the RHS of
the above equation, the constraint operators always stand to the right. This
guarantees that $\hat{C}_I(\skhatA\ket\psi)=0$ for all physical states
$\ket\psi$, i.e.\ that $\skhatA\ket\psi$ is a physical state if $\ket\psi$ is
one%
\footnote{Since we are  interested only in physical states, I will henceforth
 drop the qualifier  ``weakly'' when applied to the vanishing of the commutator
 of physical operators with the  constraints.}.
The collection of all such operators forms a sub-algebra of \cA, the algebra of
physical observables. We denote it by \cAp.

{}From the $\star$-relation on \cAs, we {\it induce an involution} (denoted
again
by $\star$) on the physical algebra \cAp\ (see the final remark in section 4).
Denote the resulting $\star$-algebra of physical operators by \cAps.

\subsubsection*{7: Physical inner product}

In order to answer or even pose physically relevant questions in the quantum
theory (say about various transition amplitudes, expectation values or spectra
of interesting physical observables), we need an inner product on the space of
physical states. The general idea is to fix this inner product by requiring
that various operators are Hermitian \wrt\ it. However, since arbitrary
operators in \cA\ are not operators on the space of physical states, we cannot
require that they satisfy Hermiticity conditions \wrt\ an inner product on
\cVp. On the other hand, operators in \cAp\ {\em are} operators on physical
states. Hence, we fix the physical inner product in the following manner:
\begin{quote} {\it Introduce on \cVp\ a Hermitian inner product so that the
abstract $\star$-relations on \cAps\ --which have been ignored so far-- are
represented as Hermitian adjoint relations on the resulting Hilbert space.}
\end{quote}
In other words, if $\hat{F}^\star=\hat{G}$ in the abstract algebra \cAps, then
the inner product on physical states should be chosen such that the
corresponding explicit operators in the representation satisfy
$\hat{F}^\dagger=\hat{G}$, where $\dagger$ is the Hermitian adjoint \wrt\ the
physical inner product. In the Dirac bra-ket notation: \be
\bra\psi\hat{G}\ket\phi = \bra\psi\hat{F}^\dagger\ket\phi :=
\ovr{\bra\phi\hat{F}\ket\psi} \quad \forall \psi,\phi\,\in\cVp. \ee This is a
key step in the algebraic quantization of constrained systems, which
distinguishes it from previous approaches. It establishes a clear criterion for
the selection of an inner product on physical states (based on the algebra of
physical observables), and thus completes the Dirac program. Note that unlike
in group theoretic quantization \cite{gtq2}, no group theoretic or symmetry
considerations are explicitly involved here. Neither do we invoke specific
techniques to find the inner product, such as K\"ahler structures in geometric
quantization \cite{gq2}.

\mysection{Summary of the program}

Now that we have considered the program in detail, and have also provided the
required definitions and established various criteria to guide our choices at
the various stages, I will summarize the steps in the algebraic approach to the
quantization of constrained systems. This summary can serve as a ready
reference for the reader through the following chapters.

\begin{enumerate}
\item Choose a complete set of (complex valued) functions on $\Gamma$ such that
the vector space they generate is closed under both Poisson brackets and
complex conjugation. Identify all algebraic relations that may exist between
its elements.
\item Construct the associative algebra \cA\ generated by the elementary
variables in \cS, with the canonical commutation relations and the
anticommutation relations imposed.
\item On this algebra, identify the $\star$-involution, corresponding to
the classical complex conjugate relations on \cS.
\item Find a linear representation of \cA\ on a vector space \cV. (Ignore
the $\star$-relations in this and the next step.)
\item Solve the constraint equations and obtain the space of physical
states, \cVp.
\item Construct the $\star$-algebra \cAps\ of physical observables.
\item Introduce on \cVp\ a Hermitian inner product so that the abstract
$\star$-relations on \cAps\ are represented as Hermitian adjoint relations on
the resulting Hilbert space.
\end{enumerate}

\mysection{Remarks}
\remarks

\remark{: Inputs to the program}

At the outset I should point out that the above ``program'' is {\em not} a
constructive technique to quantize constrained systems. As a leading researcher
in the field has so eloquently stated \cite{anon2}, {\sl ``This program is not
a
meat-grinder, into one end of which we can stick a classical system, turn the
crank, and then expect to get a quantum sausage out the other end.''} When
following this program, there are various stages at which {\em choices} have to
be made. The first crucial choice is that of the set of elementary functions
\cS. A poor choice of \cS\ may make it difficult to either represent the
constraints, or solve them, or find observables. While there is no prescription
to make a good choice, (which after all depends largely on the intuition one
has for the problem) minimum requirements that \cS\ should satisfy have been
established: These requirements are essential in order that the quantum theory
capture all the physics.

The second crucial choice is that of the representation of \cA\ on a complex
vector space \cV. A poor choice of representation may again make it difficult
to solve the constraints. If the representation is not ``general enough'' in
some loose sense, it may be impossible to find a physical inner product. Again,
there is no prescription to avoid such bad choices. However, once a
representation has been chosen and the constraints solved, the program
specifies the criteria that should be used to select a physical inner product.

\remark{: Algebraic identities and quantum theory}

There may exist algebraic identities between the elementary functions which are
of a more general form than the simple quadratic case we considered in step 2.
There is a generalization of (\ref{qp:acr}). Consider the case when there are
$N$ sets of up to $m$ elementary functions, the $(i)$th set denoted by
$(F^{(i)}_1,\dots,F^{(i)}_m)$, which satisfy the identity $\sum_{(i)}
F^{(i)}_1\cdot F^{(i)}_2\cdots F^{(i)}_m=0$. On the commutator algebra
constructed from \cS, one should impose
 \be\label{qp:acr2}
  \hat{\cal F}:=
   \sum_{(i)}\hat{F}^{(i)}_{(1}\cdot\hat{F}^{(i)}_2\cdots\hat{F}^{(i)}_{m)} =
0,
 \ee
where as usual $(1...m)$ denotes $1/m!$ times the sum of all the permutations.

Note that even though we are imposing anti-commutation relations, there is
nothing fermionic about the system. Further, I should emphasize that the
purpose of (\ref{qp:acr}) is {\it not} to resolve any factor ordering ambiguity
in the definition of the operator $\widehat{FG}\equiv\hat{H}$: since $H=FG$ is,
by assumption, an elementary variable, $\hat{H}$ is already an unambiguous and
well-defined quantum operator. Rather, the role of the ACRs is to eliminate
possible spurious sectors in the final quantum description. How do these
spurious sectors arise? In simple cases, the left hand sides of the ACRs (e.g.\
the operator $\hat{\cal F}$ in (\ref{qp:acr2})) are {\it superselected}
operators in the associative algebra, i.e., they commute with all the
elementary operators%
\footnote{More generally, they constitute a Lie ideal in the commutator
 algebra. Thus, technically, the imposition of the ACRs amounts to taking the
 quotient of the commutator algebra by this ideal. The result is the required
 algebra of quantum operators. I do not know the {\em most} general statement
 about the role of the ACRs in the algebra.}.
Thus, in these cases, had we not imposed the ACRs, the representation of \cA\
would be reducible, the irreducible representations would be carried by the
eigenspaces of the operators $\hat{\cal F}$. We would obtain the correct
classical limit {\it only} in the sector on which $\hat{\cal F}$ takes the
value zero. By imposing (\ref{qp:acr2}), we ensure that the spurious sectors
are avoided.

The algebraic identities arise due to a ``failure of coordinatization'' of the
\ps. As we saw in the $S^2$ example considered earlier, we cannot classically
solve the algebraic identity for and eliminate one of the elementary functions
{\em globally} on the \ps. The identity then has to be appropriately
incorporated in quantum theory.

Given a classical identity $F\cdot G=H$, is there any ambiguity in the quantum
condition that we wish to impose? As we will see, rather than {\em solving} a
factor-ordering problem, as one might have initially thought, the algebraic
identities {\em create} one. {\it A priori}, in the quantum theory, one does
not know what condition to impose. Suspend for a moment the ACR (\ref{qp:acr})
and consider instead a nonsymmetric condition of the form
 \be
  \alpha\hat{F}\cdot\hat{G}+\beta\hat{G}\cdot\hat{F}=(\alpha+\beta)\hat{H}.
 \ee
If we define $\delta=i\frac{\alpha-\beta}{\alpha+\beta}$ and re-arrange terms,
we can write this as
 \be\label{qp:acramb}
  \hat{F}\cdot\hat{G}+\hat{G}\cdot\hat{F}+\delta\hbar\wh{\{F,G\} } =2\hat{H}.
 \ee
Since in the limit $\hbar\rightarrow0$ the extra term on the LHS vanishes,
independent of the value of $\delta$ we obtain the correct classical limit.
Hence, there is a genuine ambiguity in the condition we wish to impose in
quantum theory. In ordinary quantum mechanics, the factor ordering ambiguity is
often resolved by invoking the Hermiticity of the new operator. In the
algebraic framework, we would similarly require that the quantum condition
corresponding to the classical identity be compatible with the
$\star$-relation. Assuming in the simplest case that $F,G,H$ are real
functions, this implies only that $\delta\in\real$.

In the quantization program we spelled out in section 2, we made the specific
symmetric choice $\delta=0$ (and hence refer to the quantum condition as an
anticommutation relation). The justification for this comes from ordinary
quantum mechanics on a manifold (see \cite{aa:cmp2} or \cite[appendix
C]{newbook2}), where the representation {\em identically} satisfies the quantum
condition in the specific, symmetrized form (\ref{qp:acr}). One could work with
a family of algebras, parametrized by as many $\delta$s as there are algebraic
identities. However, it is much simpler to work with a specific choice,
meanwhile retaining the existence of ambiguity in the back of one's mind, in
case e.g.\ there are subtle topological issues involved, and one wants to relax
the condition $\delta=0$. Henceforth, in this thesis, I will always work with
the symmetric choice.

\remark{: Inner product}

Note that for constrained systems the above quantization program neither
requires nor introduces an inner product on the large representation space \cV.
(In step {\bf4}, the $\star$-relations are ignored.) From a physical point of
view, we need to introduce a Hermitian inner-product {\it only} on the physical
subspace. Indeed, as simple examples show, physical states in \cVp\ are often
{\it not} normalizable in the fiducial inner product one may be tempted to
introduce on \cV. For example, if one considers a constraint linear in momenta,
defined by a vector field on configuration space, physical states are constant
along the vector field. If the vector field is, say, a translational vector
field in $\real^3$, then physical states do not fall-off in those directions
(unless the states are identically zero) and then, \wrt\ the usual Euclidean
measure, they are not normalizable. On the other hand, consider a constraint
which is a function only of the configuration variables. This defines a surface
in the configuration space. In the configuration representation, physical
states have support only on the sub-manifold defined by the constraint in
configuration space. Thus, they are either of zero norm \wrt\ the usual
Euclidean measure, or, they are all ``normalizable'' to Dirac
$\delta$-distributions, in which case the quantum theory is not well-defined
unless one introduces a {\em new} inner product.

Hence, for constrained systems, it seems inappropriate to attempt to introduce
a preferred inner product on all of \cV\ (except perhaps as a technical device,
e.g., for regularization of the constraint operators in field theories).

\remark{: Quantum observables}

Quantum mechanically, the true degrees of freedom of the constrained system are
coded in \cAp, the algebra of observables. Recall that in classical theory
there is gauge freedom on the constraint surface: not all points on
$\bar\Gamma$ are physically distinguishable. A somewhat similar situation
occurs in the quantum theory too. To see this, let us explore the space of
physical operators. An operator $\skhatA$ in \cA\ will leave \cVp\ invariant if
and only if it commutes with the constraints (see (\ref{qp:physop})). Let
$\skhatA$ and $\hat{B}$ be physical operators: $[\skhatA,\hat{C}]=\skhatf_A
\hat{C}$ and $[\hat{B},\hat{C}]=\skhatf_B \hat{C}$. Then
 \bea
  [\skhatA\hat{B},\hat{C}]&=&
\skhatA\skhatf_B\hat{C}+\skhatf_A\hat{C}\hat{B}\cr
   &=& \skhatA\skhatf_B\hat{C} + \skhatf_A(\hat{B}\hat{C} -
\skhatf_B\hat{C})\cr
   &=& (\skhatA\skhatf_B + \skhatf_A\hat{B} - \skhatf_A\skhatf_B)\hat{C} .
 \eea
Thus, $\skhatA\hat{B}$ also commutes weakly with the constraint: the physical
operators constitute an associative sub-algebra of \cA, with the induced
commutator bracket. Denote this sub-algebra by \cAp$'$.

Since the constraints are by assumption first class, the constraint operators
belong to \cAp$'$ too. This represents the gauge freedom quantum mechanically.
However, since constraints annihilate all physical states, the algebra \cAp\ of
physical observables can be obtained ``by setting the constraint operators to
zero'' in \cAp$'$. More precisely: one constructs the ideal ${\cal I}_C$ of
\cAp$'$ generated by the constraints%
\footnote{In general an ideal in an associative algebra generated by a set $\{
 \hat{C}\}$ consists of elements of the form $\skhatA\hat{C}\hat{B}$ for all
 $\skhatA,\hat{B}$ in the algebra. However, due to (\ref{qp:physop}), in
 \cAp$'$ the ideal can equivalently be considered to consist of elements of the
 form $\skhatA\hat{C}$ for all $\skhatA\in\cAp'$.}.
One then takes the quotient of \cAp$'$ by this ideal: $\cAp:=\cAp'/ {\cal
I}_C$. For physical operators quadratic or higher order in momenta, one first
makes a factor-ordering choice and then quotients by ${\cal I}_C$.

One may be tempted to shorten the procedure and quotient by the ideal generated
by the constraints in the ``large'' algebra \cA\ itself. However, it is easy to
see that this ideal is too large. In simple cases, it is the whole algebra.
Consider the constraint $\hat{p}=0$. Since $\hat{q}$ is in \cA\ and
$[\hat{q},\hat{p}]=\hat{1}$, $\hat{1}$ must belong to the ideal ${\cal I}_p$.
But now, since $[\skhatA,\hat{1}]=\skhatA$ for all $\skhatA\in\cA$, ${\cal
I}_p=\cA$. Thus the quotient is just the identity operator, and all physical
information has been lost.

How do we know that the set of physical operators captures {\it all} the
physical degrees of freedom? To answer this question, one has no recourse but
to resort to the classical theory. Consider the classical functions (classical
Dirac observables) corresponding to the generators of \cAp. These are functions
on the constraint surface $\bar\Gamma$ which are Lie-derived along the orbits
generated by the constraints, and are thus isomorphic to functions on the
reduced \ps\ $\hat\Gamma$. Since $\hat\Gamma$ represents the true classical
degrees of freedom, we can require that the set of functions corresponding to
\cAp\ should be complete on $\hat{\Gamma}$. As in the definition of
completeness of \cS\ (see (\ref{qp:complete})), this means that the gradients
of the functions on $\hat\Gamma$ (corresponding to the generators of the
physical operator algebra) should span the cotangent space of the {\it reduced}
\ps, at every point on $\hat\Gamma$.

However, in the Dirac theory, one is not necessarily interested in constructing
the reduced \ps; indeed, it may be quite difficult to do so. In the absence of
$\hat{\Gamma}$, we would like to formulate the condition for completeness on
the constraint surface itself. First consider the simplest case when the
constraint surface is of codimension 1, i.e. there is only one first class
constraint. On $\bar\Gamma$, the gradients of the Dirac observables are
orthogonal to the Hamiltonian vector field of the constraint function. Hence,
the pull-backs of the gradients of the observables to $\bar\Gamma$ span a
$2n-2$ dimensional subspace of the cotangent space to $\bar\Gamma$. In
practice, it may not be possible to coordinatize the constraint surface, and
thus one may not be able to evaluate the pull-backs in a straightforward
fashion. Thus, a more practical strategy is to explicitly include the
constraint function in the set (thus the set corresponds to the set of
generators of \cAp$'$, as opposed to \cAp), and then require that at each point
of $\bar\Gamma$ the gradients of this set span a $2n-1$ dimensional space,
which is automatically orthogonal to the gauge vector fields. Since the
pull-back of the gradient of the constraint (which is normal to the constraint
surface) vanishes, this condition is clearly equivalent to the previous one. In
the general case, when the constraint surface is of codimension $m$, we require
that
 \be\label{qp:rank}
  {\rm Rank}\left.\left(d_\mu O^{(i)}\right)\right|_{\bar\Gamma} =2n-m,
 \ee
where ($\mu=1...2n$) $d_\mu O^{(i)}$ are the $2n$ components of the gradients
of the functions corresponding to the generators of \cAp$'$, labelled by $(i)$.

This condition is perhaps easier to visualize in terms of the Hamiltonian
vector fields of the observables: we require them to span the tangent space to
$\bar\Gamma$. However, since there is no natural subspace of the tangent space
{\em orthogonal} to the gauge vector fields, one has to explicitly include the
gauge vector fields. Since the symplectic structure is nondegenerate, it is
easy to see that the condition (\ref{qp:rank}) is equivalent to
 \be\label{qp:rank2}
  {\rm Rank}\left.\left( X^\mu_{O_{(i)}}\right)\right|_{\bar\Gamma} =2n-m,
 \ee
where $X^\mu_{O_{(i)}}$ is the Hamiltonian vector field of the function
corresponding to the $(i)$th generator of \cAp$'$.

As in the case of $\Gamma$ itself, if the generators of \cAp\ are overcomplete,
or if one is unable to properly factor out the constraints classically, then
there are algebraic relations between the physical observables.

\remark{: Physical states}

Since there is no inner product on the representation space \cV, how does one
know in step 5 that one has a ``sufficiently large'' set of solutions to the
constraints? We now have the machinery to formulate a criterion: {\it the space
of physical states \cVp\ should be large enough to carry a faithful
representation of \cAp}. This is already an indication that the program cannot
always be implemented as sequentially as it has been presented, since for some
systems one may have to find the full physical algebra and then go back and
check that \cVp\ is large enough.

\remark{: $\star$-involution on \cAp}

In step 7, we find the physical inner product using the $\star$-relations on
\cAps, and it is necessary and sufficient to implement the Hermiticity
conditions on only a set of generators of \cAp.

Now, if $\skhatA$ in \cA\ belongs to \cAp$'$, its $\star$-adjoint in \cA,
$\skhatA^\star$, may not belong to \cAp$'$. Hence, in general the
$\star$-relation on \cA\ does {\it not} induce an involution on \cAp. In this
case, no prescription is available to select the physical inner product. If, on
the other hand, $\skhatA^\star\in\hbox{\cAp}\>\forall\> \skhatA\in\hbox{\cAp},$
then we do obtain an involution on \cAp\ (denoted again by $\star$) and hence a
physical $\star$-algebra \cAps, which can now be used to select the inner
product. Are there physical systems for which the above condition is guaranteed
to be satisfied? The answer is in the affirmative. Consider, in particular, the
case when the constraint operators satisfy: $\hat{C}_I{}^\star = \hat{C}_I,
\>\forall I$. Now, if a physical operator $\skhatA$ commutes strongly with the
constraints, i.e. if $[\skhatA, \hat{C}_I] =0$, then $\skhatA{}^\star$ is also
a physical operator. We will see that this situation occurs in a number of
model systems.  In these cases, the quantization program can be completed
successfully. Note however, that it is {\it not essential} that the operators
commute strongly with the constraints for the $\star$-relations to be
well-defined on the physical algebra. For example, if a set of generators of
\cAp\ are their own $\star$-adjoints (i.e. if they correspond to {\it real}
functions on the \ps), the $\star$-relations on \cA\ induce an unambiguous
$\star$-relation on \cAp.

\mysection{(\dag) Reduced space quantization}

As I mentioned in the introduction, there is an alternative approach to the
quantization of (first class) constrained systems. In {\it reduced space
quantization} one first solves the constraints classically and restricts
attention to the allowed states, which lie in the constraint surface
$\bar\Gamma$. Recall that since the constraint surface is first class, not all
points in it are physically distinguishable: points related to each other by a
\ct\ generated by the constraints are {\em gauge equivalent}. Thus one has to
find the orbits of the \ct s, by integrating the gauge flats spanned by the
Hamiltonian vector fields of the constraints. The space of gauge orbits
$\hat\Gamma$ represents classically the true degrees of freedom of the system.

Now one can construct a quantum theory on $\hat\Gamma$. Namely, as in step 1,
introduce a subspace \hcas\ of the space of complex-valued functions on
$\hat\Gamma$: \hcas\ is overcomplete on $\hat\Gamma$, and closed under Poisson
brackets and complex conjugation. The variables in \hcas\ will have unambiguous
quantum analogs. As in steps 2 and 3 above, one now constructs the abstract
$\star$-algebra \hcas. Since there are no longer any constraints to worry
about, the final step here is to find a linear representation of the above
algebra, and to select an inner product such that the $\star$-relations on
\hcas\ are represented by Hermitian adjoint relations in the representation.

One is naturally led to ask for the relationship between the two quantum
theories. In general this is a difficult problem, complicated by the fact that
the answer is not unique and depends on precisely how the question is framed.
However, the algebraic approach to quantization is a natural arena in which to
consider this issue, and it does provide some answers. Since the generators of
the physical operator algebra \cAp\ correspond to a complete set of functions
on $\hat\Gamma$ (e.g.\ the generators of \hcas) one is naturally led to expect
that there is some correspondence between \cAp\ and \hca. If there is a
complete correspondence between the {\it generators} of \cAp\ and \hca, then
one expects at least the ``local'' kinematical structure of the two quantum
theories to be identical.

There are, however, a number of potential ambiguities. First,  while there is a
correspondence at the classical level, there is no guarantee that such a
correspondence exists at the level of the abstract operator algebras. Remember
that the operators in \cAp\ are obtained by first factor-ordering in \cAp$'$
and then quotienting by the constraints. Thus there is a {\em factor-ordering
ambiguity}. Second, while the overcompleteness of \cAp\ is {\it necessary}, it
is {\it not sufficient} to pick out a unique inner product in all cases,
precisely because it is a local condition on $\bar\Gamma$ and not sensitive to
any {\it large} \ct s on the constraint surface. For most simple systems the
generators of \cAp\ and \hca\ may still be in $1-1$ correspondence with each
other inspite of the above observation. However,  there do exist
counter-examples. Some of the systems we will examine admit {\it discrete
physical operators} (corresponding to large canonical transformations on
$\bar\Gamma$) which have {\it no action} on $\hat\Gamma$ whatsoever. Such
``superselected'' discrete operators belong to \cAp\ and not to \hca.
Therefore, even kinematically, there is no guarantee that the two quantization
procedures yield equivalent results.

Are there {\it any} situations in which the two quantum theories are {\em
equivalent}? Consider the case when $\Gamma$ has the structure of a cotangent
bundle. Then, if {\it all} the generators of the observable algebra are linear
or independent of momenta, there is no factor ordering ambiguity and the
theories are kinematically equivalent: there is a unitary transformation
between the two Hilbert spaces which preserves the action of the {\em
generators} of the two observable algebras \hca\ and \cAp. However, interesting
dynamical variables, such as the Hamiltonian, may be quadratic or higher order
in momenta. For these, in general there will be a factor-ordering ambiguity,
and depending on how it is resolved the two methods may lead to {\em
dynamically inequivalent} theories. In fact, in a number of examples,  the
prescriptions for factor-ordering which are natural to each approach do lead to
such an inequivalence \cite{jr:rst2}.

\newpage\mbox{}
\chapter{SIMPLE EXAMPLES}
\pagestyle{myheadings}
\markboth{{\sf Chapter 3}}{{\sf Simple Examples}}

\def\hq{\hat{q}}
\def\hp{\hat{p}}
\def\hz{\hat{z}}


\def\mss{minisuperspace}
\def\mmss{mini-\mss}

\def\bfb{{\rm B}}

\def\Mscr{{\cal M}}
\def\cone{\hbox{${\cal L}^-$}}
\def\halfcone{\hbox{${\cal L}^-_R$}}
\def\halfconel{\hbox{${\cal L}^-_L$}}
\def\halfplane{\hbox{$\Sigma_R$}}

\newcommand{\bb}{\bar{\beta}{}}
\newcommand{\bbz}{\bar{\beta}{}^0}
\newcommand{\bbp}{\bar{\beta}{}^+}    \newcommand{\bbm}{\bar{\beta}{}^-}
\newcommand{\bpz}{\bar{\pi}_0}
\newcommand{\bpp}{\bar{\pi}_+}          \newcommand{\bpm}{\bar{\pi}_-}

\newcommand{\tb}{\tilde{\beta}{}}     \newcommand{\tbz}{\tilde{\beta}{}^0}
\newcommand{\tbp}{\tilde{\beta}{}^+}  \newcommand{\tbm}{\tilde{\beta}{}^-}
\newcommand{\tp}{\tilde{\pi}{}}       \newcommand{\tpz}{\tilde{\pi}{}_0}
\newcommand{\tpp}{\tilde{\pi}{}_+}    \newcommand{\tpm}{\tilde{\pi}{}_-}

\newcommand{\hQ}{\hat{Q}}                 \newcommand{\hP}{\hat{P}}
\newcommand{\dP}{{\rm I\! P}}     
\newcommand{\dB}{{\rm I\! B}}     

\newcommand{\rp}{{\rm p}}     
\newcommand{\rpz}{{\rm p}^0}
\newcommand{\rpp}{{\rm p}^+}                     \newcommand{\rpm}{{\rm p}^-}
\newcommand{\rb}{{\rm b}}     
\newcommand{\rbp}{{\rm b}_+}                     \newcommand{\rbm}{{\rm b}_-}

\newcommand{\rB}{{\rm B}}         
\newcommand{\hB}{{\hat{B}}}
\newcommand{\hA}{{\skew4\hat{A}}}

\mysection{Introduction}

The algebraic approach to quantization that I presented in the previous
chapter has several new features. In this chapter I will quantize some simple
models, each of which allows us to focus on and understand one of these new
features, in a familiar setting.

In step 1 of the quantization program, one has to select a complete set of
functions on the \ps. As I explained in the previous chapter, one is forced to
consider {\it over}complete sets --i.e., sets in which there are more functions
than the number of dimensions of the \ps-- if the \ps\ is a non-trivial
manifold, which does not admit global coordinates. This overcompleteness leads
to certain algebraic identities between the elementary functions, which cannot
be solved globally on the \ps. The algebraic relations are incorporated in
quantum theory, via (2.2.3). In section 1 I will consider the simplest
non-trivial manifold, $S^1$, as the configuration space for a particle, and
explicitly find the algebraic relation. We will see that the algebraic relation
has to be appropriately incorporated in the quantum theory, in order to
eliminate unphysical sectors.

Another new feature is the use of the Hermiticity conditions on observables to
{\it select} an inner product on physical states. As I discussed in chapter 1,
this extension of Dirac quantization was motivated by the hybrid nature of the
new variables for \gr\ (see e.g. \cite{newbook3}), where one variable is real
and the other is complex. The canonically conjugate variables are a real triad,
$E$, and a complex connection, $A$, which satisfies the reality condition
 \be\label{ex:rc}
  \bar{A}=2\Gamma(E)-A.
 \ee
Therefore, an immediate question that had arisen was whether the connection
representation is consistent: can the momentum conjugate to a complex variable
be itself Hermitian? (The gravitational \ps\ is constrained, and the new
canonical variables are not observables; however, the above question is about
{\em mathematical} consistency, and hence still important.) In section 2 we
will consider the harmonic oscillator in hybrid variables analogous to the new
variables for \gr. I will quantize the oscillator in these variables (following
\cite[(chapter 10)]{newbook3}), and we will  find that implementing the
Hermiticity conditions leads to a quantum theory unitarily equivalent to the
usual one. Thus there is no {\it a priori} obstruction to using hybrid
variables.

In section 3 I will apply the approach to a more physically motivated problem:
the quantization of the Bianchi type II cosmology. This is a constrained
system, and is exactly soluble. In the usual metric variables, a set of
solutions to the quantum constraint has long been available \cite[see Table
III.1, pp. 91]{mpr3}. However, to my knowledge there has not been a full
discussion of the completeness of the set of solutions, a physical inner
product or the Dirac observables. I believe that at least in part this is due
to a poor choice of the elementary variables. In terms of certain new
geometrodynamical variables introduced by Uggla \cite{cu:pc3,atu:II3}, which
are adapted to the symmetries of the problem, and using the quantization
program, one can find the physical inner product and the above issues can be
understood. This example illustrates  the importance of a good choice of
elementary variables and subsequent representation. Since this is a constrained
system, one can compare the Dirac and reduced space quantum theories. As it
turns out, in the representation we have chosen, the Dirac quantum theory is
kinematically equivalent to the reduced space theory.

\mysection{Particle on a circle}

Consider a particle whose configuration space is the unit circle $S^1$ in the
Euclidean plane. The \ps\ is just the cotangent bundle over $S^1$, and thus the
symplectic structure is given by $\Omega=d p\wedge d\theta$, where $p$ is
the momentum variable. Note that even though $\theta$ is {\em not} a
well-defined function on $S^1$, the 1-form $d\theta$ is globally defined, and
so is the symplectic structure.

Since $\Gamma=T^\ast S^1$ is a nontrivial manifold, there is no global chart.
However, one can choose the following overcomplete set of elementary functions:
$\cS=(1, q_1:=\sin\theta, q_2:=\cos\theta, p)$. The commutation relations
between the corresponding operators are:
 \be \label{s1:ccr}
  [\hq_1,\hp]=i\hq_2,\quad [\hq_2,\hp]=-i\hq_1\quad\hbox{and}\quad
   [\hq_1,\hq_2]=0.
 \ee
Note that these are the commutation relations of the 2-dimensional Euclidean
group. Classically, the functions $q_1$ and $q_2$ satisfy $q_1^2+q_2^2=1$. Due
to the nontriviality of the manifold, this equation cannot be solved globally
for one variable. We impose the ACR:
 \be \label{s1:acr}
  (\hq_1)^2+(\hq_2)^2-\hat{1} =0
 \ee
on the quantum operators. Note that the operator on the LHS, which we wish to
set to zero, commutes with all the generators of the algebra, and hence with
any Hamiltonian that one may construct. $(\hq_1)^2+(\hq_2)^2$ is the Casimir
invariant of the above algebra, and the ACR requires that we restrict ourselves
to the sector on which its value is 1.

Let us choose as our representation space the space of (smooth) functions
$\psi(\theta)$ on a circle of radius $r$. A representation of the commutation
relations (\ref{s1:ccr}) is given by:
 \bea  \label{s1:rep}
  \hq_1\circ\psi(\theta)&=& r\sin(\theta)\psi(\theta)\cr
  \hq_2\circ\psi(\theta)&=& r\cos(\theta)\psi(\theta)\cr
  \hp\circ\psi(\theta)  &=& \dfrac\hbar{i}\dfrac\d{\d\theta}\psi(\theta).
 \eea
Note again that while the ``coordinate'' $\theta$ is bad, the vector field
$\d/\d\theta$ is well-defined. However this representation is still manifestly
wrong, e.g. the spectrum of $\hq_1$ lies between $\pm r$, while classically
$|q_1|\le 1$. In order to obtain a quantum theory with the correct classical
limit, we have to satisfy the algebraic relation, and choose the $r=1$
representation only. Thus, while the commutator algebra is correctly
represented on functions on {\it any} $S^1$, the correct quantum theory
corresponds to a representation only on the functions on the {\it unit} $S^1$.
Finally, the inner product is simply $\IP\psi\phi=\int_{S^1}d\theta\,\bar\psi
\phi$.

One could have represented the commutator algebra on the space of functions on
the two plane. While this representation contains the physical one, it also
contains infinitely many spurious sectors, unless the algebraic condition is
also imposed. The representation is reducible, and since the LHS of the
algebraic relation (\ref{s1:acr}) is a superselected operator, its eigenspaces
carry the irreducible representations. Note that this representation points out
the importance of imposing the algebraic condition {\em before} selecting an
inner product. For, suppose that we assume an inner product of the form
$\IP\psi\phi=\int_{\real^2}\mu(r,\theta)\bar\psi\phi$. Then the Hermiticity
conditions on $\hp$ imply that $\d\mu/\d\theta=0$ \ie, $\mu=\mu(r)$. Now,
however, to regain the physical sector, the support of the wavefunctions has to
be restricted to $r=1$, and the corresponding states are not normalizable,
unless we choose $\mu$ to be a {\it distribution}, $\mu=\delta(r-1)$.

\subsection{(\dag) Fractional statistics on $S^1$}

In group theoretic quantization \cite{gtq3}, one constructs (a unitary
representation of) a certain group with a {\it global} action on the \ps\
$\Gamma$. The global properties of $\Gamma$ are built into the procedure right
from the start. In algebraic quantization, on the other hand, one represents
the Poisson bracket algebra of a set of functions on $\Gamma$. This algebra
``knows'' only the (local) symplectic geometry of $\Gamma$, and {\it a priori},
is not always sensitive to the global properties of $\Gamma$. Thus, even after
imposing the correct ACRs, and the Hermiticity conditions on an (over)complete
set of observables, there may still remain some ambiguity in the quantum
theory. An example is provided by the quantum theory of the particle on $S^1$,
if we choose a more general representation space than the space of functions on
the unit circle. In this case, as we will see, the new sectors at least have a
simple physical interpretation.

In step 4 of the quantization program, the only requirement is that we find
{\it a} representation of \cA. Instead of choosing, as before, the carrier
space of the representation to be the space of smooth complex functions on the
unit circle, I will now make an alternate choice.

Consider the (non-trivial) {\it $Z_2$ bundle over $S^1$} (see Fig3.1)
\cite[pp.\ 362]{wald3}; call it $B$.
\begin{figure}
\vspace{2.5in}
\caption{$B$: $Z_2$ bundle over $S^1$}
\end{figure}
The group $Z_2$ consists of two elements: the identity ${\bf 1}$, and
``parity'' $\dP$, with the group law $\dP^2={\bf 1}$. The base space is $S^1$
and the fibre over any base point on $S^1$ consists of just two points. Note
that because of the ``twist'', the bundle is topologically $S^1$, and not
$S^1\oplus S^1$. Let $\phi\in[0,4\pi)$ denote a point on the bundle. The action
of $\dP$ is given by:
 \be \label{fs:par}
  \dP\circ\phi=\phi+2\pi.
 \ee
The projection mapping $\Pi$ from the bundle to the base space $S^1$
`coordinatized' by $\theta\in[0,2\pi)$ is simply
 \be \label{fs:proj}
  \theta=\Pi\circ(\phi):=pr.val.(\phi)
 \ee
where $pr.val.(\phi)$ denotes the principal value of $\phi$.
Thus the two points $\phi$ and $\dP\circ\phi=\phi+2\pi$ in the bundle are
mapped
to the same point in the base space.

Let us use the space of functions on $B$ as the carrier
space \cV, and represent the operators by:
 \bea \label{fs:rep}
  \hq_1\circ\Psi(\phi)&=&\sin\phi\cdot\Psi(\phi) \cr
           \hq_2\circ\Psi(\phi)&=&\cos\phi\cdot\Psi(\phi) \cr
    \hp\circ\Psi(\phi)&=&\dfrac\hbar{i}\dfrac{\d}{\d\phi}\Psi(\phi).
 \eea
These are representations of the commutation relations (\ref{s1:ccr}). Note
that since the base space is the unit circle, the ACR plays no further role.
The inner product on \cV\ is $\IP\Psi\chi= \frac{1}{4\pi} \int_B
d\phi\,\bar\Psi\chi$. With respect to this inner product, a convenient o.n.
basis is $\{\ket{m} = e^{im\phi/2}\},\, m\in integers$. The spectra of
$\hq_1,\hq_2$ are the expected ones. However, the spectrum of $\hp$, which is
diagonal in this basis, is $\{m/2\},\>\forall\,m\in integers$. We obtain not
only the expected integer eigenvalues, but also {\it half}-integer eigenvalues
in the spectrum of $\hp$. Let us see how this arises.

First, notice that the representation defined above is {\it reducible}. There
exists a {\it superselected operator}
 \be \label{fs:parop}
  \hat{\dP}\circ\Psi(\phi):=\Psi(\phi+2\pi)
 \ee
which is the quantum analog of the parity element in $Z_2$. $\hat\dP$ commutes
with all the elementary observables, i.e., the elementary variables are {\em
even} under parity. Each eigenspace of $\hat\dP$ will carry an irreducible
representation of the algebra (\ref{s1:ccr},\ref{s1:acr}). Since
$\hat\dP{}^2=\hat1$, its eigenvalues are just $\pm1$, and the respective
eigenspaces are
 \bea \label{fs:v+}
  \cV_+&=&\{\ket{2m}\},\hphantom{+1}\quad \forall\,m\in {\rm integer} \\
       \label{fs:v-}
    \hbox{and}\quad \cV_-&=&\{\ket{2m+1}\},\quad \forall\,m\in {\rm integer}.
 \eea
On the `bosonic' sector $\cV_+$, $\hp$ has integer eigenvalues, whereas on the
`fermionic' sector $\cV_-$, $\hp$ has half-integer eigenvalues. Each of these
sectors carries a unitarily inequivalent representation of the 2-dimensional
Euclidean algebra (\ref{s1:ccr}). Note that the even representation is
naturally isomorphic to the one considered in the first part of this section,
since the parity even functions on $B$ satisfy $\Psi(\phi+2\pi)=\Psi(\phi)$ and
are thus projectible to functions on $S^1$ itself.

Since the spectra of $\hp$ in the two representations are just `shifted'
w.r.t.\ each other, one could argue that they are physically indistinguishable.
In the first place, however, one can easily construct a `parity even' operator.
say $H=p^2/2m$, whose spectra in the two representations are not related by
just a shift.  Furthermore, these representations are distinguishable if, e.g,
the Hamiltonian contains an interaction term corresponding to a magnetic field
coupled to the `spin'. {\it A priori} one must consider both representations,
and allow an experiment to pick the correct physical sector.

There is a straightforward generalization of the above effect. One can
construct a representation of the algebra (\ref{s1:ccr}) on the $Z_n$ bundle
over $S^1$. Since in this case $\hat\dP{}^n=\hat1$, there are $n$ irreducible
representations, labelled by the $n$th roots of unity. In each sector, the
eigenvalues of $\hp$ are integer plus  $\frac{k}{n} , \> k<n$. Clearly, this is
a description of `fractional statistics' for the particle.

Thus, exploiting the fact that in the algebraic approach to quantization the
representation need not be tied to the \ps, we have obtained new
representations in the quantum theory for the above system. This occurs inspite
of the facts that: {\it i}) it is a finite dimensional system; {\it ii}) there
is a faithful representation of the algebraic relations; and {\it iii}) all
Hermiticity conditions have been satisfied. These new representations are not
just of  mathematical interest, but they enable one to obtain physically
important sectors of the theory, corresponding to particles with fractional
statistics. In a canonical approach to \qg\ based on the new variables, the
most useful representation appears to be one which is not based on a
polarization of \ps, but in which the states are labelled discretely by loops.
``New'' sectors of the theory, analogous to those in the above example, could
play an important role.

\mysection{Harmonic oscillator in the $(q,z)$ variables}

Consider the \ps\ of a harmonic oscillator of unit mass and spring constant.
$\Gamma$ is coordinatized by the real functions $(q,p)$ and the symplectic
structure is given by $\Omega= d p\wedge d q$. In analogy with \gr\ in the
connection variables, let us introduce the complex variable $z=q-ip$. Note that
the \ps\ is still real, and that $(q,z)$ are {\it not} coordinates on $\Gamma$.
However, we choose as \cS\ the vector space spanned by the (complex) functions
$(1,q,z)$. Note that this is complete and closed under Poisson brackets. \cS\
is also closed under complex conjugation, since $\bar{q}=q$ is real and
$\bar{z}=2q-z\in\cS$. These hybrid $\star$-relations are analogous to the ones
(\ref{ex:rc}) for the new variables. The canonical commutation relations
between the corresponding operators are
 \be \label{qz:ccr}
  [\hq,\hz]=1,
 \ee
and the $\star$-relations are:
 \be  \label{qz:star}
  \hq^\star=\hq,\quad\hz^\star=2\hq-\hz.
 \ee

We next have to find a representation of this algebra. The hybrid variables
suggest a new approach. We can represent the above operators on the space of
{\it holomorphic} functions of one complex variable. Note that we have
introduced complex coordinates on $\Gamma$ in order to define $z$, and we can
use holomorphic functions on $\Gamma$ since the notion $\d\psi/\d\bar{z}$ would
be well-defined. However, this might be confusing since $\bar{z}$ is {\em not}
one of the elementary variables in \cS. In order to avoid this possible source
of confusion, and to further emphasize the independence of the representation
space from the \ps, we can consider a fiducial 1-dimensional complex space with
coordinates $(\zeta,\bar\zeta)$. Let \cV\ be the space of holomorphic functions
$\psi=\psi(\zeta)$ (\ie, $\d\psi/\d\bar\zeta=0$), and represent the operators
by:
 \be\label{qz:rep}
  \hz\circ\psi(\zeta)=\zeta\psi(\zeta)\quad\hbox{and}\quad
  \hq\circ\psi(\zeta)= \frac{d}{d\zeta}\psi(\zeta)
 \ee
(Since $\hz$ is diagonal in this representation; one can loosely
identify $\zeta$ with $z$.)

A natural ansatz for the inner product on these holomorphic states is
 \be\label{qz:ip}
  \IP\psi\chi=\lint{d\zeta\wedge d\bar\zeta \over2\pi
   i}\>e^{\mu(\zeta,\bar\zeta)}\>\bar\psi\,\chi,
 \ee
where $\mu=\bar\mu$ is some as yet undetermined measure. We now have to impose
the Hermiticity conditions (\ref{qz:star}) on these operators. This is the
crucial step: it is quite counterintuitive to have a real momentum conjugate to
a complex variable; the whole formalism could fall apart right here. However,
we have
 \bea
  \bra\psi\hq^\dagger\ket\phi
  = & {\disp\int}\>e^\mu\>\left(\dfrac{d}{d\bar\zeta}
                     \bar\psi(\bar\zeta)\right)\,\phi
  & = -\int\>e^\mu\,\left(\dfrac{d}{d\bar\zeta}\mu\right)\,\bar\psi\phi \cr
  \hbox{and}\quad
  \bra\psi\hq\ket\phi
  = & {\disp\int}\>e^\mu\>\bar\psi\dfrac{d}{d\zeta}\phi\hphantom{(\zeta)(aa)}
  & = -\int\>e^\mu\,\left(\dfrac{d}{d\zeta}\mu\right)\,\bar\psi\phi,
 \eea
where we have used the holomorphicity of the wavefunctions to integrate by
parts. Thus the Hermiticity condition on $\hq$ is solved by
$\mu=\mu(\zeta+\bar\zeta)$. Similarly, the Hermiticity condition on $\hz$
yields a differential equation for $\mu$:
 \be
  \frac{d}{d\zeta}\mu=-\half(\zeta+\bar\zeta),
 \ee
which is solved by:
 \be\label{qz:meas}
  \mu(\zeta,\bar\zeta)=-{(\zeta+\bar\zeta)^2\over4}.
 \ee
The Hermiticity conditions do determine the measure uniquely. Note that even
though the measure does not ``fall-off'' as expected, there exist normalizable
states, which are of the form $\psi(\zeta)=e^{\zeta^2\over4}f(\zeta)$, where
$f(\zeta)$ are polynomials in $\zeta$. Now, if we note that the $z$ we have
defined is $\sqrt2$ times the usual Bargmann variable, and identify $f(\zeta)$
with the Bargmann states, the unitary equivalence of the two quantum theories
is easily established%
\footnote{Alternately, one could have represented $\hq$ by $\hq\circ\psi(\zeta)
 = (\frac{d}{d\zeta} + \frac\zeta{2})\psi(\zeta)$. Then, the commutation
 relations are still satisfied, but the inner product is the usual Bargmann
 inner product $e^{-\frac{\zeta\bar\zeta}{2}}$. The unitary equivalence is then
 manifest.}.

Thus, we have completed the quantization of the 1-dimensional oscillator in the
$(q,z)$ variables. The Hermiticity conditions on the elementary operators can
be implemented, and they fix the inner product. This indicates that there is no
obstruction {\it a priori} to the connection representation for \gr\ and the
use of a hybrid set of variables.

\mysection{Bianchi II model}

For a certain class of spatially homogeneous (SH) models (see
\cite{m:mini3,mpr3,RandS3}) one can choose a ``diagonal'' gauge in
which the \st\ metric is given by
\be
  ds^2=-(N(t))^2dt^2 + \sum_{i=1}^3 q_{ii}(t)(\omega^i)^2,
\ee
where $N(t)$ is the lapse function; $q_{ii},i=1,2,3$ are the diagonal
components of the spatial metric and the $\omega^i$ are basis 1-forms. The
components of the metric depend only on the time coordinate $t$. The
homogeneous basis forms $\omega^i$ are Lie-derived by the generators of the
spatial homogeneity group, and satisfy $d\omega^i = -\fr12 C^i_{jk}\omega^j
\wedge\omega^k$, where $C^i_{jk}$ are the structure constants of the spatial
homogeneity group.

For the above class of diagonal SH models, the scalar constraint can be
obtained directly by applying the ADM procedure \cite{adm3}, assuming that the
spatial manifold is compact. The diffeomorphism constraint is automatically
solved, in the gauge in which the metric is diagonal. Thus there is a single
first class constraint.

For Bianchi type II models, one can choose a basis in the Lie algebra such that
the only non-vanishing structure constants are $C^3_{12}=-C^3_{21}=1$. One can
make certain initial choices of variables to simplify the expression of the
scalar constraint. Starting with the Misner parametrization of the metric
components \cite{m:mini3}, one can make a point transformation, and thus
parametrize the metric in the following manner \cite{ruj:vac3}. Let
$q_{ii}=\exp(2\beta^i),\> i=1,2,3$. Define new variables $\bb^A,\>A=0,+,-$ by
 \be \label{b2:param}
  \left( \begin{array}{c}
         \bbz \cr
         \bbp \cr
         \bbm \cr
         \end{array} \right) = \dfrac{1}{2\sqrt3}
  \left( \begin{array}{ccc}
          1 &  1 &  2 \cr
          0 &  0 & -2 \cr
          1 & -1 &  0 \cr
         \end{array} \right)
  \left( \begin{array}{c}
          \beta^1 \cr
          \beta^2 \cr
          \beta^3 \cr
         \end{array} \right)\ .
 \ee
For reference, the inverse of this transformation is given by
 $$
  \left( \begin{array}{c}
          \beta^1 \cr
          \beta^2 \cr
          \beta^3 \cr
         \end{array} \right) = \sqrt3
  \left( \begin{array}{ccc}
          1 &  1 &  1 \cr
          1 &  1 & -1 \cr
          0 & -1 &  0 \cr
         \end{array} \right)
  \left( \begin{array}{c}
          \bbz \cr
          \bbp \cr
          \bbm \cr
         \end{array} \right)\ .  \eqno(\ref{b2:param}^{-1})
 $$

To further simplify matters, one usually chooses the ``Taub time
gauge'', i.e., one chooses the lapse function $N_T=
12\exp{(2\sqrt3\bbz+\sqrt3\bbp)}$ \cite{taub513}. (This time gauge is
also known as Misner's supertime gauge \cite{m:mini3}.) With this
choice the scalar constraint for Bianchi II takes the form:
\be\label{b2:barcon}
 C=\half (-\bpz^2+\bpp^2+\bpm^2) + 6e^{-4\sqrt3\bbp}
\ee
where the $\bar{\pi}_A$ are the momenta canonically conjugate to
$\bb^A$; the Poisson brackets are given by $\{\bb^A,\bar{\pi}_B\} =
\delta^A_B$. In these variables, the model is kinematically equivalent
to a relativistic particle moving in 3-dimensional Minkowski space
under the influence of a potential. (Of course, interesting physical
and dynamical observables will be completely different.) Clearly, the
constraint is separable. One could now choose $(1,\bb^A,\bar{\pi}_A)$
as the set of elementary variables and proceed with the quantization
of the model. While it is by no means trivial, the quantum constraint
(in the $\bb^A$ representation) {\it can} be solved by elementary
methods \cite{mpr3}. However, there is no obvious inner product on the
space of solutions (but see \cite{atu:I3}), and one has to resort to
the algebraic approach to find one. It is here that a poor choice of
elementary variables makes further progress difficult. First, a
complete set of observables is not easy to find. Second, the
complicated form of some of the observables makes it difficult, if not
impossible, to factor-order them in this representation; one cannot
therefore hope to impose Hermiticity conditions.

However, Uggla \cite{cu:pc3} noticed that this model admits a \ct, under the
action of which the Hamiltonian constraint is in the form of that of a
relativistic {\em free} particle. This is achieved by absorbing the potential,
which depends on only $\bbp$, into a new momentum. I will first outline his
\ct\ and then use the new variables for quantization of the model.

Postulate a new momentum, defined by
 \be\label{b2:ct}
  \tpp = +\sqrt{\bpp^2 + 6e^{-4\sqrt3\bbp} }.
 \ee
Note that, by definition, $\tp_+ > 0$. We would now like to find the
canonically conjugate variable. Consider first the intermediate \ct\
$(\bbp,\bpp)\mapsto(\bbp,\tpp)$. In terms of these variables, the symplectic
structure is
 \be\label{b2:ss1}
  \Omega=d\bpp\wedge d\bbp = \frac\tpp{(\tpp^2-6e^{-4\sqrt3\bbp})}\cdot
   d\tpp\wedge d\bbp.
 \ee
Now we would like to perform a second \ct, $(\bbp,\tpp)\mapsto(\tbp,\tpp)$,
such that $(\tbp,\tpp)$ are canonical variables. Thus we want
 \be
  d\tpp\wedge d\tbp = d\tpp\wedge \left(\frac{\d\tbp}{\d\tpp}\cdot d\tpp +
  \dfrac{\d\tbp}{\d\bbp}\cdot d\bbp\right)
  = \frac{\d\tbp}{\d\bbp}\cdot d\tpp\wedge d\tbp \equiv \Omega.
 \ee
Comparing with (\ref{b2:ss1}), we obtain a differential equation for $\tbp$:
 \be
  \frac{\d\tbp}{\d\bbp}= \frac\tpp{(\tpp^2-6e^{-4\sqrt3\bbp})}.
 \ee
Since there is no condition on $\d\tbp/\d\tpp$, the above p.d.e.\ can be
readily solved:
 \be \label{b2:ctm}
   \tbp = -\fr{1}{2\sqrt3}(\ln[-\bpp + \sqrt{
           \bpp^2 + 6e^{-4\sqrt3\bbp}}] -
           \ln[\sqrt6e^{-2\sqrt3\bbp}]) .
 \ee
Thus (\ref{b2:ct}) and (\ref{b2:ctm}) together define a \ct\ which yields a
constraint of the desired form. The inverse of this \ct\ is given by
 \be\label{b2:cti}
   e^{-2\sqrt3\bbp} = \frac\tpp{\sqrt6\cosh(2\sqrt3\tbp)}\ ,\qquad
   \bpp = \tpp\tanh(2\sqrt3\tbp)\ ,
 \ee
from which we can see that the canonical transformation is globally defined on
the \ps, since $\bpp$ takes all values.

Since the other variables are invariant under the transformation, $(\tbz =\bbz,
\tbm=\bbm, \tpz=\bpz, \tpm=\bpm)$. the constraint now takes the simple form
\be  \label{b2:con}
   C = \fr12(-\tpz{}^2 + \tpp{}^2 + \tpm{}^2) = 0\ .
\ee
However, as mentioned above, this canonical transformation leads to
a nonholonomic constraint
\be \label{b2:nonhol}
   \tpp > 0.
\ee
on the new momentum. (As before $\tpz < 0$ since the Taub time has been chosen
to be future directed.)

\subsection{Dirac quantization}

In the quantum theory the presence of the non-holonomic constraint $\tpp > 0$
and the requirement that $\tpz < 0$ means that the \ps\ cannot be considered as
the cotangent bundle over the configuration space $\{(\tb)\}$, and thus makes
it inconvenient to use the $\tilde\beta^A$-representation. The form of the
Hamiltonian constraint function naturally suggests that we view the $\{
\tilde{p}_A \}$ as the configuration variables since this leads to a constraint
function which is independent of the momenta ${\tilde\beta}^A$. Since the
constraint is quadratic in the configuration variables, which form a vector
space, it naturally defines a {\it covariant} metric on the configuration
space. The \ps\ again has a cotangent bundle structure, and the configuration
space is the 3-dimensional Minkowski space $M^3$, coordinatized by $\{
\tilde{p}_A \}$.

We choose for \cS\ the set of all functions on \ps\ independent of or linear in
the momenta $\tb^A$. The set of elementary quantum operators is constructed as
follows \cite[(see appendix 2)]{aa:cmp3,aa:rg3}: With any function $f(\tp)$ on
the configuration space $M^3$ associate the configuration operator $\hQ(f)$.
Functions linear in momenta $\tb^A$ are determined by a vector field $V^A(\tp)$
on $M^3$; we associate with $V^A$ the momentum operator $\hP(V):=\sum_A
V^A\tb^A$. The above set of operators satisfy the commutation relations:
 \bea
  [\hQ (f(\tp)), \hQ(g(\tp))] &=& 0, \cr
  [\hP(V), \hQ (f(\tp))] &=& i\hbar \hQ(\Lie{V} f)(\tp), \cr
   \label{b2:ccr:pp}
   [\hP(U),\hP(V)]            &=& \> i\hbar\hP([U,V]) \>,
 \eea
which can be obtained from the classical Poisson brackets. They also satisfy
the anti-commutation relations:
 \bea
   \{\hQ(f(\tp)),\hQ(g(\tp))\}_+ - 2\hQ(f\cdot g(\tp))  &=& 0 \cr
   \label{b2:acr:qp}
  \hbox{and} \quad \{\hQ(f(\tp)),\hP(V)\}_+ - 2\hP(fV) &=& 0 \ ,
 \eea
which follow from the algebraic relations between functions and vector fields
on ${\cal C}$. One can easily check that the LHS of the above relations
constitute a Lie ideal in \cA. For steps 1-5 in the quantization program, we
could have chosen $(1,\tp,\tb)$ as the set of elementary variables. It is to
implement step 6 --i.e., to find a complete set of Dirac observables-- that I
have chosen a more general set of elementary operators than the set
$(1,\tp,\tb)$.

Let the representation space \cV\ consist of functions on $M^3$,
$\Psi=\Psi(\tp)$. The action of the above operators on these wavefunctions is
given by:
\bea \label{b2:rep}
   \hQ(f(\tp))\circ\Psi(\tp) & := & f(\tp)\cdot\Psi(\tp) \cr
   \hbox{and} \quad \hP(V)\circ\Psi(\tp) & := & i\hbar\left(\Lie{V}\Psi(\tp)
    +\fr12 \xi(V) \Psi(\tp_A)\right)\ ,
\eea
where $\xi(V)$ is a real function on \cC, linear in the vector field $V$, i.e.
$\xi(U+V)=\xi(U)+\xi(V)$. In order to satisfy the CCRs (\ref{b2:ccr:pp}) and
the
ACRs (\ref{b2:acr:qp}), $\xi(V)$ also has to satisfy the following
conditions:
 \bea \label{b2:xi}
     \xi([U,V]) &=& \Lie{U}\xi(V)-\Lie{V}\xi(U)      \\
  \label{b2:x2}
       \hbox{and}\quad \xi(fV) &=& \Lie{V}f+f\cdot\xi(V)
 \eea
Note first that since (\ref{b2:x2}) is {\it not} linear in $\xi$, the inclusion
of the multiplicative term $\xi(V)$ is {\it necessary} to satisfy
(\ref{b2:acr:qp}). Second, we have not yet introduced any inner product on \cV.
However, in order to {\em eventually} find an inner product on the physical
states, the ``extra'' term $\xi(V)$ in the representation of the momentum
operators (\ref{b2:rep}) is again crucial.

The constraint is easily solved in this representation. Physical wavefunctions,
\ie\ solutions to the quantum constraint
\be\label{b2:qcon}
   \hat{C}\circ\Psi(\tp):=
   \fr12(-\tpz{}^2+\tpp{}^2+\tpm{}^2)\Psi(\tp)=0\ ,
\ee
(as well as the nonholonomic constraints) are simply those wavefunctions with
support only on the right half ($\tpp>0$) of the past ($\tpz<0$) null cone
\halfcone. Let us denote physical states by $\psi$ and the vector space of
solutions by \cVp. Next we find the Dirac observables, i.e.\ those operators
which commute weakly with the constraint. Configuration observables correspond
to functions on \halfcone; a complete set consists of
$\wh{\dP}{}^\pm\circ\psi:=\hQ(\tp_\pm)\circ\psi=\tp_\pm\psi(\tp)$. Momentum
observables correspond to vector fields on \halfcone. Note that since
\halfcone\ is a manifold with boundary, one must be careful to select only
those vector fields which are tangential to the boundary. Any such vector field
corresponds to a momentum observable which leaves the space of physical states
invariant, and is a good Dirac operator. In order that the corresponding
momentum operators form an overcomplete set closed under Poisson brackets, the
vector fields should form a Lie algebra, and span the tangent space to
\halfcone\ (almost) everywhere. In order to select an inner product, I will
make a specific choice of physical momentum observables.

Two vector fields, which span the tangent space to \halfcone\ everywhere, and
happen to commute with each other, are:
 \bea \label{b2:dvf}
    V_-^A&=& \left(\dfrac\d{\d\tpm}\right)^A+\dfrac\tpm\tpz\left(
            \dfrac\d{\d\tpz}\right)^A \cr
    V_+^A&=& \tpp\left(\dfrac\d{\d\tpp}\right)^A + \dfrac{\tpp^2}\tpz \left(
            \dfrac\d{\d\tpz}\right)^A.
 \eea
Note that $V_-^A$ is proportional (by a factor of $1/\tpz$) to the boost vector
field in the $\tpm$ direction and is the ``lift'' to \halfcone\ of the {\it
translational} vector field $(\d/\d\tpm)$ (see (\ref{b2:diffeo}). $V_+^A$, on
the other hand, is more complicated. It is the lift to \halfcone\ of the {\it
dilation} vector field in the $\tpp$ direction. In place of $V_\pm$, we might
have been tempted to choose the usual Lorentz vector fields corresponding to a
rotation and two boosts. However, unlike $V_\pm^A$, the Lorentz vector fields
are {\it not} tangential to the $\tpp=0$ boundary of \halfcone, and thus the
corresponding momentum operators do not leave the space of physical states
invariant.

Thus, a complete set of momentum observables is generated by the operators:
 \bea \label{b2:obs}
   \wh\dB_- &:=& \hP(V_-) \equiv \tbm + \dfrac{\tpm}{\tpz}\tbz \cr
  \hbox{and}\quad\wh\dB_+ &:=& \hP(V_+)
\equiv\tpp\tbp+\dfrac{\tpp^2}{\tpz}\tbz,
 \eea
which are represented by (\ref{b2:rep}). The algebra \cAp\ of Dirac
operators is then given by the commutation relations
 \bea \label{b2:obsccr}
                   [\wh\dB_-,\wh\dP{}^-] &=& i\hbar \cr
  \hbox{and} \quad [\wh\dB_+,\wh\dP{}^+] &=& i\hbar \wh\dP{}^+\ .
 \eea

Since the physical states are functions with support only on \halfcone,
a natural ansatz for the physical inner product is
 \be\label{b2:dip}
  \IP\psi\phi:=\int_{\scriptstyle\halfcone}\>\mu\> \bar{\psi}\,\phi\,\ ,
 \ee
for some two-form $\mu$ on \halfcone. We now require that all physical
operators be self-adjoint. Since the $\wh\dP^\pm$ are already symmetric, the
only nontrivial condition is that the momentum observables (\ref{b2:obs}) be
represented by self-adjoint operators. This condition is satisfied \iff, in the
representation of the momentum operators (\ref{b2:rep}),
 \be
  \xi(V)=\Div\mu V,
 \ee
where $\Div\mu V$ is the divergence of the vector field $V$ \wrt\ to the
measure $\mu$. Note that the Hermiticity conditions have {\em not} fixed the
inner product. However, the {\em relation} between the representation and the
inner product has been fixed. Alternate choices of $\mu$ yield unitarily
equivalent theories. One way to fix the inner product is to require that the
vector fields (\ref{b2:dvf}) be divergence free. This yields a set of
differential equations which can be solved for the measure
 \be\label{b2:dmeas}
  \mu=\frac1\tpp\> d\tpp\wedge d\tpm.
 \ee
With this choice of $\mu$, the momentum observables (\ref{b2:obs}) are
represented by just the Lie derivative terms in (\ref{b2:rep}).

Thus, a good choice of elementary variables and representation has enabled us
to complete the (kinematical) quantization of this model. We have found a
complete set of Dirac observables, and an inner product on physical states such
that the observables are Hermitian operators.

At this stage, I would like to emphasize again the importance of the choice of
elementary variables. To find the Dirac operators in the ``old'' variables
$(\bb^A,\bar{\pi}_A)$,  one would have to formally solve the Heisenberg
equations of motion (with the constraint as the generator of evolution) and
find the constants of motion, either classically or quantum mechanically.
Assuming this could have been done, the formal operators would then have to be
factor-ordered. The enormity of this task can be seen by using the Dirac
observables $(\wh\dP{}^\pm,\wh\dB_\pm)$ (\ref{b2:obs}) and expressing them in
terms of $(\bbp,\bpp)$ by substituting the inverse (\ref{b2:cti}) of the \ct.

\subsection{(\dag) Reduced space quantization}
In chapter 2.4 I outlined an alternative approach to the quantization of
constrained systems. Recall that to construct the reduced space quantum theory,
we have to find the space of the true degrees of freedom of the system, and
then quantize. In the Dirac quantum theory we constructed in the previous
section, we viewed the $\tp$ as the configuration variables. Thus the
constraint was independent of momenta. As was pointed out in section 2.5, for a
constraint of this form, the two quantum theories are kinematically equivalent.
In this section, as an example I will explicitly construct the reduced space
theory, and demonstrate the unitary equivalence to the Dirac theory of the
previous section.

Obviously, it is most convenient to use the ``tilde'' variables introduced
earlier. To construct the reduced \ps, we have to solve the constraint
(\ref{b2:con}) and find the constants of motion classically. The constraint is
most easily solved for $\tpz$
 \be\label{b2:ham}
  \tpz=-\sqrt{\tpp^2+\tpm^2}.
 \ee
Let $\tau$ be an affine parameter along the dynamical trajectories generated by
the Hamiltonian constraint in the Taub time gauge. ($\tau$ is a ``time''
function on $\bar\Gamma$ rather than on \st. In each solution, $\tau$ can be
identified with the Taub time in the \st\ picture.) The Hamiltonian equations
of motion, for any function $f$ on the constraint surface are given by $\fr{\d
f}{\d t}\equiv \dot{f}=\{ f, C \}$. Due to the simple form of the constraint,
these can be readily solved to yield
 \bea \label{b2:clsol}
   \tp_\pm &=& \rp^\pm \hphantom{+ \rp^\pm \tau} \quad \hbox{and} \quad
   -\tpz = \rpz := \sqrt{(\rpp)^2 + (\rpm)^2} \ ,\cr
   \tb^\pm &=& \rb_\pm + \rp^\pm \tau \quad \hbox{and} \quad
   \hphantom{-} \tbz  =\rpz \tau + \tbz(0)\ ,
 \eea
where $(\rp^\pm, \rb_\pm, \tbz(0)\>)$ are five constants of motion which
correspond to the initial values of $(\tp_\pm, \tb^\pm, \tbz)$ respectively.
Since the space of orbits is only 4-dimensional, the five constants of motion
are not all independent. Clearly, of the five constants of motion, one should
be chosen as a parameter which serves only to fix the ``initial time''. This
corresponds to fixing a particular cross-section of the gauge orbits on which
the rest of the initial data is specified. Of all the constants of motion, let
us select $\tbz(0)$ as the ``gauge-fixing'' parameter, since the constraint is
solved for its conjugate momentum. Without any loss of generality, I will set
$\tbz(0)=0$. Now we can consider the five functions $(\rp^\pm,\rb_\pm,\tau)$ as
coordinates on the constraint surface: four of them ($\rp^\pm,\rb_\pm$) are
coordinates on the reduced \ps\ which label the orbits, and $\tau$ is by
definition the affine parameter along the orbits. We can invert
(\ref{b2:clsol}) and explicitly obtain these five functions on $\bar\Gamma$.
Substituting (\ref{b2:clsol}) into the canonical symplectic structure one sees
that the $\rp^\pm$ are canonically conjugate to the $\rb_\pm$:
 \be\label{b2:rpb}
  \{\rb_\hA,\rp^\hB\}=\delta_\hA{}^\hB,
 \ee
where $\hA,\hB=+,-$ are indices on $\hat\Gamma$. The nonholonomic constraint
(\ref{b2:nonhol}) implies that $\rpp>0$.

Now we want to quantize the system whose \ps\ is $\hat\Gamma$, with the Poisson
brackets (\ref{b2:rpb}). Note that due to the presence of the nonholonomic
constraint on $\rp^+$, it is convenient to choose $\rp^+$ (and $\rp^-$) as the
configuration observables. The reduced phase space $\hat\Gamma$ is then a
cotangent bundle over the haf plane $\halfplane=\{(\rpp>0,\rpm)\}$, with the
$\rb_\pm$ as the new momenta (note that an upper (lower) index position is used
for configuration (momentum) observables). Momentum observables $\rB(v)
=v^\hA\rb_\hA$ correspond to vector fields $v^\hA(\rp)$ on the reduced
configuration space $\halfplane$. Note however that since \halfplane\ is a
manifold with boundary, it admits only one translational vector field, (the
vector field $(\d/\d\rpp)^A$ is incomplete) and thus momentum variables are
somewhat trickier to select. We need a Lie algebra of vector fields which span
the tangent space to \halfplane\ everywhere. Such an algebra (which happens to
be Abelian) is generated by the translational vector field in the $\rpm$
direction and a dilation in the $\rpp$ direction:
 \be\label{b2:vf}
   v_{(-)}^\hA := \left(\frac{\d}{\d\rpm}\right)^\hA \quad \hbox{and} \quad
   v_{(+)}^\hA := \rpp\left(\frac{\d}{\d\rpp}\right)^\hA\ .
 \ee
Note that these are just the projections to \halfplane\ of the vector fields
$V_\pm^A$ on \halfcone\ considered in the Dirac theory. While one could have
chosen any pair of linearly independent vector fields which are tangential to
the boundary, I have selected this pair in order to make the isomorphism with
the Dirac theory manifest (see (\ref{b2:diffeo}).

With these particular vector fields, we associate the momentum observables
 \be\label{b2:redobs}
  \rB_-=\rbm\quad\hbox{and}\quad\rB_+=\rpp\rbp
 \ee
The commutation relations can be easily obtained from the Poisson brackets
(\ref{b2:rpb}), the algebra ${\cal A}_{red}$ of Dirac observables is
given by
\be \label{b2:redccr}
   [\wh\rB_-,\wh{\rp}^-] = i\hbar\quad\hbox{and}\quad
   [\wh\rB_+,\wh{\rp}^+] = i\hbar\wh{\rp}^+\ .
\ee
Note that this algebra is isomorphic to the algebra of quantum Dirac
observables (\ref{b2:obsccr}).

In the configuration representation, states are functions
$\psi = \psi(\rpp,\rpm)$ on
\halfplane. A representation of the reduced space
observable algebra is
 \bea \label{b2:redrep}
   \wh{\rp}^\pm\circ\psi &=& \rp^\pm\psi \cr
   \hbox{and} \quad \wh\rB_\pm\circ\psi &=& i\hbar(\Lie{v_\pm}\psi +
   \fr12 \xi(v))\psi
 \eea
where $\xi(v)$ satisfies the same properties as in (\ref{b2:xi}).
A natural ansatz for the inner product is
\be\label{b2:redip}
   \IP\psi\phi=\int_{\scriptstyle\halfplane}\>\epsilon\>\bar\psi\,\phi
\ee
where $\epsilon$ is some two-form on \halfplane. We see immediately that the
momentum observables (\ref{b2:redobs}) are Hermitian with respect to the inner
product \iff\ $\xi(v)=\Div\epsilon{v}$. Different choices of $\epsilon$
give rise to unitarily equivalent quantum theories.

As in the Dirac theory, a simplifying choice for the inner product can be made
by requiring that the vector fields (\ref{b2:vf}) corresponding to the momentum
observables (\ref{b2:redobs}) be divergence free with respect to $\epsilon$.
The resulting set of differential equations on $\epsilon$ can be solved to
yield
 \be\label{b2:meas}
   \epsilon = \frac{1}\rpp \> d\rpp\wedge d\rpm
 \ee
which is unique up to the usual scale factor. The momentum operators
(\ref{b2:redobs}) are now represented in (\ref{b2:redrep}) by just the Lie
derivative terms.

To explicitly see the equivalence between the reduced and the Dirac quantum
theories, note that there is a natural isomorphism, a projection mapping along
the orbits of the vector field $(\d/\d\tpz)^A$, between \halfcone\ (defined in
section (3.1)) and \halfplane\ (coordinatized by $\tp_\pm$). This induces a
unitary mapping between $\cV_{phy}$, the space of functions $\psi$ on
\halfcone, and $\cV_{red}$, the space of functions on \halfplane:
 \be\label{b2:map}
  \psi(\tpp,\tpm)\longleftrightarrow\psi(\rpp,\rpm)
 \ee
Note that the inner product is the same, since $\mu=\epsilon$. Since the
configuration observables $\wh{\rp}^\pm$ and $\wh{\dP}^\pm$ are obviously
transformed into each other, we only need to check the momentum operators. The
above diffeomorphism between \halfplane\ and \halfcone\ induces the following
transformation between vector fields $V$ tangential to \halfcone\ and vector
fields $v$ tangential to \halfplane:
 \bea\label{b2:diffeo}
   V^A(\tp_\pm) &=& V^+(\tp_\pm)
   \left(\dfrac\d{\d\tpp}\right)^A+ V^-(\tp_\pm)\left(\dfrac\d{\d\tpm
   }\right) +
   \dfrac{(\tpp V^+(\tp_\pm) + \tpm V^-(\tp_\pm))}\tpz
   \left(\dfrac\d{\d\tpz}\right)^A \cr
     \longleftrightarrow \quad  & &
   v^A(\rp^\pm)= V^+(\rp^\pm)\left(\dfrac\d{\d\rpp}\right)^A +
                   V^-(\rp^\pm)\left(\dfrac\d{\d\rpm}\right)^A\ .
 \eea
The vector fields $V_\pm^A$ (\ref{b2:dvf}) are just the ``lifts'' of the
vector fields (\ref{b2:vf}). Since both sets of vector fields are also
divergence free with the respective inner products, we see that the
corresponding momentum operators are unitarily transformed into each other.
Thus, the Dirac and reduced quantum theories are manifestly equivalent to each
other: the physical states carry unitarily equivalent representations of the
isomorphic observable algebras \cAp\ and ${\cal A}_{red}$.

\mysection{Remarks}\remarks

Each of the models quantized in this chapter has clarified the role of some
essential aspect of algebraic quantization. In this section I will summarize
the lessons one has learnt from these examples.

\remark{}
The quantization of the particle on $S^1$ illustrates the importance of the
ACRs. Before we imposed the ACR (\ref{s1:acr}), there were ambiguities in the
theory, which corresponded to a large number of spurious sectors. In this
example, since the single ACR is a Casimir invariant of the algebra \cA, each
eigenspace carries an irreducible representation of \cA. Only one of these
spaces, in which the Casimir invariant has value 1, corresponds to the correct
physical quantum theory. The ACR identifies the ambiguities, and imposing
(\ref{s1:acr}) on the quantum theory got rid of the spurious sectors. Of course
in this model it is trivial to ``see'' the ambiguities and resolve them; the
point however was not just to solve the quantum theory of the particle on
$S^1$, but rather to understand the role of the ACRs in a more general context.

Further, as I already mentioned at the end of section 2, in order to be able to
find a good inner product, it is important to impose the ACRs and eliminate the
spurious sectors {\it before} the $\star$-relations are imposed.

A key feature of algebraic quantization is the fact that it frees up the choice
of representation from ties to the \ps. As we saw in subsection 2.1, this
allows us to consider representations which are unusual but physically
significant; and which could not have been obtained in other quantization
schemes where the representation of \cA\ is constructed on functions on the
\ps.

\remark{}
{}From the construction of the quantum theory for the harmonic oscillator in
the
$(q,z)$ variables we see that it is consistent to use a hybrid set of functions
(one real, one complex) on a real \ps\ as the elementary variables for
quantization. On the face of it, it appears counter-intuitive that a real
momentum, represented by a holomorphic derivative operator $d/dz$, can be
canonically conjugate to a complex variable. However, there is an inner
product, selected uniquely by implementing the $\star$-relations on these
variables, with respect to which $d/dz$ is in fact Hermitian. This indicates
that the hybrid connection-dynamical variables are a consistent choice for \qg.

\remark{}
The homogeneous cosmological model Bianchi type II is the only constrained
system quantized in this chapter. In the algebraic approach, real physical
observables should be represented by operators which are Hermitian with respect
to the physical inner product. This model demonstrates that the above criterion
can be used to explicitly construct an inner product on physical states.

In the $\bar\beta$ variables, the constraint ({\ref{b2:barcon}) is separable.
The quantum constraint equation has been previously ``solved'': the solutions
are sinister looking Bessel functions in $\bbz$ \cite[Table III.3, pp.
91]{mpr3}, and it is not at all obvious what if anything one can learn from
them. The set of solutions was in some ways considered complete. However, the
Hilbert space structure necessary in quantum theory, and a physical
interpretation, were absent. Until recently \cite{atu:I3}, there was no known
inner product on these states. The algebraic framework provides a powerful tool
to find the physical inner product; and it clarifies and emphasizes the role of
physical operators. Since the problem is after all separable, once the
importance of the observables was understood, it was relatively straightforward
to construct them, and then to use the Hermiticity relations on them to find
the physical inner product. As we will see in chapter 6, the existence of a
kinematically complete quantum theory does open up some possibilities to
extract a physical interpretation.

This model serves a few other purposes.
\begin{itemize}
\item It illustrates the importance of making good choices, for the algebra of
elementary variables and its representation. As I already mentioned, the
solutions in the $\bar\beta$-representation are of a horrendous form. More
importantly, it is difficult to see how the the physical operators $\tpp$
(\ref{b2:ct}) and $\wh\dB_+$ (\ref{b2:obs}) could even have been found, let
alone factor ordered and represented explicitly.
\item We began with a fairly general representation of the algebra \cA. Recall
that in (\ref{b2:rep}) $\xi(V)$ is an undetermined function in the
representation of the momentum variables. It is necessary to include this term
in order to satisfy some of the ACRs. Later, we make an ansatz for the physical
inner product (\ref{b2:dip}), and find that the physical operators are
Hermitian if and only if $\xi(V)$ is the divergence of the vector field with
respect to the measure $\mu$. However, neither $\xi(V)$ nor the measure is
determined. Different choices of the measure $\mu$ change the explicit
representation of the momentum observables, but yield unitarily equivalent
quantum theories. Thus the Hermiticity conditions determine the {\it relation}
between the representation and the inner product, leaving us with the freedom
to make unitarily equivalent choices, which can simplify the expressions of
some of the interesting physical operators.
\item In this model, Dirac and reduced space quantization can be explicitly
compared. I showed in this case that the two quantum theories are unitarily
equivalent, verifying the general claim (made in section 2.6) that this would
be so for constraints independent of momenta.
\end{itemize}
\hphantom{a}

\newpage\mbox{}
\chapter{ASHTEKAR-HOROWITZ MODEL}
\pagestyle{myheadings}
\markboth{{\sf Chapter 4}}{{\sf Ashtekar-Horowitz Model}}


 \def\Cbar{\bar{\cal C}}    
 \def\tcbar{\tw{\cal C}}    

 \def\rhat{\hat{r}} \def\tthat{\hat{\theta}} \def\phihat{\hat{\phi}}
 \def\prhat{\hat{p}_r} \def\ptthat{\hat{p}_\theta} \def\pphihat{\hat{p}_\phi}

 \def\phiz{\phi_0}          

 \def\eip{\,\exp(+\fr{i}\hbar \sqrt{R}\theta)}
 \def\eim{\,\exp(-\fr{i}\hbar \sqrt{R}\theta)}

 \def\Pphihat{\wh{\rm P}_\phi}      
 \def\Pphihatp{\wh{\rm P}_\phi^+}
 \def\Pphihatm{\wh{\rm P}_\phi^-}

 \def\kop{k_1^+} \def\kom{k_1^-}
 \def\ktp{k_2^+} \def\ktm{k_2^-}
 \def\kopb{\ovr{\kop}} \def\komb{\ovr{\kom}}

 \def\Pphi{{\rm P}_\phi}
 \def\Pphip{{\rm P}_\phi^+}
 \def\tmu{{\tw\mu}}

 \def\bcV{\bar{\cal V}}         
 \def\tcV{\tw{\cal V}}          



\mysection{Introduction}

In this chapter I will quantize a model, considered by Ashtekar and
Horowitz \cite{aa:gth4}, that mimics some of the features of \gr.
It was previously believed to display
certain unexpected behaviour in quantum theory. Since this
behaviour arose from precisely those features of this model that it possesses
in common with \gr, the authors speculated that similar surprises might occur
in a quantum theory of gravity. In this chapter I will re-analyse this problem
in the context of the algebraic quantization program.

Let us recall certain features of \gr\ in the geometrodynamical variables.
In the ADM formulation, the basic \ps\ variables are the 3-metric $q_{ab}$ and
its canonically conjugate momentum. The constraint surface%
\footnote{For the purposes of this chapter, the diffeomorphism constraint can
 be ignored.}
is specified by the vanishing of the scalar constraint function. The scalar
constraint is the sum of two terms: a kinetic term, quadratic in the momenta,
the coefficient of which
defines a ``supermetric'' on the configuration space; and a potential term
--proportional to the 3-dimensional Ricci scalar-- which depends only on the
configuration variables. Due to the complicated form of the constraint,
the structure of the constraint surface, $\Gbar$, and the dynamical structure
of \gr\ are not fully understood. There are,
however, many features which are well known.
Of interest to us are the following:
\begin{itemize}
\item First, the constraint surface defines a ``classically allowed'' region in
the configuration space. More precisely, the image in
the configuration space of the constraint
surface $\Gbar$ (under the natural projection map)
is a {\it proper subset\ } of the configuration space $\cC$.
\item Second, in the asymptotically flat case, the Hamiltonian is not
constrained to vanish. On $\Gbar$, the Hamiltonian reduces to a surface
integral at
spatial infinity, called the ADM energy. The ADM energy depends only on the
3-metric and its spatial derivatives%
\footnote{I would like to point out, to readers who are somewhat familiar with
 the Hamiltonian formulation of \gr\ and the specific form of the scalar
 constraint, that the surface integral is independent of the momenta since in
 the scalar constraint, all terms containing the momenta are ultralocal.}.
\item Finally, the positive energy theorems of classical \gr\ state that on
the allowed regions of $\cC$ defined by the projection of the constraint
surface,
the ADM energy is positive. In the ``forbidden'' regions, where the constraint
cannot be satisfied, the ADM energy can be negative.
\end{itemize}

Ashtekar and Horowitz constructed a finite dimensional model which shares the
above features of \gr (I will henceforth refer to this model as the A-H
model.): Consider a particle in 3-dimensional Euclidean space,
subject to the constraint
 \be \label{ah:con}
  C\equiv p_\theta^2-R(\phi)=0.
 \ee
The ``potential'', $R(\phi)$, is a smooth function, which is not everywhere
positive. As in \gr, the constraint surface $\Gbar$ projects down to
a proper subset of the configuration space $\cC$. The
{\em classically allowed region} $\Cbar$ corresponds to those sectors where
(\ref{ah:con}) has solutions, i.e., where
$R(\phi) \ge0$.
Now introduce a Hamiltonian via
 \be  \label{ah:ham}
  H=C+E(\phi), \quad \hbox{with} \quad E(\phi)\cdot R(\phi)\ge0;
 \ee
we assume that $E$ is bounded.
On the constraint surface $C=0$, the Hamiltonian reduces to the ``ADM energy''
$E(\phi)$, and depends only on the
configuration variable $\phi$. Since $E(\phi)$ is positive in the classically
allowed regions, where $R(\phi)$ is positive,
this function satisfies a classical positive energy theorem, as does
the Hamiltonian in \gr.

Now consider the Dirac quantum theory of this model, say in the Schr\"odinger
representation,
where states are functions of the configuration variables. Ashtekar and
Horowitz raised the following question: In the Dirac quantum theory, do there
exist physical states that penetrate the classically forbidden region ($R<0$)?
Let us consider the situation if such states do exist.
Now, physical states are solutions to the quantum
constraint equation, on which the Hamiltonian acts via a multiplication by
$E(\phi)$. Recall that --as in \gr-- in this model, the ``ADM energy'' $E$ is a
function only of the configuration variables. In this model $E$ is negative in
the {\em
classically forbidden region} $\tcbar=\cC - \Cbar$. Thus, on some physical
states which penetrate into $\tcbar$, the energy will be negative. Due
to the close analogy with \gr, such behaviour in this model might indicate
similar tunnelling in quantum gravity.

The key
result of the previous quantization was that in the Dirac quantum theory, there
are indeed physical quantum states which tunnel into the
forbidden region. Further, some states
had support entirely in the forbidden region $\tcbar$! Since
the usual Euclidean measure on $\real^3$ was used, this led to the conclusion
that a large number of physical quantum states in fact possessed negative
energies: for this model the classical positive energy theorem was violated in
quantum theory.

Ashtekar and Horowitz also compared the reduced space quantum theory with the
Dirac quantum theory. They concluded that the former
contained only some of the states present in the Dirac theory and was thus
incomplete. This comes about as follows: Recall that in the reduced space
approach, one first solves the constraint classically. In the A-H model,
clearly one can solve the constraint only in the region $R>0$.
The resulting reduced \ps\ does not contain any classically forbidden sectors,
and the energy is positive everywhere on $\hat\Gamma$. The analogs of the
Dirac states
with negative energy are absent from the quantum theory on $\hat\Gamma$.

In the quantum theory constructed previously, the standard Euclidean measure on
$\real^3$ was assumed to define the inner product on physical
states. In terms of the present algebraic quantization program, the steps
missing from
the construction of this quantum theory are {\it precisely} the ones that
require a
physical inner product which realises the $\star$-relations on the Dirac
observables as Hermiticity conditions. As we will see in detail, the
Hermiticity conditions are satisfied only if the measure has support only on
the classically allowed region $\Cbar$.
Thus, while it is true that physical states can have support in the forbidden
region, since this region is a set of measure zero in the physical inner
product,
they do not define unique states in the Hilbert space. States in the Hilbert
space are specified completely by their support in the allowed region itself.

Thus, at least for this model, a complete and consistent quantum theory
displays no extreme quantum tunnelling.
A careful application of the $\star$-relations eliminates the
pathological negative energy states. Further, as I will discuss in a later
subsection, the Dirac and reduced quantum theories are (almost) equivalent.

\mysection{Dirac quantization}

Let $(\rhat, \tthat, \phihat)$ denote the operators corresponding to the usual
spherical polar coordinates in 3-dimensional Euclidean space, and let $(\prhat,
\ptthat, \pphihat)$ be the canonically conjugate momenta. In the configuration
representation, states are functions of the coordinates,
$\Psi=\Psi(r,\theta,\phi)$, and the operators are represented by
 \bea \label{ah:rep}
  \rhat\circ\Psi&=r\Psi, \hphantom{\dfrac\hbar{i}\dfrac\d\d} \quad
   \tthat\circ\Psi=\theta\Psi \hphantom{\dfrac\hbar{i}\dfrac\d\d} \quad
   & {\hbox{and}} \quad \phihat\circ\Psi=\phi\Psi; \cr
  \prhat\circ\Psi&=\dfrac\hbar{i}\dfrac\d{\d r}\Psi,\quad
   \ptthat\circ\Psi=\dfrac\hbar{i}\dfrac\d{\d\theta}\Psi \quad & {\hbox{and}}
   \quad \pphihat\circ\Psi=\dfrac\hbar{i}\dfrac\d{\d\phi}\Psi .
 \eea
These operators satisfy the usual CCRs. (Although $(r,\theta,\phi)$ are not
globally well-defined coordinates, this issue is not important here.)
Since we have no inner product, the familiar
divergence terms have been left out of the representations of the momentum
operators%
\footnote{The quantization of a classical system is not a linear or
well-defined {\em process}, particularly when the system is constrained. One
can only demand that the final theory constructed is internally consistent with
its initial hypotheses. Hence I would urge readers to set aside their
objections until after I have constructed the final quantum theory. In remarks
at the end of this section, I will try and answer some objections that have
been raised \cite{jl:pc4}.}.
In this representation, the constraint equation satisfied by physical states
$\psi$ is
 \be \label{ah:qcon}
  \hbar^2\fr{\d^2}{\d\theta^2}\psi+R(\phi)\psi=0.
 \ee
Since this is a quadratic differential equation, there are two linearly
independent sets of solutions
 \be  \label{ah:qsol}
  \psi_\pm=k_\pm(r,\phi)\,\exp({\pm\fr{i}\hbar\sqrt{R}\theta})
 \ee
where $\sqrt{R}$ denotes the principal value: $\sqrt{R} =+\sqrt{R}$, if $R>0$,
and, $\sqrt{R} =+i\sqrt{|R|}$ if $R<0$. The functions $k_\pm(r,\phi)$ are
arbitrary. In particular, their support is {\em, not} restricted to the allowed
region $\Cbar$.
and have support everywhere in configuration space, including the
classically forbidden region $\tcbar$. Let the linear vector space of physical
states be denoted by $\cV$. Then $\cV=\cV^+\oplus\cV^-$, where $\psi_\pm \in
\cV^\pm$ respectively, and we can write a general state as
 \be  \label{ah:gsol}
  \psi=\left(\ba{c} \psi_+ \\ \psi_- \ea \right).
 \ee
Note that since we have not yet defined an inner product, this is {\em not} an
orthogonal decomposition of $\cV$.

Let us construct the Dirac
observables. Since the constraint is first class, there are two true degrees of
freedom, and thus we expect four independent Dirac observables. From their
representations, it is clear that $\rhat, \prhat, \ptthat, \phihat$ commute
with the constraint and leave $\cV$ invariant, their only effect on the
physical states is to change the coefficients $k_\pm$ of the corresponding
exponential terms. However, on physical states $\ptthat =\pm
\wh{\sqrt{R(\phi)}}$, and
thus so far we only have three {\em independent} observables. In order for the
set to
be complete, we need another observable. In correspondence with
the classical Dirac observable constructed by Ashtekar and Horowitz, consider
the operator
 \be  \label{ah:tp1}
  \Pphihat= \left(\ba{ll} \Pphihatp & 0 \\ 0 & \Pphihatm \ea \right), \quad
   \hbox{where}\quad \Pphihat{}^\pm=\pphihat \mp
                        \half\wh{\left(\frac{R'}{\sqrt{R}}\theta\right)}.
 \ee
To see that this is a physical operator, let us concentrate on one component,
acting on $\cV^+$. Let $'\equiv\d/\d\phi\,$ denote the partial derivative \wrt\
$\phi$. Then,
 \be  \label{ah:tpyes}
  \Pphihatp\circ\psi_+=\frac\hbar{i}\, k'_+\eip \> \in\> \cV^+.
 \ee
Similarly, $\Pphihatm\circ\psi_-\in\cV^-$. Hence $\Pphihat$ as defined above
is a physical operator. We now have a ``sufficient number'', i.e.\ a locally
complete set, of physical operators. For future reference, note that
 \be  \label{ah:tpno}
  \Pphihatm\circ\psi_+=\frac\hbar{i}k'_+\eip \>+\> 2(\sqrt{R})'\theta k_+\eip
         \>\not\in\>\cV,
 \ee
is not a physical state, since the coefficient of the exponential in the second
term is no longer
a function only of $(r,\phi)$.

We now have to find the $\star$-relations on the algebra of observables.
Clearly, $\rhat^\star=\rhat$, $\prhat^\star=\prhat$, $\ptthat^\star=\ptthat$,
$\hat\phi^\star=\hat\phi$ and are physical operators. In order to analyse
the $\star$-relation on $\Pphihat$, let us further decompose $\cV^+$ into the
set
of states $\bcV^+$ with support only on the classically allowed region
$\Cbar$ and the set of states $\tcV^+$ with support entirely in the
classically forbidden region $\tcbar$. Now, on the sector of states
$\bcV^+$, $(\wh{\sqrt{R}})^\star=\wh{\sqrt{R}}$ and hence
 \be   \label{ah:tpystar}
  (\Pphihatp)^\star=\left(\pphihat
   -\frac12\wh{\left(\frac{R'}{\sqrt{R}}\theta\right)}\right)^\star
   =\pphihat-\frac12\wh{\left(\frac{R'}{\sqrt{R}}\theta\right)}
   =\Pphihatp
 \ee
is a physical operator on $\bcV^+$. However, on the ``forbidden sector''
$\tcV^+$, $(\wh{\sqrt{R}})^\star=-\wh{\sqrt{R}}$ and hence
 \be  \label{ah:tpnostar}
  (\Pphihatp)^\star=\left(\pphihat
   -\frac12\wh{\left(\frac{R'}{\sqrt{R}}\theta\right)}\right)^\star
   =\pphihat+\frac12\wh{\left(\frac{R'}{\sqrt{R}}\theta\right)}
   =\Pphihatm,
 \ee
which, as we see from (\ref{ah:tpno}), is {\em not} a physical operator.
In terms of the matrix notation, on the sector of
``allowed'' states $\bcV$, since $(\Pphihat{}^\pm)^\star=\Pphihat{}^\pm$,
$\,\Pphihat^\star=\Pphihat$ is a physical operator; whereas on the sector of
forbidden states $\tcV$,
 \be   \label{ah:tpnostar2}
  (\Pphihat)^\star=
    \left(\ba{cc} (\Pphihat^+)^\star & 0 \\ 0 & (\Pphihat^-)^\star \ea\right)
    = \left(\ba{cc} \Pphihat^- & 0 \\ 0 & \Pphihat^+ \ea\right)
 \ee
does not leave the space of physical states invariant.

Clearly something peculiar is happening here. We have a complete set of
physical states, and a complete algebra $\cAp\subset\cA$ of physical operators
generated by $\{\rhat, \prhat, \ptthat, \phihat, \Pphihat\}$. Further, the
$\star$-involution on \cA\ has a well-defined action on \cAp\, and induces a
map from \cAp\ into \cA. What fails however, is that the induced $\star$ is
{\em not an involution} on \cAp; its action on one of the generators of
$\cAp$, namely $\Pphihat$, takes it out of \cAp. (This can be understood in
terms of the Hamiltonian vector field of the observable, see Fig.4.2.)
The algebra of physical
observables, \cAp, does not admit a $\star$-involution induced from \cAs. Thus
there is no sensible way to formulate the Hermiticity conditions on physical
operators in terms of an inner product on physical states.

In remark 6, section 2.4, I noted that if the constraint is real and if the
physical operators commute strongly with the constraint, then a
$\star$-involution is induced on \cAp. However, for the A-H model, as we have
just seen, there is no $\star$-involution on \cAp, inspite of the fact that it
{\it appears} to satisfy the above sufficiency conditions. What goes wrong
here? While it is true that the constraint is real, in the representation we
have chosen the physical states are solutions not just of the constraint
operator, but of one of its two square-roots. These square-root constraints are
not necessarily real, thus violating the above-mentioned conditions which
guarantee an induced $\star$-involution on \cAp.

On the face of it, due to the above mathematical inconsistency, i.e.\ the lack
of a $\star$-involution on \cAp, one cannot proceed with the quantization
program. Since the difficulty arises due to the sector of ``forbidden'' states,
one way out would be to discard them on mathematical grounds. However, this is
somewhat unsatisfactory since there appears to be no compelling physical reason
to do so. We would be ruling out, {\it by fiat}, precisely the ``tunnelling''
states that we were looking for.

An alternative approach would be to try and implement the $\star$-relations on
the other physical observables, and then see if it leads to a mathematically
and physically sensible framework. Forward-flashing to the denouement,
basically we will find that the measure on the forbidden region $\tcbar$
is 0, and on the remaining states, $\Pphihat{}^\star$ is an observable. I will
now
show this in detail.

For the purposes of the analysis above, we had decomposed $\cV$ into two
linearly independent parts $\cV^+$ and $\cV^-$. Each of these sectors carries
an irreducible representation of $\cAp$, and we are tempted to consider an
inner product in which $\cV^+$ and $\cV^-$ are mutually {\em orthogonal}. This
would be justified {\em if} we knew that $\cV^\pm$ were the eigenspaces of some
operator which is expected to be Hermitian or unitary. Thus, we are led to look
for a physical operator
which corresponds to a discrete symmetry of the constraint.

An obvious symmetry is reflection in the $x$-$y$ plane, $I_z:\theta\mapsto
\pi-\theta$. In quantum theory, the corresponding operator is represented by
$\hat{I}_z\circ\Psi(r,\theta,\phi)=\Psi(r,\pi-\theta,\phi)$. It is manifestly a
physical operator, and since classically $I_z^2=1$, it should be both Hermitian
and unitary, and its eigenspaces should be orthogonal. Unfortunately,
$\psi_\pm$ are not eigenspaces. In fact, the even and odd physical eigenstates
(with eigenvalues $+1$ and $-1$ respectively) are of the form
 \bea  \label{ah:ve}
  \psi_e&=k_e(r,\phi)\cos[{\textstyle \frac{\sqrt{R}}\hbar} (\theta-\pi/2)]
        &=\left(\ba{c}
          \hphantom{-} \frac12 k_e e^{-\frac{i\pi}{2\hbar}\sqrt{R}}\eip \\
          \hphantom{-} \frac12 k_e e^{+\frac{i\pi}{2\hbar}\sqrt{R}}\eim
                                                             \ea\right) \\
       \label{ah:vo}
  \psi_o&=i k_o(r,\phi)\sin[{\textstyle \frac{\sqrt{R}}\hbar} (\theta-\pi/2)]
        &=\left(\ba{c}
          \hphantom{-} \frac12 k_o e^{-\frac{i\pi}{2\hbar}\sqrt{R}}\eip \\
                      -\frac12 k_o e^{+\frac{i\pi}{2\hbar}\sqrt{R}}\eim
                                                             \ea\right).
 \eea
Hence, $\cV=\cV_e\oplus\cV_o$ should be an orthogonal decomposition in the
desired inner product.

Is $\hat{I}_z$ ``super-selected'' in the sense that it commutes with all other
observables? If it were, then its eigenspaces would carry irreducible
representations of the algebra \cAp\ of (continuous) Dirac operators. However,
while it does commute with $\rhat, \prhat, \phihat, \Pphihat$, it does not
commute with $\ptthat$. In fact, they anticommute, $[\ptthat,\hat{I}_z]_+=0$.
Since $\ptthat$ is not diagonal in the even/odd states, I will continue to use
the $+/-$ decomposition, even though $\cV^+$ and $\cV^-$ are
not obviously orthogonal.

Let us consider as an ansatz, an inner product of the form
 \bea  \label{ah:ip1}
  \IP{\psi_1}{\psi_2}=&\hphantom{+}
{\displaystyle\int_{\scriptstyle\Cbar}}\>(\mu_+\kopb\ktp
        + \mu_-\komb\ktm + \mu_{+-}\kopb\ktm + \ovr\mu_{+-}\komb\ktp) \cr
                     & +
{\displaystyle\int_{\scriptstyle\tcbar}}\>(\tmu_+\kopb\ktp
        + \tmu_-\komb\ktm + \tmu_{+-}\kopb\ktm + \ovr\tmu_{+-}\komb\ktp)
 \eea
where the integration is \wrt\ $d r\wedge d\theta\wedge d\phi$; the
$\mu=\mu(r,\theta,\phi)$ have arbitrary dependence on all three coordinates;
and where factors of $\exp(-\frac{2i}\hbar\sqrt{R}\theta)$,
$\exp(-\frac{2}\hbar\sqrt{|R|}\theta)$, and
$\exp(\frac{2}\hbar\sqrt{|R|}\theta)$ have been absorbed into $\mu_{+-}$,
$\tmu_+$ and $\tmu_-$ respectively. The inner product can be positive definite
\iff\ the ``diagonal'' measures are positive and the ``off-diagonal'' measures
satisfy
 \be  \label{ah:ip>0}
  |\mu_{+-}|<\sqrt{\mu_+\mu_-} \quad\hbox{and}\quad
            |\tmu_{+-}|<\sqrt{\tmu_+\tmu_-}.
 \ee

(In terms of the matrix notation, the inner product corresponds to
 \be  \label{ah:ipmat}
  \IP{\psi_1}{\psi_2}= \ba{cc} (\>\kopb & \komb\>) \\ & \ea
      \left(\ba{ll} \mu_+ & \mu_{+-} \\ \ovr{\mu}_{+-} & \mu_- \ea \right)
      \left(\ba{c}  \ktp \\ \ktm \ea\right)
 \ee
Suppose we write a general physical state as $\psi=\psi_+ + \psi_-$, and
postulate the natural inner product $\IP{\psi_1}{\psi_2}=\int_{\Cbar} \mu
\>\ovr{\psi_1}\,\psi_2 +\int_{\tcbar} \tmu \>\ovr{\psi_1}\,\psi_2$. In the
above
notation, this corresponds to choosing a matrix all of whose components are
equal up to phase. A short calculation, attempting to impose the Hermiticity
conditions on $\ptthat$, shows that there is in fact no such inner product.
Therefore one {\em has} to work with the general form (\ref{ah:ip1}).)

Now let us impose the Hermiticity conditions on the observables. Clearly,
$\rhat$ and $\phihat$ are essentially self-adjoint. Requiring that $\prhat$ is
symmetric leads to the condition that $\d\mu/\d r=0$ for all terms in the inner
product. Next, let us consider in detail the Hermiticity of $\ptthat$. Since
physical states are specified entirely by their coefficients $k_\pm$, we can
represent operators entirely by their action on $k_\pm$:
 \bea  \label{ah:krep}
  \rhat\circ k_\pm = & r k_\pm, \hphantom{\dfrac\hbar{i}\dfrac\d\d} \quad
              \hat\phi\circ k_\pm = & \phi k_\pm; \\
  \prhat\circ k_\pm = & \dfrac\hbar{i}\dfrac\d{\d r} k_\pm,\quad
  \Pphihat\circ k_\pm = & \dfrac\hbar{i}\dfrac\d{\d\phi} k_\pm\ ;
 \eea
and, in particular,
 \bea  \label{ah:krep2}
  \ptthat\circ \left(\ba{c} k_+ \\ k_- \ea\right)
     = & \left(\ba{c} +\sqrt{R} k_+ \\ -\sqrt{R} k_- \ea\right) \quad
           & \hbox{on}\quad \Cbar \cr
     = & \left(\ba{c} +i\sqrt{|R|} k_+ \\ -i\sqrt{|R|} k_- \ea\right) \quad
           & \hbox{on}\quad \tcbar.
 \eea

We want to find an inner product such that $\bra{\psi_1}\ptthat\ket{\psi_2} =
\bra{\psi_1}\ptthat{}^\dagger\ket{\psi_2}$  for all $\psi_1,\psi_2$. Consider
states in $\cV_+$, i.e. $\kom=\ktm=0$. Then
 \be  \label{ah:iptp}
  \bra{\psi_1}\ptthat\ket{\psi_2} =
       \int_{\Cbar}\>\mu_+  \>\kopb\,\sqrt{R}\,\ktp
     +i\int_{\tcbar}\>\tmu_+\>\kopb\,\sqrt{|R|}\,\ktp.
 \ee
On the other hand, by definition
 \be  \label{ah:iptpstar}
\bra{\psi_1}\ptthat^\dagger\ket{\psi_2}:=\ovr{\bra{\psi_2}\ptthat\ket{\psi_1}}
   =   \int_{\Cbar}\>\mu_+  \>\kopb\,\sqrt{R}\,\ktp
     -i\int_{\tcbar}\>\tmu_+\>\kopb\,\sqrt{|R|}\,\ktp.
 \ee
Clearly, $\ptthat$ can be Hermitian on $\cV_+$ if and only if $\tmu_+=0$.
Similarly, one can conclude that $\tmu_-=0$. Thus, both the diagonal terms in
the forbidden region vanish. But now, due to the condition (\ref{ah:ip>0}), in
the forbidden region the cross term also vanishes, $\tmu_{+-}=0$. {\em In the
inner product {\rm (\ref{ah:ip1})}, the only terms that survive are the
integrals
over the classically allowed region!} Thus, while physical solutions can indeed
have support in the classically forbidden region $\tcbar$, this support is of
measure zero in the physical inner product.

Let us return to the Hermiticity of $\ptthat$. Consider states
$\psi_1=\psi_{1+}$ and $\psi_2=\psi_{2-}$, i.e. $\kom=\ktp=0$. Then,
 \be
  \bra{\psi_1} \ptthat \ket{\psi_2}
          = \int_{\Cbar}\> \mu_{+-}\, \kopb(-\sqrt{R}) \ktm.
 \ee
However,
 \be
  \bra{\psi_1} \ptthat^\dagger \ket{\psi_2} := \ovr{\bra{\psi_2}
        \ptthat \ket{\psi_1}} = \int_{\Cbar}\>\mu_{+-}\, \ktm(\sqrt{R})\kopb.
 \ee
Equating
the two, we find that $\ptthat$ is Hermitian if and only if $\mu_{+-}=0$. Thus,
it is the Hermiticity of a continuous operator which implies that the subspaces
$\cV_+$ and $\cV-$ are orthogonal. (Since they are in fact orthogonal, clearly
the system admits another discrete symmetry, albeit a hidden one.)

Imposing the $\star$-relations on $\rhat,\prhat,\ptthat,\phihat$ as Hermiticity
conditions on the representation, we have reduced the form of the physical
inner product (\ref{ah:ip1}) to
 \be  \label{ah:ipred}
  \IP{\psi_1}{\psi_2}=\int_{\Cbar}\>(\mu_+\,\kopb\,\ktp + \mu_-\,\komb\,\ktm),
 \ee
where $\mu=\mu(\theta,\phi)$. Since the measure in the classically forbidden
region vanishes, as elements in the Hilbert space physical states can be
specified
entirely by their support on $\Cbar$. In other words, by restricting the
support of physical states to the classically allowed region, in this problem
we do not lose any elements of the physical Hilbert space. Thus, as we saw
earlier (\ref{ah:tpyes}), on states with support only on $\Cbar$,
$\Pphihat^\star$ is a physical operator, in fact $\Pphihat^\star=\Pphihat$.
Since, in terms
of $k_\pm$, the action of $\Pphihat$ is
 \be
  \Pphihat\circ k_\pm = \frac\hbar{i} \frac\d{\d\phi} k_\pm .
 \ee
it is now trivial to check that $\Pphihat$
is symmetric if and only if $\d\mu_\pm/\d\phi =0$.

Thus, the measure depends only on $\theta$, and furthermore, this dependence is
undetermined by any Hermiticity conditions. How has this come about? In
the quantization of
ordinary unconstrained systems, the dependence of the measure on a coordinate
$\theta$ is determined by requiring that the corresponding momentum operator
$\ptthat$,
which is represented by $\frac\hbar{i}\d/\d\theta$,
should be Hermitian. In the quantum theory of constrained systems,
one is finally only interested in the measure on physical states. In this
example, the momentum operator $\ptthat$ is represented {\em on physical
states} by a multiplication (\ref{ah:krep2}). Hence there is no condition that
determines the $\theta$-dependence of the measure $\mu$.

Now, since the coefficients $k_\pm$ do not depend on $\theta$ either,
we can choose $\mu_\pm(\theta)$ such that the integral over $\theta$ can be
performed trivially, and the inner product is thus reduced to
 \be \label{ah:ipred2}
  \IP{\psi_1}{\psi_2}= \mu_+\int_{\ovr\phi}\>d r\wedge d\phi \, \kopb\ktp
                     + \mu_-\int_{\ovr\phi}\>d r\wedge d\phi \, \komb\ktm,
 \ee
where $\ovr\phi$ indicates that the integral is performed only over classically
allowed values of $\phi$.

Is there a criterion that will fix the relative weights of the two terms? Note
that we have not yet imposed the Hermiticity of the discrete symmetry
$\hat{I}_z$. Since its eigenspaces must be orthogonal, we have
 \be  \label{ah:orth}
  0=\IP{\psi_e}{\psi_o}=\sfr14[\mu_+ - \mu_-]\int_{\ovr\phi}\>\ovr{k}_e\,k_o.
 \ee
Thus we can choose $\mu_+=\mu_-=1$, and the final form for the inner product is
 \be  \label{ah:ipfin}
  \IP{\psi_1}{\psi_2}= \int_{\ovr\phi}\> d r\wedge d\phi \>
       [\kopb\,\ktp + \komb\,\ktm].
 \ee

\subsection{Remarks}

Let us first briefly review the process by which we have obtained a complete
quantum theory for this model. We chose a representation of the elementary
operators, on some vector space of complex valued functions on the
configuration space. In this representation, the constraint equation is a
second order partial differential equation, which we solved
explicitly. This set of solutions is ``large'' in the
sense that it includes
the tunnelling solutions, which penetrate the forbidden region.
Next, we constructed a set of generators of \cAp, the algebra of physical
observables. These operators act on the
``large'' space of solutions and leave it invariant. Then we attempted to
induce the $\star$-involution on \cAp, from \cA. The $\star$s of most of the
generators of\cAp\ were also in \cAp. However, the $\star$ (evaluated in \cA)
of one of the physical operators was no longer a physical operator itself.
Thus, we were unable to induce on \cAp\ the structure of a $\star$-algebra.
This appeared to be an impass\`e, in terms of constructing the physical inner
product {\it a la} the prescription of the algebraic approach.

In desperation, at this point we attempted to implement part of the
$\star$-relations as Hermiticity conditions on {\it some} of the physical
operators. Quite unexpectedly, these conditions forced us to rule out the
tunnelling solutions. The Hermiticity condition on one of the observables
leads to the conclusion that {\it all solutions} to the constraint equation
{\it cannot} be
considered as {\it physical states}. Finally, the algebra of operators on
the ``smaller'' space of physical
states --which does not include the tunnelling solutions-- {\it does} admit a
$\star$-involution. Hence we were able to complete the quantization program.

Now, let us consider the Hamiltonian (\ref{ah:ham}) in quantum theory.
It is manifestly positive. On physical states the
Hamiltonian operator is $\wh{H}\circ\psi=E(\phi)\cdot\psi$. Since the states
have (measurable) support only in $\Cbar$, where $E(\phi)\ge0$, $\langle\wh{H}
\rangle\ge0$ for all physical states. As promised, a careful re-analysis of the
problem has removed the peculiarities that were present in the previous
quantization. While the final result seems somewhat disappointing (there are no
extreme
tunnelling states in the classically forbidden region, and the positivity
theorem is not violated in quantum theory), this problem clearly
demonstrates the power of the quantization program of chapter 2. While in
retrospect our final
result could have been obtained by a number of other quantization
procedures, in other schemes the elimination of the spurious tunnelling states
would almost certainly be an {\it ad hoc} step.

In fact, based on an analysis of a similar model, which is perhaps even more
peculiar than the A-H model,
Gotay \cite{mjg:ag4} proposes exactly such a requirement for quantum theory:
{\it by fiat}
the measure is restricted to the classically allowed region. On the other hand;
in the algebraic
approach to quantization, this result is {\em derived} from a much more
general hypothesis: namely that real physical operators should be Hermitian in
the physical inner product.

In another re-analysis of this model, Boulware
\cite{db:ag4} assumes the usual Euclidean inner product on the representation
space, and requires that $p_\theta$ be a Hermitian operator on the space of
{\it all} $L^2(S^2)$ states%
\footnote{$r$ and $p_r$ can be safely ignored in this model, since they ``just
go along for the ride''. Thus the initial configuration space is $S^2$.},
{\em before} solving the constraint equation
and isolating the physical states. Then, physical states have support only on
the classically allowed region, and there is no tunnelling in this theory.
As in the case of Gotay's analysis, the absence of tunnelling is simply an
input. Furthermore, now a new problem arises.
A symmetric extension of $\hat{p}_\theta$ was chosen by defining its domain
to consist of states which satisfy the following condition:
 \bdm
   \left. \frac{\d}{\d\theta}\Psi(\theta,\phi)\right|_{\theta=0,\pi}=0.
 \edm
That is, the $\theta$-derivative, of states in the domain of $\hat{p}_\theta$,
vanishes on the $z$-axis. $\hat{p}_\theta$ so-defined, clearly
has discrete (integer) eigenvalues. Let us return to the constraint equation
(\ref{ah:con},\ref{ah:qcon}), $(\hat{p}_\theta^2-R(\phi))\circ\Psi=0$.
Due to the discrete spectrum of $\hat{p}_\theta$, the support of
physical states is further restricted to those regions where
$R(\phi)$ is the square of an integer! However, these physical states, which
have support only on a
discrete set of longitudes on $S^2$, are {\em not normalizable}
\wrt\ the implicit measure on $S^2$. Hence, this quantum theory is incomplete;
a new inner product has to be introduced. Furthermore,
since the $\theta$ dependence of the physical states is
determined by the constraint, these states are labelled by functions of a
{\em discrete set of points on $S^1$} (where $R(\phi)=n^2$) and, for generic
choices of $R(\phi)$,  the resulting
Hilbert space is {\em finite dimensional}! However, as we
will see,
the reduced \ps\ is a perfectly well-defined co-tangent bundle over $S^1$. Thus
the above quantum theory does not appear to capture all the physics in the
model.

In the quantum theory we constructed here, since $p_\theta$ is an
observable (as much as $\phi$ is) we too required it to be Hermitian, {\it but
only on physical states}. In the context
of the algebraic approach, the most important facet of the Hermiticity of
$p_\theta$ is the following: as we saw in (\ref{ah:tpnostar}), unless we impose
the Hermiticity of $p_\theta$, the rest of the formulation is {\em not
mathematically well-defined}.

I will now address the question: ``How well-defined is the constraint operator
(\ref{ah:qcon}) in the representation (\ref{ah:rep})?''.  First of all, in
ordinary quantum mechanics, there is an inner product  available to make the
``elementary'' operators like (\ref{ah:rep}) symmetric and to define their
domains. However, in the quantum theory of a constrained system, we are
eventually interested only in the physical operators. In general the elementary
operators do not correspond to physical observables, and we may be able to
ignore or postpone the issue of whether or not they are well-defined. At the
very least though, to find physical states, we need a reasonably well-defined
constraint operator, which --as is usually the case-- may be quadratic in
momenta and thus may pose even more difficulties. Recall that there are
considerable ambiguities in constructing operators quadratic in momenta. Again,
in ordinary quantum mechanics, one is able to eliminate these ambiguities, to a
large degree, by using the inner product to define symmetric extensions or to
specify the domains of the operators.  However, in the quantum theory of a
constrained system, one does not have an inner product on the representation
space \cV. One has considerable freedom, though, to define \cV, and the
operators thereon. Hence, one has to make choices, based on physical intuition
for the system. Now one can proceed as follows: define a representation of the
operators on a certain vector space, and  if necessary enlarge the
representation space to include the range of these operators. As far as the
constraint operator is concerned, we are finally interested only in its kernel,
and as long as this is large enough, the enlargement of \cV\ is immaterial. As
far as other physical operators are concerned, since we are finally interested
only in their actions on physical states, we need not define them too precisely
until after the constraint has been solved.

The quantization of a constrained system is inherently an ambiguous process.
One
can only require that the final quantum theory is ``complete and consistent'',
in the sense that one has a faithful $\star$-representation of a suitably large
algebra of observables and that one recover the classical description in a
suitable
limit. To achieve this end, one is justified in riding
roughshod over the initial stages of the road to quantization.

Thus, one can view the quantization of the A-H model in
the following way: In the early part of section 2, I exploited the freedom
to define the representation space and the operators,
and was intentionally obscure about the specification of \cV.
We then made some choice of factor-ordering for the constraint and solved it on
this space. On physical states, we defined the actions of
operators which formally had vanishing commutators with the constraint. Then we
imposed Hermiticity conditions on these and found appropriate self-adjoint
extensions of the physical observables.

The only remaining thing is to check the subsidiary result, whether Dirac
quantization and reduced space quantization indeed
yield inequivalent answers.

\mysection{(\dag) Reduced space quantization}

Recall from section 2.5 that in reduced space quantization,
one solves the constraint classically and
finds the reduced \ps\ \Ghat, and then quantizes the resulting system, free of
constraints. There are two steps involved in constructing the reduced \ps.
First one has to find, i.e. construct an explicit parametrization of, the
constraint surface \Gbar. Then, one has to factor out \Gbar\ by the gauge
orbits generated by the constraint. The space of orbits on the constraint
surface is the reduced \ps. Let us use this approach to quantize the above
model.

As has been mentioned before, the projection of the constraint surface to the
configuration space is a proper subset of the configuration space. The
constraint (\ref{ah:con}) can only be solved in the classically allowed regions
where the
``potential'' $R(\phi)$ is positive. Let us focus on one such region. In this
region, the constraint $C=p_\theta^2-R(\phi)=0$ can be solved for $p_\theta$ in
terms of $\phi$,
 \be \label{ah:clsol}
   p_\theta^\pm=\pm\sqrt{R(\phi)}
 \ee
while the other \ps\ coordinates $r,p_r,\theta,p_\phi$ are free. \Gbar\ appears
to have two ``sectors'', $\bar\Gamma_\pm$,  corresponding to the choice of sign
in
(\ref{ah:clsol}). However, these two sectors are obviously connected at the
points (denoted by $\phiz$) where $R(\phi)=0$. The question then is whether or
not the two halves of the constraint surface are {\em smoothly} joined at
$\phiz$. To
answer this question, let us consider the intersection of the constraint
surface with a $(p_\theta,\phi)$ plane, and compare the slopes of the two
halves of the constraint surface at $\phiz$. We see that $(\d p_\theta /\d
\phi)
|_{\Gbar_\pm} = (\sqrt{R})' = \pm R'/2 \sqrt{R}$. At $\phiz$, both slopes are
infinite. Since the $R>0$ region is obviously bounded on both sides by $\phiz$,
and the two halves are `smoothly' (at least $C^1$) joined there, there is in
fact just one constraint surface. Thus, the topology of the constraint surface
is non-trivial, it contains one $S^1$ component%
\footnote{I am assuming that $\phiz$ is not a critical point, i.e. $R'(\phiz)
 \not= 0$. If $\phiz$ is a first order critical point, $R'=0,\>
 R''\not=0$, the analysis becomes much more complicated. The constraint surface
 is no longer a (Hausdorff) manifold. In fact the corresponding cross-section
 is a ``figure-eight''; or worse if $\phiz$ is a higher order critical point.}
(see Fig.4.1).
Clearly, this is a wholly different picture from the one we obtained in Dirac
quantization.
\begin{figure}
\vspace{3in}
\caption{A-H model: Constraint surface}
\end{figure}

Now let us analyse the gauge orbits.
The Hamiltonian vector field of the constraint is
 \be  \label{ah:hvf}
  X_C=2p_\theta\left(\frac\d{\d\theta}\right) + R'\left(\frac\d{\d
          p_\phi}\right).
 \ee
Let $t$ be an affine parameter along the Hamiltonian vector field. It is easy
enough to find the gauge orbits, since $r, p_r, p_\theta, \phi$ are constant.
The orbits are then given by
 \bea
  \label{ah:orbit1}
    \theta &=& 2p_\theta t + \Theta \\
  \label{ah:orbit2}
    p_\phi &=& R't + \Pphi.
 \eea
The orbits start on the positive $z$-axis and follow the lobes of the
constraint surface down to the negative $z$-axis. Since the Hamiltonian vector
field has no components in the $p_\theta$ or $\phi$ directions, it does not
close on itself around the $S^1$ component of \Gbar. Thus the reduced \ps\ (the
space of orbits) also has nontrivial topology, with one $S^1$ component. (In
fact, as we will see later, $\wh\Gamma=T^*(S^1\times\real_+)$.) Our objective
now is to find reduced space coordinates which make this structure
manifest. In the absence of a specific form for $R(\phi)$, this is difficult to
do. Let us then do the next best thing, which is to split \Gbar\ into the two
``halves'' $\Gbar_\pm$, and concentrate henceforth on $\Gbar_+$.

On the constraint surface $\Gbar_+$, one can specify an orbit by the initial
values of the {\em five} independent functions $r,p_r,\phi,p_\phi,\theta$. (The
initial value of $p_\theta=+\sqrt{R(\phi)}$ is determined by that of $\phi$.)
However, the space of orbits is only four dimensional. Thus, of the five
constants of motion corresponding to the initial values of the independent
functions on \Gbar, we have to choose the classical Dirac observables, i.e.
four of these five constants which will coordinatize the reduced \ps. The
function whose initial value is not included in the $\wh{\Gamma}$ coordinates
will play the role of a parameter along the orbit, and should thus ideally be
monotonic in the affine parameter $t$. For this model we have only two choices,
$\theta$ or $p_\phi$. Since the constraint is solved for
$p_\theta$, a natural choice would be to consider $\theta$ as an affine
parameter and use $(r,p_r,\phi,\Pphi)$ as the coordinates on (a half of)
$\wh{\Gamma}$.
Since in fact $\theta$ is monotonic in $t$ almost everywhere on
\Gbar\ (the exception is at the points $\phiz$), this is a reasonable choice.
The value of the fifth constant of motion $\Theta$ then
serves to fix the $t=0$ cross-section of the gauge orbits.
Setting $\Theta=0$, we can solve the orbit equation (\ref{ah:orbit1}) for $t$
in terms of
$\theta$, $t=\theta/2p_\theta$. Substituting this expression in
(\ref{ah:orbit2}), we can solve for the only non-trivial Dirac observable,
 \be   \label{ah:clobs}
  \Pphi^+=p_\phi-\frac{R'\theta}{2\sqrt{R}}.
 \ee

(Now, I can explain classically why --in the Dirac approach-- $\Pphihat^\star$
is
not an observable. Recall that physical observables are functions on $\Gbar$
whose Hamiltonian vector fields are tangential to the constraint surface.
Consider the Hamiltonian vector field of $\Pphi^+$:
 \be
  X_{\Pphi^+}=\frac\d{\d\phi}+ (\sqrt{R})'\frac\d{\d p_\theta}
    +(\sqrt{R})''\theta\frac\d{\d p_\phi},
 \ee
and let us focus on its projection to the $p_\theta-\phi$ space (see Fig.4.2.)
\begin{figure}
\vspace{3in}
\caption{A-H model: Constraint surface and Hamiltonian vector field of
$\Pphi^+$ in $p_\theta-\phi$ space}
\end{figure}
Note that in the $R>0$ region, $X_{\Pphi}$ is real, and hence its complex
conjugate is also tangential to $\Gbar$. However, in the forbidden region
$R<0$, $X_{\Pphi}$ is {\it complex}, and as we can see from Fig.2, its complex
conjugate is {\it not} tangential to $\Gbar$. Hence, the corresponding
generating function, and the operator $\Pphihat^\star$, are not observables.)

The reduced \ps\ $\wh{\Gamma}_+$ is coordinatized by $(r,p_r,\phi,\Pphi^+)$.
The symplectic structure (or equivalently, the Poisson brackets) can be easily
calculated, and we find that $p_r,\Pphi^+$ are the momenta canonically
conjugate
to $r,\phi$ respectively.

Since the Dirac observables (both classical and quantum) are at most linear in
momenta, there is no factor ordering ambiguity, and the reduced space quantum
theory in the above coordinatization of $\wh{\Gamma}$ will be equivalent to the
Dirac quantum theory. In the obvious configuration representation the
states are represented by $k_\pm = k_\pm (r,\phi)$. Let us label the two
sectors
by $\cV_\pm$ respectively. On each sector
the operators corresponding to the reduced space coordinates
can be represented by
 \bea  \label{ah:redrep}
  \rhat\circ k_\pm = & r k_\pm, \hphantom{\dfrac\hbar{i}\dfrac\d\d{}_\phi}
\quad
              \hat\phi\circ k_\pm = & \phi k_\pm; \\
  \prhat\circ k_\pm = & \dfrac\hbar{i}\dfrac\d{\d r} k_\pm,\quad
  \Pphihat\circ k_\pm = & \dfrac\hbar{i}\dfrac\d{\d\phi} k_\pm .
 \eea
Note that the algebra of observables and the representation is equivalent to
the
quantum Dirac operator algebra (\ref{ah:krep2}).

Note that in the reduced space theory, there is no question of any cross terms
(in the inner product) between the two sectors $\cV_\pm$: the states simply
live on different spaces. Thus, in the reduced space theory, $\cV_+$ and
$\cV_-$ are naturally mutually orthogonal. On the other hand, it is the
symmetry $I_z$, which was so natural in the Dirac theory, that is ``hidden''
here. It shows up as the natural map between $\hat{\Gamma}_+$ and $\hat{
\Gamma}_-$. The spaces $\cV_\pm$ each carry irreducible representations of the
algebra of
functions on one half of the reduced \ps, and the relative scale factor
between the terms in the inner product is fixed by requiring that $I_z$ is
Hermitian. A brief calculation shows that the
inner product is exactly that given by (\ref{ah:ipfin}).

However, as I indicated above, this is an incomplete way to carry
out reduced space quantization of this model. On each half of the constraint
surface
$\Gbar_\pm$, the Hamiltonian vector field of the classical Dirac observable
of $\Pphi$ is given respectively by
 \be\label{ah:phvf}
  X_{{\rm P}_\phi}=\frac\d{\d\phi} \pm \frac{R'}{2\sqrt{R}}
       \frac\d{\d p_\theta} + (\sqrt{R})''\theta\frac\d{\d p_\phi}.
 \ee
Its projection to the $\phi-p_\theta$ plane is then simply
 \be\label{ah:phvfp}
  X_{{\rm P}_\phi}=\frac\d{\d\phi} \pm \frac{R'}{2\sqrt{R}}
       \frac\d{\d p_\theta}.
 \ee
To see the incompleteness, consider
the following quantization of a particle whose configuration space is the unit
$S^1$ in the $x-y$ plane.
(In terms of the variables we are using, the analogy is $\phi\leftrightarrow x,
\, p_\theta\leftrightarrow y$.)
Now, the analogous quantization of the particle on $S^1$ would be the
following:
First split $S^1$ into the top and bottom halves ($x>0,x<0$); and use in
each half, $x$ as the configuration variable and $\frac\d{\d
x} \pm \frac{x}{y}\frac\d{\d y}$ as the vector field defining the momentum in
each half (in
analogy with the Hamiltonian vector field (\ref{ah:phvfp})
of the observable $\Pphi$).
The correct matching boundary conditions at $x=\pm 1$
are difficult, if not impossible to state in terms of this chosen momentum.
Even if one used the angular coordinate in the $x-y$ plane,
and its conjugate momentum, a naive
quantization would yield a spectrum for the momentum operator consisting of
only the even eigenvalues; because of
the two sectors, this spectrum would also be doubly degenerate. However, as we
well know, the spectrum for the angular momentum consists of all the integers,
and is nondegenerate.

For definiteness, consider a model where, in a neighbourhood of the
constraint surface, the ``potential'' is given by
$R(\phi)=a^2-(\phi-\pi)^2$, and $a<\pi$.
(The specific form of $R(\phi)$, as long as it is smooth, will not change any
of the
topological considerations.) On the $(p_\theta,\phi)$ cylinder,
the projection of the constraint surface is a circle. In a neighbourhood
of the constraint surface, introduce angular and radial coordinates
 \bea \label{ah:polar}
  \rho   = & \sqrt{p_\theta^2+(\phi-\pi)^2} \\
  \alpha = & \tan^{-1}(p_\theta/(\phi-\pi)
 \eea
and their corresponding conjugate momenta,
 \bea
  p_\rho&=(-\theta p_\theta +(\phi-\pi) p_\phi)/\rho \\ \label{ah:mom}
  p_\alpha&=-\theta(\phi-\pi)-p_\theta p_\phi.
 \eea
Then the constraint surface is
specified by $\rho-a=0$ and the Hamiltonian vector field of the constraint is
$X_C=-(\d/\d p_\rho)$. The classical Dirac observables are
$r,p_r,\alpha,p_\alpha$, and thus the reduced \ps\ is a cotangent bundle, over
the configuration space $S^1\times\real^+$. In the quantum theory, the
eigenvalues of $p_\alpha$ are integer multiples of $\hbar$.

In the Dirac quantum theory, even if one is lucky enough to find the operators
in the above form, or if one simply {\em uses} the above classical functions to
motivate the corresponding operators, it is clear from the expressions
(\ref{ah:polar}-\ref{ah:mom}) that in the representation of section 2 a good
factor ordering would be extremely difficult to construct. In fact, the
simplest or the only consistent representation and factor ordering may well be
the one from the reduced quantum theory.

In conclusion, while the Dirac and reduced space quantum theories for this
model are ``locally'' equivalent, in the sense that they are unitarily
equivalent representations of the continuous physical operators, subtle
differences arise due to the nature of the discrete symmetries. In addition,
there are gross differences between the two since the topology of the reduced
\ps\
is nontrivial.

%
\chapter{COUPLED OSCILLATORS: CONSTRAINED ENERGY DIFFERENCE}
\pagestyle{myheadings}
\markboth{{\sf Chapter 5}}{{\sf Coupled Oscillators}}


\def\rmfrac{{\rm frac}}

\def\ha{\hat{a}} \def\hc{\hat{c}} \def\hJ{\skew4\hat{J}} \def\hN{\hat{N}}
\def\hC{\hat{C}}  \def\hL{\hat{L}}

\mysection{Introduction}

In this chapter I will consider the system consisting of two harmonic
oscillators with the same frequency, set to 1 for simplicity. In the main
portion of this
chapter (sections 2 to 5) I will consider the model in which the two
oscillators are coupled to each other via a first class constraint on the
energy difference. This model mimics some of the features of \gr\ in the
geometrodynamical variables: i) the constraint is quadratic in momenta, ii) the
kinetic piece of the constraint is of indefinite signature and iii) the
potential
is of indefinite sign. Due to these similarities, some of the results of the
quantization of this model
are of qualitative interest in \qg\ or quantum cosmology.

As a matter of fact, this model arises when one
conformally couples a massless scalar field to the Friedman-Robertson-Walker
universe with $S^3$ spatial topology (section 4). Though for this cosmological
model the
energy difference --of the two ``effective'' oscillators-- is fixed to be zero,
we will consider generic (and possibly more interesting) models in which the
constrained energy difference is {\it any} real number.

The \ps\ of the system is described by
position and momentum coordinates $(x_I,p_I,\>I=1,2)$ and the first class
constraint is
 \be \label{dif:con}
  \textstyle{1\over 4}(p_1^2+x_1^2-p_2^2-x_2^2)=\delta,
 \ee
where $\delta$ can be viewed as the (real) difference in energies of the two
oscillators. This model has a number of interesting features:
\begin{enumerate}
\item This model is an example in which it is important to impose the
Hermiticity conditions only on physical states. (The A-H model of the previous
chapter is another such example.) If one imposes the Hermiticity
conditions on the elementary operators, prior to solving the quantum
constraint, then generically one does not obtain
{\em any} physical states (see section 2).
\item Unlike in the constrained systems we have considered so far,
the reduced \ps\ for this model is {\it not} a cotangent bundle.
Another novel feature is that the algebra of observables is {\it
over}complete, inspite of
the trivial topology of the reduced \ps;
it is the requirement that the set of generators of \cAp\ be closed
under the commutator Lie bracket that forces one to include an ``extra''
element in the set of generators of \cAp. Thus,
there is an algebraic relation on the physical observables,
\item In most of the models considered so far, \cAp\ has comprised of operators
which
correspond to generators of {\it continuous} canonical transformations on the
constraint surface. In this model, as in the A-H model,
there is in addition a {\it discrete}
classical symmetry of
the constraint (call it ``parity''), which maps entire gauge orbits to entire
gauge orbits. A
quantum operator corresponding to this symmetry must be included in the algebra
of observables, inspite of the fact that the set of continuous physical
observables is {\it locally} complete. The parity operator plays a role in
obtaining an inner product on the physical states.
\item One can introduce a first class Hamiltonian
which satisfies a classical positive energy theorem.
In the canonical quantum theory (section 2), there exist representations of
the physical
operator algebra in which the above Hamiltonian acquires negative energy
eigenvalues (section 3). However, going beyond the quantization program and
using additional physical conditions, on semi-classical grounds
one can rule out most --but not all-- of
these representations.
\item Classically, the parity transformation discussed above squares to unity.
Also, in classical theory we can impose any real value for the energy
difference $\delta$. These two conditions are incompatible
in the Dirac quantum theory (see section 3).
\item In section 4 I will construct the reduced space quantum theory. As we
will then
see, the two quantum theories are even {\it kinematically}
inequivalent. This inequivalence arises from
the specific form of the algebraic condition that the observables in the two
theories satisfy.
\end{enumerate}

In section 7, for the sake of completeness, I will quantize a related model,
in which the sum of the energies is imposed as a constraint.

\mysection{Dirac quantization}
Choose as the set \cS\ of elementary classical variables the standard
`creation' and `annihilation' functions on $\Gamma$
 \be \label{dif:var}
 z_I={\textstyle{1\over\sqrt 2}}(x_I-ip_I)\quad\hbox{and}\quad \bar{z}_I=
  {\textstyle {1\over\sqrt 2}}(x_I+ip_I),
 \ee
as well as the constant function. There are no ACRs on the elementary
operators. In these variables, the constraint function is
 \be\label{dif:zcon}
  C=\half(z_1\bar{z}_1-z_2\bar{z}_2)-\delta.
 \ee
The quantum $\star$-algebra \cAs\ is straight forward to construct. To make the
notation transparent we will denote the elementary quantum operators
$\hat{z}_I$ by $\hc_I$ and $\hat{\bar{z}}_I$ by $\ha_I$. \cAs\ is then
generated by the set of elementary quantum operators
($1,\ha_1,\hc_1,\ha_2,\hc_2$) satisfying the canonical commutation relations:
 \be \label{dif:ccr}
  [\ha_I,\ha_J]=0=[\hc_I,\hc_J]\quad\hbox{and}\quad [\ha_I,\hc_J]=
   \delta_{IJ},\quad   I,J=1,2;
 \ee
and subject to the $\star$-relation
 \be\label{dif:elemstar}
  \ha_I^\star=\hc_I.
 \ee
In terms of these operators, the quantum constraint we wish to impose is
 \be  \label{dif:qcon}
  \hat{C}\ket{\psi}_{phy}:=\left[\half(\hc_1\ha_1
  -\hc_2\ha_2)-\delta\right]\ket{\psi}_{phy}=0,
 \ee
where we have used normal ordering to resolve the ordering ambiguity. Actually,
since the constraint is the {\it difference} between the energies of the two
oscillators, then as long as we use the same ordering for each term
$z_I\bar{z}_I$, there is no ambiguity in the constraint operator.

The next step in the quantization program is to represent the algebra \cA\ by
means of concrete operators on a vector space \cV. Recall that the
$\star$-relations are ignored at this stage. Indeed, had we imposed them, i.e.,
had we introduced a $\star$-representation of \cAs\ by operators on a Hilbert
space, we would have run into the following difficulty: The number operators
$\hN_1:=\hc_1\ha_1$ and $\hN_2:=\hc_2\ha_2$ would have taken on only integral
values, and if $\delta$ were {\it not} an integer or half-integer, the only
state in the kernel of the constraint operator would have been the zero state.
Thus, we would have been led to the conclusion that \cVp\ is zero
dimensional.
This quantum theory is clearly incomplete, since in this case the reduced
  phase space is a 2-dimensional (non-compact) manifold; the system has one
  ``true'' degree of freedom.
Thus, our strategy of holding off the imposition of the $\star$-relations until
after the physical states are isolated is {\it essential} in this example to
obtain an acceptable quantum theory.

Let us choose the vector space representation of \cA\ as follows. Since any
complete set of commuting operators consists of only two of the elementary
operators, let us choose as $V$ the complex vector space spanned by states of
the form $\ket{j,m}$, where (to begin with) $j$ and $m$ are any {\it complex}
numbers, and represent the elementary quantum operators as follows:
 \bea \label{dif:rep}
  \ha_1\ket{j,m} &=& \alpha_1(j+m)\ket{j-\half,m-\half}, \cr
                \hc_1\ket{j,m} &=& \gamma_1(j+m+1)\ket{j+\half,m+\half}, \cr
                \ha_2\ket{j,m} &=& \alpha_2(m-j)\ket{j+\half,m-\half} \cr
   \hbox{and}\quad
  \hc_2\ket{j,m}&=&\gamma_2(m-j+1)\ket{j-\half,m+\half};
 \eea
where the coefficients, $\alpha_I(k)$ and $\gamma_I(k)$, functions only
of their argument $k$, are subject to the conditions
 \be  \label{dif:cond}
  \alpha_1(k)\gamma_1(k)=k \quad\hbox{and}\quad\alpha_2(k)\gamma_2(k)= k
 \ee
It is straightforward to check that the commutation relations (\ref{dif:ccr})
are satisfied by this choice of representation. The notation $\ket{j,m}$ to
represent the kets may seem strange at first. Note, however, that
$\hN_1\ket{j,m}=(m+j)\ket{j,m}$ and $\hN_2\ket{j,m}=(m-j)\ket{j,m}$. Each
$\ket{j,m}$ is an eigenket of the total number operator $\hN =\hc_1\ha_1 +\hc_2
\ha_2$ with eigenvalue $2m$, as well as of the constraint operator $\hat{C}$
with eigenvalue $j-\delta$. (Thus, had we represented states as Bargmann type
wave functions,
we would have $\psi(z_1, z_2):=\IP{z_1,z_2}{j,m}\equiv z_1^{m+j}z_2^{m-j}$.)
These angular momentum like states arise naturally
because the the constraint surface $\ovr\Gamma$ is the group manifold of
$SO(2,1)$ and the Poisson bracket algebra of physical observables is the Lie
algebra of $SO(2,1)$
(see section 5). This is the reason I chose the angular momentum like
$\ket{j,m}$ representation instead of the perhaps
more familiar $\ket{n_1,n_2}$ representation.

Since $\hC$ is diagonal in this representation,
with eigenvalues $j-\delta$, the quantum constraint is easy to solve.
A basis for the {\it physical} subspace
\cVp\ is given simply by the kets $\{\ket{\delta,m}\}$. Physical operators
are the elements of ${\cal A}$ that map \cVp\ to itself and should thus
maintain the difference in energies of the two oscillators. Clearly, operators
that raise and lower the energy of each oscillator by a unit, and an operator
that measures the total energy, are physical operators. Hence, consider the
algebra generated by the set $\{1,\hJ_+,\hJ_{-},\hJ_z\}$; where
$\hJ_+:=\hc_1\hc_2$ raises the energy of each oscillator by a unit,
$\hJ_-:= \ha_1\ha_2$ lowers the energy of each oscillator by a unit, and
$\hJ_z:=\half(\hN+1)$ is half the total energy.
The non-vanishing commutation relations between these operators are
 \be  \label{dif:occr}
  [\hJ_z,\hJ_\pm]=\pm\hJ_\pm\quad\hbox{and}\quad[\hJ_+,\hJ_-] =-2\hJ_z,
 \ee
from which it is clear that they generate the Lie algebra of $SO(2,1)$. $\hJ_+$
and $\hJ_-$ are the (angular momentum) raising and lowering operators,
respectively.
Note further that it is in order to obtain the $SO(2,1)$ commutation relations
in (\ref{dif:occr}) that I have chosen the definition $\hJ_z:=
\half(\hN+1)$, as opposed to $\hJ_z=\half\hN$.

The classical analogs of $\{\hJ_z,\hJ_\pm\}$ are the functions
 \be  \label{dif:obs}
  J_z=\half(z_1\bar{z}_1+z_2\bar{z}_2),\quad J_+=z_1z_2,\quad
       J_-=\bar{z}_1\bar{z}_2
 \ee
on \ps. (Note that there is an ambiguity in the correspondence between the
operator $\hJ_z$ and the classical function $J_z$. This ambiguity is resolved
by requiring the Poisson bracket algebra between the classical functions to
be the Lie algebra of $SO(2,1)$.)
One can easily check, using (2.\ref{qp:rank}), that the set $(J_+,J_-)$ is by
itself
{\em complete}, the set coordinatizes the reduced \ps. It is in order to
ensure
that the algebra of observables is {\em closed} that one has to include $J_z$
in the set of generators of \cAp, and thus make it {\em over}complete. As in
some of the previous examples, the algebraic relation satisfied by this
overcomplete set fixes a value of the
Casimir invariant of \cAp:
 \be
  \hJ^2:=-\hJ_z^2+\half[\hJ_+,\hJ_-]_+ \ .
 \ee
Using the definitions of $\hJ_\pm$ and the commutation relations
(\ref{dif:ccr}), one finds that
 \be\label{dif:oacr0}
  [\hJ_+,\hJ_-]_+=\hat{1}+\hN_1+\hN_2+2\hN_1\hN_2 \ .
 \ee
Thus, $\hJ^2=\frac14[1-(\hN_1-\hN_2)^2]$. On physical states, therefore,
 \be  \label{dif:oacr}
  \hJ^2={\textstyle\frac14}-\delta^2.
 \ee
Equivalently, the algebraic identity can be expressed as
 $$
  \hJ_+\hJ_-=(\hJ_z-\half)^2-\delta^2, \eqno(3.\arabeq ')
 $$
which is a useful form for future calculations.

The $\star$-relation induced on \cAp\ is given by:
 \be  \label{dif:o*}
  \hJ_+^\star=\hJ_- \quad\hbox{and}\quad \hJ_z^\star=\hJ_z.
 \ee

Using (\ref{dif:rep},\ref{dif:cond}), we can evaluate the action of the
physical operators on
the physical states. Doing so, we get
 \bea \label{dif:orep}
  \hJ_z\ket{\delta,m} &=& (m+\half)\ket{\delta,m}, \cr
              \hJ_+\ket{\delta,m} &=& \lambda_+(m+1)\ket{\delta,m+1} \cr
    \hbox{and}\quad
  \hJ_-\ket{\delta,m}&=&\lambda_-(m)\ket{\delta,m-1},
 \eea
where $\lambda_\pm$ are functions of their arguments only. The
algebraic identity (\ref{dif:oacr}$'$) is satisfied if and only if
 \be   \label{dif:ocond}
  \lambda_+(m) \lambda_-(m)= m^2-\delta^2.
 \ee
With this condition, the CCRs are also identically satisfied. Note that
$\lambda_\pm$ are just combinations of the coefficients $\alpha_I, \gamma_I$ in
(\ref{dif:rep}), so if one uses (\ref{dif:cond}), (\ref{dif:ocond}) is
trivially satisfied. However, as we will see, it is
simpler to work with the $\lambda_\pm$. This notation is
suggestive and also useful later for comparison to the reduced space
quantization
and the related model in section 6.

The last step in the program is to select an inner product by requiring that
the $\star$-relations (\ref{dif:o*}) become Hermitian adjointness relations on
the
resulting Hilbert space. The Hermiticity condition on $\hJ_z$
requires that its eigenvalues $m$ must be real, and its eigenkets orthogonal to
each other. Before we proceed any further, let us recall from
representation theory that the Hermiticity conditions should be implemented
only
on {\it irreducible} representations of the algebra. The representation
(\ref{dif:orep}), however, is reducible: Note that the physical operators
either leave the value of $m$ unchanged, or change it by an {\it integer}.
Thus,
the fractional part of $m$ --denoted by $\epsilon={\rm frac}(m)$-- is invariant
under the action of the $\hJ_z,\hJ_\pm$. Consider ${\cal V}_{phy}^\epsilon$,
the vector
space of states with the same fixed value of $\epsilon$. Each ${\cal
V}_{phy}^\epsilon$
carries an irreducible representation of the $SO(2,1)$ Lie algebra
(\ref{dif:occr}).

Naturally, one is lead to ask whether there is some ``hidden'' symmetry in this
problem, and whether the eigenvalues of the corresponding operator label the
irreducible sectors ${\cal V}_{phy}^\epsilon$. In the classical theory, one
notices that
the function (\ref{dif:zcon})
$\half(z_1\bar{z}_1-z_2\bar{z}_2)-\delta$ corresponding to the constraint and
the functions (\ref{dif:obs}) $\{1, \half(z_1\bar{z}_1+z_2\bar{z}_2),
z_1z_2,\bar{z}_1\bar{z}_2\}$ corresponding to the set of observables,
are all invariant
under the {\it parity} map $(z_I)\longmapsto {\bf P}(z_I):=(-z_I)$.
In
quantum theory, this discrete symmetry should correspond to a {\em
super-selected} physical operator, i.e.\ one which commutes with all other
operators in \cAp. In spite of the fact that the functions (\ref{dif:obs}) are
a locally
(over)complete set on $\Gbar$, the set of operators $\{1,\hJ_z,\hJ_\pm\}$ does
{\it not} generate the full physical algebra; \cAp\ is in fact generated by the
set $\{1,\hJ_z,\hJ_\pm,\hat{\bf P}\}$. In the representation we have chosen,
from the action of the
operators $\hJ_z,\hJ_\pm$ (\ref{dif:orep}), it is clear that on physical states
the Parity operator must be diagonal, and its eigenvalues must depend only on
$\epsilon$. From the Bargmann-type representation, in which this discrete
symmetry corresponds to the operation
$\psi(z_I)\mapsto\psi(-z_I)$, we see that in the $\ket{j,m}$ representation
(\ref{dif:rep}) the action of the parity operator is given by
 \be \label{dif:par}
  \ket{j,m} \longmapsto\hat{\bf P}\ket{j,m}:=(-1)^{2m}\ket{j,m},
 \ee
where, to evaluate the right hand side we will take the principal value,
namely, $(-1)^{2m}=\exp(i2\pi\epsilon)$, and as before
$\epsilon=\rmfrac(m)$ is the
fractional part of $m$.

Let us now return to the question of the reducibility of the representation on
\cVp. Let $m=n+\epsilon$, $n=\cdots -2,-1,
0,1,2\cdots$. Recall that the physical operators change $m$ in
integral steps only and do not affect the fractional part $\epsilon$.
Consequently, \cVp\ is reducible,
and {\em each eigenspace ${\cal V}_{phy}^\epsilon$ of the parity operator
provides an
irreducible
representation of the algebra \cAp}.
Each ${\cal V}_{phy}^\epsilon$ has a countable basis, labelled by $n$,
the integer part of $m$, and it is on these {\it irreducible} representations
that one implements the Hermiticity conditions on \cAp.
Note that at this stage it appears that we have a 1-parameter
family of ambiguities in quantization of the system, labelled by the parameter
$\epsilon\in [0,1)$.

Henceforth, for definiteness, consider a representation with a fixed value of
$\epsilon$. The Hermiticity of $\hJ_z$ implies that on ${\cal
V}_{phy}^\epsilon$ there
exists an inner product in which the above basis is orthogonal; without any
loss of generality, we can choose it to be orthonormal. Hence the inner product
can be chosen to be:
 \be   \label{dif:ip}
  \IP{\delta,m'=n'+\epsilon}{\delta,m=n+\epsilon}=\delta_{n',n},
 \ee
where both states on the left have the same fractional part of $m$. Note that
it is only because we implement the Hermiticity conditions on an {\em
irreducible} sector
--with a countable basis-- that we can postulate a Kronecker
$\delta$ inner product on ${\cal V}_{phy}^\epsilon$ as opposed to a Dirac
$\delta$ on
\cVp.

Now, the first of the $\star$-relations (\ref{dif:o*}) implies that
$\lambda_+(m)=\ovr{\lambda_-(m)}$. Substituting this in (\ref{dif:ocond}),
the condition
on the undetermined coefficients, yields
 \be\label{dif:ocond2}
  |\lambda_+(m)|^2= m^2-\delta^2.
 \ee
For solutions to exist, we require that physical states satisfy
 \be \label{dif:condsol}
  m^2\ge\delta^2.
 \ee
Without loss of generality, we can choose the phase of $\lambda_+(m)$ to be
zero, and solve
(\ref{dif:ocond2}). Then we have
 \bea \label{dif:fixrep}
  \hJ_z\ket{\delta,m} &=& (m+\half)\ket{\delta,m}, \cr
  \hJ_+\ket{\delta,m} &=& \sqrt{(m+1)^2-\delta^2}\>\ket{\delta,m+1} \cr
  \hbox{and}\quad\hJ_-\ket{\delta,m} &=& \sqrt{m^2-\delta^2}\>\ket{\delta,m-1}.
 \eea
Under what conditions do states with $m^2\ge\delta^2$ form an invariant
subspace? Consider for example a state satisfying (\ref{dif:condsol})
with arbitrary
$m\ge|\delta|$. Using $\hJ_-$ repeatedly, one can lower $m$ until condition
(\ref{dif:condsol}) is violated, unless there exists a state $\ket{\delta,m_0}$
annihilated
by $\hJ_-$. From (\ref{dif:fixrep}), we see that this will occur iff
$m_0=\pm|\delta|$.
Acting with $\hJ_+$ repeatedly on the ``ground'' state $m_0=+|\delta|$,
we see that an allowed representation consists of states labelled by
 \be \label{dif:m+}
  m=|\delta|+n,\quad n=0,1,2...
 \ee
This corresponds to a representation with a fixed eigenvalue of $\hat{\bf P}$,
$\epsilon=\rmfrac(|\delta|)$.

Similarly, starting with arbitrary
$m\le|\delta|$ one can use $\hJ_+$ repeatedly to raise $m$ until
(\ref{dif:condsol}) is violated, unless there exists a ``top'' state,
$m_0=\pm\delta-1$ annihilated by $\hJ_+$. Thus, one obtains the inequivalent
representation
 \be   \label{dif:m-}
   m=-|\delta|-1-n,\quad n=0,1,2...,
 \ee
with $\epsilon=1-\rmfrac(|\delta|)$. Thus for each
value of $\delta$ one obtains the two representations (\ref{dif:m+},
\ref{dif:m-}).

If $0<|\delta|<\half$ we have the additional
representations
 \be \label{dif:m+'}
  m=-|\delta|+n,\quad n=0,1,2...\qquad 0<|\delta|<\half\ ,
 \ee
for which $\epsilon=1-\rmfrac(|\delta|)$; and,
 \be \label{dif:m-'}
  m=|\delta|-1-n,\quad n=0,1,2...\qquad 0<|\delta|<\half\ ,
 \ee
with $\epsilon=\rmfrac(|\delta|)$. In fact, as Louko pointed out \cite{jl:pc5},
for $|\delta|\in[0,\half)$, the above representations are only special cases.
There is a whole slew of representations, one for each
$\epsilon\in[|\delta|,\half)$ or $\epsilon\in(-\half,-|\delta|]$. However, not
that in all but the representation (\ref{dif:m+'}), which corresponds to
$\epsilon=|\delta|$, $m$ is unbounded below.

Note that the coefficients
$\alpha_I, \gamma_I$ in (\ref{dif:rep}) are left undetermined, and we have {\it
not}
obtained an inner product on the original representation space. However,
(\ref{dif:ip}) provides us with an inner product on \cVp.
Unexpectedly, we have two representations of the physical
observable algebra, labelled by the eigenvalue of $\hat{\bf
P}$.

This completes the quantization of the coupled oscillator, and I would like to
discuss a number of interesting features.

\mysection{Classical expectations and quantum theory}
\remarks

\remark{: Positivity of energy}
In the classical theory, the function   $ H(x_I, p_I) := \half(x_1^2 + p_1^2 +
x_2^2 + p_2^2),$ corresponding to $\hat{H}=2\hJ_z$%
\footnote{ In the case of two oscillators this corresponds to the total
  energy, and thus I will henceforth refer to $2\hJ_z$ as the Hamiltonian,
  though it may have nothing to do with dynamics.},
is non-negative, i.e. $H\ge0$. In the quantum theory however, we have obtained
representations (\ref{dif:m-},\ref{dif:m-'}) of \cAp\ in which the
corresponding operator $\hat{H}$ is {\it
unbounded below} (its eigenvalues are $\pm 2|\delta|-2n-1,\>n=0,1,2...$). Since
$\hat{H}$ is an {\it elementary physical observable}, it is not of the form
$\hat{O}
^\star \hat{O}$ for any $\hat{O}\in$\cAp, and we cannot impose positivity.
Thus mathematically, one is stuck
with the above representation. However, in the representations
corresponding to (\ref{dif:m-}) and (\ref{dif:m-'}), there are {\it no} states
with positive energy
eigenvalues. Therefore, on {\it physical} grounds,
in order to obtain a good classical limit of the
theory, one may have to rule out the representations corresponding to
$m<0$. One cannot use the above argument to rule out a representation
  in which the Hamiltonian, though non-positive, is
  bounded below. For such a case, one could construct semi-classical states
with
  the right behaviour. This occurs for the representations (\ref{dif:m+'}), in
which there is only one negative-energy eigenstate.

\remark{: Implications for quantum gravity?}
Let us return to the general case.
Quantum mechanically, as we have seen, for any value of $\delta$ there exist
representations in which the Hamiltonian operator $\hat{H} = \hbar(\hc_1\ha_1
+\hc_2\ha_2 + 1)$ is {\it unbounded from below}! It thus appears that in the
presence of non-trivial constraints, extreme quantum tunneling can occur in
which ``half'' the (reducible) physical Hilbert space corresponds to states
that
are classically forbidden. In addition, when $\delta$ is less than half, we are
allowed the representations (\ref{dif:m+'}) in which the energy has one
negative eigenvalue, but is {\it bounded} below.
In the context of quantum geometrodynamics it is
extremely important to find out if the scalar constraint of non-perturbative
canonical gravity allows such a phenomenon to occur. If it does, should one
rule out the negative energy states on classical grounds, or will there in fact
be some physical interpretation for them?

\remark{: Inner product on \cV}
Suppose we restrict ourselves to positive energies, and fix for
definiteness, the representation (\ref{dif:m+}). Note that the total energy of
the
system ($2n+2|\delta|+1$) is allowed to be
{\sl any} positive real value, bounded below by $2|\delta|+1$. This is a
somewhat surprising result in light of the quantization we would have obtained
had we chosen to implement the $\star$-relations on the entire quantum algebra
\cA. In that case the energies of each individual oscillator, $H_I=N_I+\half$,
which are physical observables, would have been positive half integers only,
and the total energy a positive {\it integer}. Here on the other hand, from the
definitions of the number operators we have:
 \bea\label{dif:nums}
  \hN_U\ket{\delta,m} &=& (m+|\delta|)\cdot\ket{\delta,m}=(2|\delta|+n)
   \ket{\delta,m} \cr \hbox{and}\quad
  \hN_L\ket{\delta,m} &=& (m-|\delta|)\cdot\ket{\delta,m}=n\ket{\delta,m},
 \eea
where we have used (\ref{dif:m+}), and $U$ and $L$ refer to the oscillators in
the
``upper'' and ``lower'' states respectively. One can see that only
the energy of the oscillator in the lower energy state is positive half
integer, the oscillator in the higher state can have any energy! How does this
come about? If one imposes reality conditions before solving the constraints,
it is the requirement that the energy of each oscillator be positive definite
and the fact that there exists a lowering operator that necessitates the
half-integer energy eigenvalues. Here, on the other hand, since we impose
$\star$-relations only on the space of physical states, once the energy of the
`lower' oscillator is positive definite (and thus half integer), the constraint
{\em guarantees} that the energy of the `higher' oscillator is positive
definite too. This is no longer an independent requirement.

\remark{: Classical conditions incompatible in quantum theory}
Consider the following two features of the classical theory.
First, classically, the parity transformation ${\bf P}(z_I)=(-z_I)$ satisfies
 \be  \label{dif:feat1}
  ({\rm I})\qquad {\bf P}^2=1.
 \ee
Second, since for any value of $\delta$ the reduced \ps\ (see section 5) is a
non-compact two
dimensional manifold (in fact topologically $\real^2$), the classical theory is
well-defined for {\it any} real $\delta$,
 \be \label{dif:feat2}
  ({\rm II})\qquad \delta\in\real.
 \ee
However, the eigenvalues of $\hat{\bf P}$ are given by $\exp(i2\pi\epsilon)$
(see (\ref{dif:par})), where in the physical representation,
$\epsilon=\rmfrac(|\delta|)$.
If we require $\hat{\bf P}^2=1$ in the quantum theory, there are non-trivial
representations only for
$\half$-integer or integer values of $\delta$.

Thus, {\it there are two conditions {\rm (I)} and {\rm(II)}, satisfied in the
classical theory, which cannot be satisfied simultaneously in quantum
theory}.

Within quantum mechanics, there seems to be {\it no} compelling reason to
restrict ourselves to states with
eigenvalues $\pm 1$ of $\hat{\bf P}$. However, to allow other eigenvalues seems
to violate
classical intuition. This is, however, not entirely unfamiliar. Indeed,
a similar situation occurs in systems
of identical particles in 2-spatial dimensions where the use of eigenvalues
other than $\pm 1$ for the parity or the permutation operators leads to the
interesting quantum phenomena of fractional statistics.

\mysection{FRW universe with conformally coupled scalar field}
The Ricci scalar for the closed ($k=+1$) Friedman-Robertson-Walker universe is
 \be
  R= \frac6{a^2} \left( a\frac{\d^2a}{\d\tau^2} + \left(\frac{\d
    a}{\d\tau}\right)^2 + 1\right),
 \ee
where $a$ is the scale-factor of the universe and $\tau$ is the proper time.
Hence, the gravitational part of the Lagrangian (up to a factor of
$\frac{4\pi}{3}$, and after an integration by parts) is:
 \be
  {\cal L}_G=\frac6{G} \left(-a\left(\frac{\d a}{\d\tau}\right)^2+a\right),
 \ee
where $G$ is the gravitational constant, and the action is $S=\int \rd\tau
{\cal L}$. The Lagrangian for
the homogeneous, conformally coupled ($\xi=\frac16$, massless) scalar field is
 \be
  {\cal L}_{KG}=8\pi \left( a^3\left(\frac{\d\phi}{\d\tau}\right)^2
   + a\left(\frac{\d a}{\d\tau}\right)^2\phi^2
   + 2a^2\phi\frac{\d a}{\d\tau}\cdot\frac{\d\phi}{\d\tau} - a\phi^2 \right)\ .
 \ee
Introduce a reparametrization of the time, $\d t=\d\tau/N$, where $N$ is the
lapse.
Let $\dot{\hphantom{a}}\equiv(\d/\d t)$. Then the total Lagrangian is
 \be \label{frw:lag}
  {\cal L}= -\frac{6}{GN}a\dot{a}^2 + \frac{6Na}{G}
   + \frac{8\pi a}{N}\dot{(a\phi)}^2 - \frac{8\pi N}{a}(a\phi)^2.
 \ee

Define the variables
 \bea  \label{frw:var}
             x_1 &:=& \sqrt{\frac{12}{G}}\,a \\
             x_2 &:=& \sqrt{\frac{16\pi}{G}}\,a\phi.
 \eea
Now the Lagrangian takes the form
 \be
  {\cal L}= -\frac{1}{2N}a\dot{X}_1^2 + \frac{1}{2N}a\dot{X}_2^2
    + \frac{6Na}{G} - \frac{N}{2a}x_2^2.
 \ee
Performing the Legendre transform, we find the canonical momenta:
 \bea
             p_1 &:=& -\sqrt{\dfrac{12}{G}} \dfrac{a\,\dot{a}}{N} \\ \hbox{and}
       \quad p_2 &:=& \tfrac{4}{\sqrt\pi N}\,a\,\dot{(a\phi)}.
 \eea
The Hamiltonian for the system is $H=\frac{2N}{x_1}C$, where the scalar
constraint is given by
 \be \label{frw:con}
  C={1\over 4}(p_1^2+x_1^2-p_2^2-x_2^2)\approx 0.
 \ee

We see that it is exactly of the form of (\ref{dif:con}), with $\delta=0$.
Note
that there is a nonholonomic constraint, $a\ge0$. A consistent approach to this
would be to consider the physical scale factor to be defined by
$a:= \frac{|x_1|} {\sqrt{12}}$, on the \ps\ defined by $(x_I,p_I)$.
The solutions then describe a periodic, bouncing
universe. Finally, since $\delta=0$ in this case, we obtain only one physical
representation ((\ref{dif:m+'}) and (\ref{dif:m+}) are the same),
which does not display any of the counter-intuitive behaviour discussed
in the previous section. For the mathematically allowed representations in
which the energy is unbounded below, no physical interpretation is immediately
forthcoming.

\mysection{(\dag) Reduced space quantization}

As has been mentioned before, the constraint surface for the general problem is
the group manifold of $SO(2,1)$, which is topologically $\real^2\times S^1$.
The orbits of the canonical transformations generated by the constraint are
closed curves in $\ovr\Gamma$, and there are {\it no fixed points}. The space
of orbits $\hat{\Gamma}$ is just the mass shell in 3-dimensional
Minkowski space. The functions $\{J_z,J_+,J_-\}$ (see (\ref{dif:obs})) are good
coordinates on $M^3$, and the ``radius'' of the mass shell is given by the
identity
 \be\label{dif:car}
  J^2:=-J_z^2+J_+J_-=\>-\delta^2.
 \ee
(Since the classical function $J_z$ is positive, the reduced \ps\ is in fact
the
{\it future} mass shell, $J_z\ge0$.)
One can easily compute the Poisson brackets between these elementary variables:
$\{J_+,J_-\}=2iJ_z \hbox{ and }\{J_\pm,J_z\}=\pm iJ_\pm.$

We choose as our set of elementary functions \cS\ = $\{1,J_z,J_\pm\}$, and
construct our algebra accordingly. The commutation relations in the reduced
space observable algebra $\cA_{red}$ are given by:
$[\hJ_+,\hJ_-]=-2\hJ_z$ and $[\hJ_z,\hJ_\pm]=\pm\hJ_\pm$, which are exactly the
same as in \cAp\ (\ref{dif:occr}). Not surprisingly the commutator algebra is
again a
representation of $SO(2,1)$. Note however that the quantum {\it algebraic
relation} (obtained from (\ref{dif:car}) via the rule (2.\ref{qp:acr})),
 \bea \label{dif:redacr}
  \hJ^2 &:=& -\hJ_z^2+\half[\hJ_+,\hJ_-]_+=-\delta^2 \cr \hbox{or,}\quad
  \hJ_+\hJ_- &=& (\hJ_z-\half)^2-{\textstyle{1\over4}}-\delta^2\ ,
 \eea
{\it is different}! (Compare to (\ref{dif:oacr}): both \cAp\ and $\cA_{red}$
are
representations of the $SO(2,1)$ algebra, but for different values of the
Casimir invariant $\hJ^2$!) Therefore the two operator algebras, \cAp, the
physical operator algebra constructed previously for constrained quantization,
and \hca, the reduced space algebra, are {\it not identical}.
Obviously, the quantum theories will also be different.

In fact, carrying out the quantization, one finds the following (the
representation of the CCRs in \hca\ is chosen as in (\ref{dif:orep}):
Corresponding
to (\ref{dif:condsol}) here one obtains the condition
 \be \label{dif:redcondsol}
  m^2\ge\delta^2+{\textstyle{1\over4}}.
 \ee
(The reduced
space quantum theory can be obtained from the Dirac theory
by making the substitution $\delta^2\mapsto\delta^2+\frac14$.)
However, since we have to impose the non-holonomic constraint $\hJ_z\ge0$, for
each value of $|\delta|$ one obtains a single representation consisting of
states with
 \be \label{dif:redm}
  m=\sqrt{\delta^2+{\textstyle{1\over4}}}+n,\quad n=0,1,2...
 \ee
The representation of the elementary operators in $\cA_{red}$ is then given
by
 \bea \label{dif:redrep}
  \hJ_z\ket{m} &=& (m+\half)\ket{m} \cr
        \hJ_+\ket{m} &=& \sqrt{(m+1)^2-{\textstyle{1\over4}}-\delta^2}\ket{m+1}
          \cr
        \hJ_-\ket{m} &=& \sqrt{m^2-{\textstyle{1\over4}}-\delta^2}\ket{m-1} \cr
   \hbox{and }\hat{\bf P}^2\ket{m} &=& \exp{i2\pi\epsilon}\ket{m}.
 \eea
Note that the quantum theory is well-defined for {\em all} real $\delta$, not
just integers.

Comparing this representation to (\ref{dif:fixrep}), one can see that the
quantum
theories cannot be made
equivalent. (The particular representation of $\hJ_z$ has been chosen to
facilitate the
comparison with the Dirac theory. This leads to just an overall shift in the
spectrum of $\hJ_z$. Now, construct e.g. the Hermitian operators
$\hJ_x=\half(\hJ_+
+\hJ_-),\quad \hJ_y={1\over2i}(\hJ_+ -\hJ_-)$, and compare spectra in the two
quantum theories. These are not related to each other by just a shift.)
This was not unexpected: due to the ACRs, the algebras \hca\
and \cAp\ are different. This arises because of the different stages at which
consistent factor-ordering is carried out in the two programs. To construct
\cAp\ one factor-orders in \cA\ {\it first} and then looks for operators that
commute with the constraints. Since the algebraic relation is calculated at the
operator level (\ref{dif:oacr0} -- \ref{dif:oacr}), and is not derived as the
``quantization'' of some classical identity, there is no ambiguity of the kind
in (2.\ref{qp:acramb}). On the other hand, to construct \hca\ one first
finds functions which have vanishing Poisson brackets with the constraints (on
$\ovr\Gamma$), and then uses these as {\em elementary} variables. Now, when one
constructs the
abstract algebra \hca, the elementary classical Dirac observables are to have
unambiguous quantum analogs, and there is no factor-ordering ambiguity at this
stage. However, the classical identity (\ref{dif:car}) has to be incorporated
in quantum theory, and we have chosen to do so by imposing the ACR
(\ref{dif:redacr}).

As in the Dirac quantum theory, there is a superselected operator
$\hat{\bf P}$ which is unitary.
However, since the classical parity operator on $\Gamma$, $\hat{\bf
P}(z_I)=-z_I$, leaves the gauge orbits invariant (it maps a point on a gauge
orbit to a diametrically opposite point on the {\em same} gauge orbit), it has
no action on $\hat\Gamma$ {\it whatsoever}. Thus in the reduced space
quantization of this problem, the operator $\hat{\bf P}$ is introduced as a
mathematical device to allow us to first decompose the representation space
into orthogonal subspaces labelled by the eigenvalues of $\hat{\bf P}$, and
then
proceed to find an inner product.
In the context of the reduced space quantum theory, there can be no physical
interpretation attached
to ${\bf P}$; in particular, there is no justification even classically
to require that ${\bf P}^2=1$.

\mysection{(\dag) Coupled oscillators: constrained energy sum}

I will briefly discuss the quantization of a model that has proven useful in
the past as a testing ground for various ideas (e.g.\ the issue of time
\cite{cr:time5}) and
has been studied extensively (see also \cite{haj5}). Consider two harmonic
oscillators
as in section 2, but unlike the previous model impose the {\it total energy}
as a first class constraint:
 \be \label{sum:con}
  C:=\half(p_1^2+x_1^2+p_2^2+x_2^2)-E\approx0,
 \ee
where $E\ge0$ in order for the classical system to be well-defined. (Note that
since the geometrodynamical scalar constraint is not of this form (see the
remarks in the introduction to this chapter), the constrained energy sum model
we consider here is {\it not} a model of \gr.)

We will use the notation of section 2. In terms of the quantum operators
defined in section 2, the
constraint equation is given by:
 \be            \label{sum:qcon}
  \hat{C}\ket\psi_{phy}\equiv (\hc_1\ha_1+\hc_2\ha_2+1-E)\ket\psi_{phy}=0.
 \ee

Consider the algebra generated by the set of operators $\{1,\hL_z, \hL_\pm\}$,
where $\hL_z=\half(\hN_1-\hN_2)$, $\hL_+=\hc_1\ha_2$ and $\hL_-=\hc_2\ha_1$ are
physical operators. The commutator algebra, given by $[\hL_+,\hL_-]=2\hL_z$ and
$[\hL_z,\hL_\pm]=\pm\hL_\pm$, provides a representation of the $SO(3)$ Lie
algebra.
The algebraic relation is $\hL_+\hL_-={1\over4}E^2-(\hL_z-\half)^2$, or
equivalently in terms of the Casimir invariant,
$\hL^2:=\hL_z^2+\half[\hL_+,\hL_-]_+=l(l+1)$ where $l={E-1\over2}$.

The quantization of this algebra is straightforward, and though this model is
not as rich as the previous one, it does have a few interesting features, most
of which are familiar from the quantization of the standard angular momentum
algebra.
\begin{enumerate}
\item Consider the parity operator corresponding to the discrete canonical
transformation ${\bf P}(z_I)=-z_I$. It is certainly superselected, but in this
case it does not serve to further reduce the physical representation. On states
in the kernel of the constraint, its action is given by
 \be  \label{sum:par}
  \hat{\bf P}\ket{m}=(-1)^{E-1}\ket{m},
 \ee
it has the same eigenvalue on the full physical subspace. The parity operator
satisfies $\hat{\bf P}^2=1$ only for integer $E$.
\item Quantum mechanically, there is however another superselected
discrete operator that one can construct, the eigenvalues of which characterize
the irreducible representations:
 \be \label{sum:dis}
  \hat{\bf T}\ket{m}=(-1)^{2m}\ket{m}
 \ee
On each eigenspace of $\hat{\bf T}$ one can introduce a discrete, countable
basis, and this
allows us to choose an inner product of the form $\IP{m}{m'}=\delta_{m,m'}$,
rather than $\IP{m}{m'}=\delta(m-m')$.
Unfortunately, this operator does not correspond to any known classical
canonical transformation on the \ps, thus it has {\it no
interpretation in classical terms}.
\item One finds that there exist non-trivial representations only when
$E$ is integer ($l$ is half integer or integer). The representation is unique,
and not surprisingly (since the
reduced \ps\ is compact), it is finite-dimensional, of
dimension $2l+1$. A basis is provided by states with $m=-l,-l+1,...l-1,l$.
The representation of the other generators of \cAp\ is
 \bea  \label{sum:orep}
  \hL_z\ket{m} &=& m\ket{m}, \cr
             \hL_+\ket{m} &=& \sqrt{(l-m)(l+m+1)}\ket{m+1} \cr \hbox{and}
        \quad\hL_-\ket{m} &=& \sqrt{(l+m)(l-m+1)}\ket{m-1},
 \eea
where as before $l={(E-1)\over2}$.
\item For general, non-integral E, there is {\em no quantum theory}.
(However, see \cite{rst:pc5}.) There are
many levels at which this can be understood, or at least seen to be not
unfamiliar. From the point of view of geometric quantization,
this is not a surprise since the reduced
\ps\ is a compact manifold, $S^2$, and it admits a K\"ahler structure only for
integer values of the radius \cite{ms:thesis5}. Related to this,
obviously, is the fact that
\hca\ is just the Lie algebra of $SO(3)$. From ordinary quantum mechanics, we
are familiar with the fact that this has only the integer and half-integer spin
representations.
\item For this model, {\it in this representation}
the reduced space quantum theory is not obviously unitarily equivalent
to the Dirac theory. However, with a suitable choice of
factor-ordering of the quantum constraint, the Dirac theory can be
made equivalent to the reduced space theory, see \cite{rst:pc5}.  (See also
\cite{cr:time5}.)
\end{enumerate}

%
\chapter{PHYSICAL INTERPRETATION FOR SYSTEMS WITH A CONSTRAINED HAMILTONIAN}
\pagestyle{myheadings}
\markboth{{\sf Chapter 6}}{{\sf Physical Interpretation}}



\def\Sch{Schr\"odinger}

\def\lo{\lambda_0}
\def\hC{\hat{C}}
\def\hH{\hat{H}}
\def\hU{\hat{U}}

\def\eiq{e^{-\frac{i}\hbar\hH q^0}}

\def\hZ{\hat{Z}}


\newcommand{\bz}{\beta^0}
\newcommand{\bp}{\beta^+}             \newcommand{\bm}{\beta^-}
\newcommand{\pz}{\pi_0}
\newcommand{\pp}{\pi_+}                 \newcommand{\pmm}{\pi_-}


\mysection{Introduction}

In classical mechanics, interpretational issues are well understood:
the physical questions one is interested in are mostly
related to predictions about the values of various \ps\ functions.
The issues of interest in a quantum theory, however,
are generally very different from
those in a classical theory. In classical mechanics, e.g.\ in the
Kepler problem, one is interested in questions about the orbits of the particle
and in predicting future positions and momenta of the particle from initial
data. In the quantum theory of the Hydrogen atom, on the other hand, one is
more interested in questions about the spectra of operators corresponding to
various \ps\ functions --e.g.\ the energy and angular
momentum operators-- and the transition amplitudes between their eigenstates.
To a certain extent, we know that these are the right questions to ask because
the full quantum theory is available to us. In general, until one has at least
the rudiments of a quantum theory, one cannot
be certain that the questions one asks will remain meaningful in the full
quantum theory.

In the quantum theory of constrained systems the interpretational problems are
further complicated
since not all operators can be considered physically interesting.
Recall that the allowed quantum states of the system are represented by the
solutions to the quantum constraint equations. Hence, we are allowed to
consider as physical only
those operators which leave the space of physical states invariant.

Now, given a vector space of physical states, it is
trivial to construct linear operators on it, say by selecting a basis and then
writing down arbitrary matrices in this basis.
In the loop
representation for \gr, for example,
one can easily construct such operators on the known
physical states: they simply break and join loops in appropriate ways.
However --and I want to emphasize this point--
this construction is not sufficient for the purposes of quantum theory.
It is true that these are physical operators, since
by construction they leave the space of physical states invariant.
However, unless one knows which functions on \ps\ they correspond to, one
cannot
relate mathematical properties of these operators to, e.g., the results of
measurements. The main difficulty is to
obtain the correct Hermiticity
conditions on these operators, and their interpretation%
\footnote{On the other hand, the corresponding classical observables do not
 have to be simple functions of some particular choice of the elementary
 variables.  Also, the representation {\it itself} does not have to be tied
 directly to the \ps. For example, unlike in the configuration representations,
 in the $\ket{n,l,m}$ representation for the H-atom or the Fock
 representation for the free Maxwell theory, the states are {\it not} functions
 on \ps. However, the states
 themselves can be identified as eigenstates of operators corresponding to
 functions on \ps; and the interesting operators on
 these states do correspond to known functions on the \ps.}.

The above discussion shows, that in contrast to ordinary quantum mechanics,
in the
quantum theory of constrained systems, one is interested in the explicit
representations on physical states of operators corresponding to a {\em
restricted} class of
functions on \ps, namely the gauge invariant functions on the constraint
surface.

Among constrained systems themselves there is a further distinction.
For {\it ordinary} constrained systems, like
gauge theories, the Hamiltonian --which generates dynamics-- is distinct from
the constraint functions and does not vanish on the constraint surface.
On the other hand there are
theories in which the vanishing of the Hamiltonian is itself a first-class
constraint.
I will refer to such theories as {\em dynamically constrained systems} since
the dynamical
trajectories are generated by a Hamiltonian which is constrained to vanish.
General relativity, in the spatially compact case, is an important example.

In order to understand some of the difficulties endemic to dynamically
constrained systems, let us first recall some features of ordinary constrained
systems. In such systems, solving the constraints
--either classically, by
constructing the reduced \ps\ (or, equivalently, a cross-section of the gauge
orbits); or in quantum theory, by constructing the physical
states, an operator algebra of observables and an inner product on these
states-- is a purely {\em kinematical} procedure. This construction is
completely independent (at least conceptually) from the dynamical structure of
the theory: one does not need the Hamiltonian to carry out the above
construction. After this kinematical construction, one can consider the issue
of dynamics. Since the (non-vanishing)
Hamiltonian is first-class, it projects unambiguously
to the reduced \ps, and all physically
interesting dynamics can be considered to occur on the reduced \ps\ itself. In
quantum theory, the corresponding Hamiltonian operator generates (unitary)
evolution on the
Hilbert space of physical states.

In contrast, for systems in which the Hamiltonian is constrained to vanish,
kinematical considerations are intimately linked with the dynamical structure
of the theory. Since in such systems ``time'' is not an external parameter, but
one of the
canonical variables, the question of the identification of the true degrees of
freedom and the physical Hilbert space takes on dynamical overtones. For
example, can one in fact complete the kinematical
construction of the Hilbert space of physical states {\em without} explicitly
isolating a ``time'' variable? An equally important issue is that of
observables. The dynamical trajectories are ``gauge'' orbits.
Hence, {\em a} set of Dirac observables (or ``gauge invariants'') is the set of
initial data.
Using these observables, or even the constants of motion,
it is difficult to see the evolution explicitly. How do we recover a dynamical
picture for such theories? Where, e.g., is
a time variable hidden, and how does one construct {\em time dependent} states
or operators?
Or, are constants of motion the only
operators about which we can or would want to formulate well-defined questions
in quantum theory?

Clearly, the dynamical treatment and physical
interpretation of the two types of constrained
systems (distinguished, recall, by the vanishing or non-vanishing of the
Hamiltonian on the constraint surface) will perforce be different.
In this chapter, I will concentrate
on some interpretational issues in the quantum theory of
dynamically constrained systems.

In order to focus on these issues, consider, as a specific example of a
dynamically constrained system, the {\it
nonrelativistic parametrized particle}. As I will discuss in detail in section
2,
the classical theory for the ordinary nonrelativistic particle can be written
as
a constrained system, by including ``time'' and its conjugate momentum as
(extended)
\ps\ variables and imposing as a first class constraint the condition that the
momentum conjugate to time equal the ``true'' Hamiltonian. (See
ADM \cite{adm6}, Kucha\v r
\cite{kk:qgII6} and references therein.)
Now, in the usual, unparametrized version of the theory, time is an {\em
external}
parameter, evolution along which is determined by the Hamiltonian. In the
process of constructing a Hilbert space of states, one need not make any
reference to the Hamiltonian or the time. In subsection 2.1, we will see that
for the parametrized theory too, the construction of the kinematical quantum
theory --i.e.\ a Hilbert space of physical states-- can be accomplished without
isolating a time variable. As far as the second issue --that of the
observables-- is concerned, recall first that in the unparametrized theory
one has no trouble
posing questions about the action of {\it instantaneous} position and momentum
operators on the quantum states; hence one expects that in the parametrized
theory too one should be able to pose such questions. Now, in the
parametrized theory, it turns out that
the
algebra of physical observables is generated by the constants of motion, e.g.\
the {\it initial} positions and momenta. However, in the quantum
theory one is not interested in
questions only about the constants of motion. As an example, consider the
quantum theory of the Hydrogen atom. One {\em is} interested in the
expectation values of the instantaneous position operator $\hat{x}$, say to
calculate the effect of a uniform electric field as a perturbation. As we will
see, in the parametrized theory, since $\hat{x}$ does not commute with the
constraint, it is not a Dirac observable, and one cannot calculate its
expectation value on physical states. How can one resolve this conflict --in
the ability to pose certain questions-- between the parametrized and
unparametrized theories?

To begin with, let us consider a related question in the context of ordinary
gauge theories, ones in which the constraints do not generate dynamics. Can one
only formulate questions about the gauge-invariant observables? In a certain
sense, the answer is ``yes''. However,
since a gauge fixing --or a cross-section of the gauge
orbits-- is related to the space of orbits (the reduced \ps\ $\hat\Gamma$) by a
symplectic diffeomorphism, the pull-back of an arbitrary function to
this cross-section naturally defines a function on $\hat\Gamma$ and is thus
manifestly
gauge-invariant. Therefore, corresponding to {\em any}
function on the \ps\ one can {\it construct} a gauge-invariant observable.

The details of the construction are useful to keep in mind during the rest of
this chapter: Let $\lambda$ be an affine parameter
along the gauge orbits, and fix a cross-section $\lambda=\lo$. Now, consider an
arbitrary function $f$ on $\Gbar$, and assume to begin with that its
Lie derivative along the gauge orbits does {\it not} vanish (if it does,
$f$, being constant along the gauge orbits, is already a Dirac
observable and we are done). We can construct a gauge invariant observable
corresponding to $f$ in the following way: First, evaluate $f$ on the
cross-section. Now construct a new function $f_{\lo}$ by Lie transporting this
``data'' on the cross-section along the gauge orbits, i.e.\
$\frac{\d}{\d\lambda}f_{\lo}=0,\>f_{\lo}|_{\lambda=\lo}=f|_{\lambda=\lo}$.
Thus,
e.g., if $f=2$ at the point of intersection of a gauge orbit with the
cross-section, then $f_{\lo}=2$ {\em everywhere} on that orbit. Thus, {\em in
this gauge}, we have obtained a classical observable corresponding to the
function $f$. In quantum theory, one would try to construct the
corresponding physical operator
$\hat{f}_{\lo}$, and formulate questions in
this gauge. This is the sense in which the connection in the Coulomb or
transverse gauge is a gauge invariant observable%
\footnote{I am not suggesting that questions about the connection
 are important in QED. However, {\em some} questions of this nature may be
 important: since we do not know in advance which ones, we need a device to
 formulate them in the quantum theory.}.
Note that the specific form of these operators will in general depend on the
value of the gauge parameter chosen to fix the cross-section. The freedom is
quite large, since we can not only choose a different value for the parameter
$\lambda$, but we can also choose entirely different foliations of the
constraint surface, instead of $\lambda=const.$ cross-sections.

This is the approach in gauge theories, such as QED. In these theories, gauge
fixing is usually not considered to be of tremendous {\em conceptual}
importance. In principle, one could work with gauge invariant observables,
i.e.\ on the reduced \ps\ itself. It is not conceptually necessary to work with
gauge-fixed observables: it is only a matter of practical and technical
comfort that we choose to work with a cross-section of the gauge orbits.

However,
dynamically constrained systems are different from ordinary gauge theories in
certain conceptually important ways.
First, recall that since the dynamical trajectories are gauge orbits {\em i)}
there is
no natural notion of dynamics on the reduced \ps\ itself, and {\em ii)}
gauge-invariant observables are constants of motion. For dynamically
constrained systems, one does not have the luxury of working on the reduced
\ps.
It is in order to see the
explicit unfolding of dynamical evolution that one is forced to work on the
constraint surface, scored by the congruence of ``gauge'' orbits.
Second, since a
gauge parameter is identified as a time variable, then unlike in ordinary gauge
theories, one is {\it not} free to gauge-fix it. Hence, on the constraint
surface, one is interested not just in one cross-section of the gauge orbits,
but a 1-parameter family of cross-sections. Now, given
a preferred foliation of
the constraint surface by such ``constant time'' slices, for each function on
$\Gbar$ one obtains a {\em 1-parameter family of observables}, not just a
single gauge-fixed observable. Each observable in this 1-parameter family
commutes (or has vanishing Poisson bracket) with the constraint and is
therefore a Dirac observable.

Up to now, in order to convey some of
the conceptual ideas, the discussion has been mostly classical. The above
general procedure can be attempted in both the
Dirac and the reduced space quantum theories. Since both methods involve
solving the (classical or quantum) Heisenberg equations of motion in order to
obtain the 1-parameter family of observables corresponding to the particular
function on the constraint surface in which we happen to be interested, I will
refer to such observables as {\it Heisenberg observables}. Note that there may
not always be a clear distinction between ``ordinary'' Dirac observables and
these Heisenberg observables.

So far, in the Dirac quantum theory, the above approach to physical and
dynamical interpretation is (even formally) well-defined only for systems which
can be deparametrized%
\footnote{In this chapter, I will use the term ``deparametrized'' to mean that
 not only has a ``time'' variable $q_0$ been identified, but also, the
 constraint is linear in the conjugate momentum $p_0$ and of the form
 $C=p_0+H(q_0, q_i,p_i)$.},
i.e.\ where the constraint can be written as a \Sch\ equation, in a form linear
in a variable which can be identified as the momentum conjugate to some time
parameter. Such systems are formally similar to the nonrelativistic
parametrized particle, which I will discuss
in section 2. On the other hand, one can construct the reduced space quantum
theory only for systems which have been completely solved classically, but then
for such systems the interpretational problem is well under control, even if
the constraint has not been written explicitly in the \Sch\ form. In section 3,
I will construct an interpretational framework for the reduced space quantum
theory of dynamically constrained systems, along the lines of the approach
discussed above for ordinary
gauge theories. This closely follows the approach discussed by Rovelli
\cite{cr:I6,cr:II6}. In section 4 I will apply this approach to the Bianchi I
cosmology, and consider the issue of the initial singularity in quantum theory.
In section 5 I will draw conclusions and outline some ideas to extend the above
interpretational framework to either Dirac quantum theories which are not in
deparametrized form, or reduced space quantum theories of classical systems
which have not been exactly solved.

\mysection{Non-relativistic parametrized particle}

Consider a non-relativistic particle moving in (3-dimensional) Euclidean space.
Dynamics is specified by a ``true'' Hamiltonian $H(q^i,p_i)$, where $p_i$ are
the momenta conjugate to the coordinates $q^i$. (For simplicity, I will
consider the Hamiltonian to be the sum of two terms: a kinetic term quadratic
in momenta, and a potential term $V$ independent of momenta.) This simple
system can be
``parametrized'' by adding to the 3-dimensional configuration space the time
variable $q^0$. Thus, the (enlarged) configuration space, ${\cal C}$, is now
{\em 4-dimensional}, coordinatized by $(q^0, q^i)$; and the phase space is
8-dimensional. There is one (first class) constraint:
 \be \label{npp:con}
  C(q,p):= p_0 + H(q^i, p_j) = 0,
 \ee
where $q$ and $p$ stand for $(q_0, q_i)$ and $(p_0, p_j)$ respectively. The
constraint reduces the fictitious 4 degrees of freedom to the original 3 ``true
degrees'': classically, the constrained system is equivalent to the original
system evolving in the 6-dimensional phase space spanned by $(q^i,p_i)$ via the
Hamiltonian $H(q^i, p_i)$.

\subsection{Kinematical quantum theory}

Let us now carry out the quantization program. (In this section, the details of
the calculation follow Ashtekar \cite{aa:pc6,aa:rst6}. Kucha\v r \cite{kk:pc6}
and
Rovelli \cite{cr:I6} have similar if not identical approaches to the physical
interpretation of such systems, and the roots probably go further back.) Let
the space \cS\ of elementary observables be the complex vector space spanned by
the 9 functions $(1,q,p)$ on the phase space $\Gamma$, with the usual
commutation relations. Choose for the representation space \cV\ the space of
smooth functions on the {\it 4-dimensional} configuration space ${\cal C}$, and
represent the operators by the usual multiplication and partial derivative
operators%
\footnote{Note that for what follows, in the choice of representation it is
 essential only that $\hq^0$ be a multiplication operator. One is free to
choose
 any representation of the $\hq^i,\hp_i$ operators, depending on the specific
form of the
 Hamiltonian.}.
The quantum constraint is now given by:
 \be \label{npp:qcon}
  \hC\circ\Psi (q)\equiv \dfrac\hbar{i}\> {\d\Psi (q)\over \d q^0} +
    \hH\circ \Psi(q)= 0\>,
 \ee
which, when we identify $q^0$ with time,
is nothing but the time-dependent \Sch\ equation, in a parametrized guise.
The space of physical states, ${\cal V}_{phy}$, now consists of solutions of
this equation, given {\it formally} by:
 \be\label{npp:qsol}
  \Psi(q)=\eiq\circ\psi(q^i),
 \ee
where the $\psi(q^i)$ are arbitrary functions which depend only on
the ``true degrees of freedom''. Note that the solutions $\Psi(q)$ to the
quantum constraint are complex valued functions on the 4-dimensional
configuration space ${\cal C}$; they are {\it not}
functions of $q_i$ alone; they necessarily depend on $q_0$ as well. In this
sense, they are ``covariant''. However, since the $q^0$ dependence is fixed by
the exponential term, physical states
are determined by the functions $\psi(q^i)$, which we can think of as
the ``initial data'' for the first order (in $q^0$) differential equation
(\ref{npp:qcon}). Clearly, none of the elementary quantum operators, $\hq^0,
\hq^i$ or $\hp_i$, corresponding to the set \cS, is a physical operator.
Therefore, the $\star$-relations between the elementary operators cannot be
directly used to single out an inner-product on ${\cal V}_{phy}$.

Our task now is to find a complete set of physical operators. Fortunately, this
is not difficult to accomplish, at least formally. The Dirac operators are
given by:
 \bea \label{npp:pop1}
  \hQ^i(0)\circ\Psi & := & \hU(0)\hq^i \hU^{-1}(0)\circ\Psi  =\eiq
     \hq^i\circ\psi \equiv \eiq\circ q^i\psi(q^i),\cr
  \hbox{and}\quad
  \hP_i(0)\circ\Psi & := & \hU(0)\hp_i \hU^{-1}(0)\circ\Psi  =\eiq
     \hp_i\circ\psi \equiv \eiq\circ \dfrac\hbar{i}\dfrac{\d}{\d q^i}\psi(q^i),
 \eea
where
 \be\label{npp:hatu}
  \hU(0):=\eiq;
 \ee
$\hq^0$ acts by multiplication and has been replaced by $q^0$; and the
Hamiltonian operator $\hH=H(\hq^i,\hp_i)$. (The reason for the notation
$\hU(0)$ will become clear later.) In (\ref{npp:pop1}), the states in the
last step are manifestly solutions of (\ref{npp:qcon}), of the form
(\ref{npp:qsol}). In fact, a simple algebraic calculation shows
that these six operators, $\hQ^i$ and $\hP_i$, commute with the constraint, and
furthermore, are their own ${}^\star$s. Since the reduced \ps\ is
6-dimensional, and the above Dirac operators are independent, they form
a complete set. Hence we can now look for an inner
product on $\cV_{phy}$ with respect to which these operators are Hermitian. For
this, let us begin by introducing a measure $\mu(q)$ on the configuration space
and set:
 \be \label{npp:ip1}
  \IP{\Psi (q)}{\Phi (q)} = \lint_{\cal C} d^4q\>
   \mu(q)\> \ovr\Psi (q)\> \Phi (q),
 \ee
for all physical states $\Psi (q)$ and $\Phi (q)$. To determine the measure, we
impose the Hermiticity requirements. The condition that $\hQ^i$ be Hermitian
does not constrain the inner product in any way. The condition that $\hP_i$ be
Hermitian requires that the measure be independent of $q_i$. (In the general
case, when the ``true'' configuration space is a non-trivial manifold or the
coordinates are not Cartesian, the Hermiticity conditions on $\hP_i$ determine
the dependence of $\mu$ on $q^i$. The important point is that the dependence of
$\mu$ on $q^0$ is left undetermined.) Thus, the inner-product can now be
calculated:
 \bea \label{npp:ip}
  \IP{\Psi (q)}{\Phi (q)} &=& {\disp \lint
  d^4q\> \mu (q^0)\> \ovr\Psi (q)\> \Phi (q)} \cr
  &=& {\disp \lint dq^0\>\mu (q^0) \lint d^3q^i \>\ovr\Psi (q^0, q^i)\> \Phi
(q^0,
      q^i) } \cr
  &=& {\disp \lint dq^0\>\mu(q^0)  \lint d^3q^i \>\ovr\psi(q^i)\>\phi(q^i) }
\cr
  &=& {\disp K \lint d^3q^i\>  \ovr\psi (q^i)\> \phi (q^i)
   \equiv K \lint d^3q^i\>  \ovr\Psi (q^0, q^i)\> \Phi (q^0, q^i) }
 \eea
where the constant $K$ is given by $K = \lint dq_0 \,\mu (q_0)$. Here, in the
third step, we have used the fact that $\Psi(q_0, q_i)$ and $\Phi(q_0, q_i)$
are physical states, i.e., they satisfy (\ref{npp:qcon}). Thus, the second
integral in the second line is independent of $q_0$. Since $\mu(q^0)$ is not
constrained in any way by the Hermiticity of the observables, we can choose it
so that $K$ is finite, say $K=1$. Thus, the reality conditions do indeed select
a unique inner product on $\cV_{phy}$ (up to the usual overall constant) and
the
resulting quantum description is completely equivalent to the quantum theory of
the original unconstrained particle moving in a potential $V$ in the Euclidean
space.

The final kinematical picture is the following: the physical Hilbert space
consists of solutions to the constraint equation (\ref{npp:qcon}), with the
Hermitian inner product given by (\ref{npp:ip}). The Hamiltonian $\hH$ is a
symmetric operator on this Hilbert space. Up to this point, the physical
operators (\ref{npp:pop1}) were formal constructs, used to find an inner
product. Now, however, we can use the the physical inner product to {\em
rigorously}
define the unitary operator (\ref{npp:hatu}), and hence the physical operators.

So far, we have been dealing essentially with the covariant states $\Psi(q)$.
Note, however, that these covariant solutions are in $1-1$ correspondence with
the
$q^0$ independent (``initial data'') states $\psi(q^i)$. In fact, there is an
obvious unitary transformation, given by (\ref{npp:qsol}), between the
covariant states
$\Psi(q)$ and the states $\psi(q^i)\equiv \psi_0(q^i)$. The inverse of the
unitary transformation is given by:
 \be \label{npp:ut0}
  \psi_0(q^i):=e^{\frac{i}{\hbar}\hH q^0}\circ\Psi(q)\equiv\left.
\Psi(q)\right|_{q^0=0}.
 \ee
With $K=1$, the inner product on these states is simply (\ref{npp:ip}).
Clearly, the
states $\psi_0(q^i)$ are not the solutions of any constraint equation.
However, they carry a faithful representation of the observable algebra. Let
$z$ denote any operator in the set $(q^i,p_i)$; and let $Z$ denote the
corresponding operator
in the set $(Q^i,P_i)$.
Under the action of the unitary transformation, the
representation of the observables (\ref{npp:pop1}) is simply
 \be \label{npp:pop2}
  \hZ(0)\circ\psi_0(q^i) = \hz\circ\psi_0(q^i).
 \ee
The physical observables have a simple action on the space of initial states
for
the constraint equation.
Now, the intuitive meaning of these operators is clear:
Since the constraint generates dynamical evolution, we know that the physical
observables correspond to constants
of motion, which in turn can be identified with the position and momentum at
some initial time. Hence, a set of Dirac operators
can be obtained by ``evolving the covariant states $\Psi$ back to $q^0=0$''
(or, via (\ref{npp:ut0}), evaluating them at $q^0=0$),
acting with
the usual ``instantaneous'' operators on the initial state $\psi(q^i)$,
and then ``evolving the {\em resulting} initial state forward to
$q^0$'', using the constraint equation.
This is exactly the procedure we have carried out, as is
obvious also from the second equalities in (\ref{npp:pop1}).

\subsection{Dynamics and interpretation}

Are these the only operators or the only states that one can define?
Recall, from the introduction, that in the classical picture, on the constraint
surface (the space of
allowed classical states), any
cross-section of the gauge orbits (say $q^0=0$, where for the relativistic
parametrized particle we have identified the
``gauge'' parameter $\lambda$ with $q^0$) is isomorphic to the
reduced \ps. There is a $1-1$ correspondence between the classical states
(points) on this cross-section, and the gauge orbits.
Hence, the pull-back of a function to this cross-section naturally
defines a classical observable. There is analogous structure on the space of
allowed quantum states $\Psi(q)$. Since they are solutions to the quantum
constraint equation, they are in $1-1$ correspondence with the initial states
$\psi_0(q^i)$ satisfying $q^0=0$. Extending the analogy, we see that quantum
observables can be (and as I emphasized in the discussion above, have been)
obtained
by evaluating their action on the states $\psi_0(q^i)$ in the ``cross-section''
$q^0=0$.

However, recall also, that for dynamically constrained systems, in order to
recover a notion of evolution it is necessary to consider not just one
particular cross-section (which would be adequate for an ordinary gauge
theory), but a {\em foliation} of the constraint surface by a 1-parameter
family
of cross-sections of the gauge orbits (corresponding say to successive values
of
$\lambda$, or in this case, $q^0$). By analogy, since the covariant
configuration space can be foliated by $q^0=constant$ surfaces, each covariant
state $\Psi(q)$ defines a 1-parameter family of \Sch\ states
$\psi_\tau(q^i)$:
 \be \label{npp:utt}
  \psi_\tau(q^i):=e^{\frac{i}{\hbar}\hH(q^0-\tau)}\circ\Psi(q)
        \equiv\left. \Psi(q)\right|_{q^0=\tau}.
 \ee
Note that this
correspondence exists {\em only} because the states $\Psi(q)$ satisfy the
constraint equation.

To construct dynamical observables, recall that in the classical theory, given
any function on the constraint
surface, we evaluate its pull-backs to successive cross-sections, and
obtain a 1-parameter family of classical observables. Analogously,
the physical operators corresponding to a particular foliation $q^0=\tau$ are
simply given by
 \be \label{npp:popt}
  \hZ(\tau)\circ\psi_\tau(q^i) = \hz\circ\psi_\tau(q^i).
 \ee
In order to see clearly the corresponding 1-parameter family of physical
operators, let us use (the inverse of) the unitary transformation
(\ref{npp:utt}) to obtain the action of the $\tau$-dependent operators on
the covariant states. Doing so, we find
 \be \label{npp:cpop}
  \hZ(\tau)\circ\Psi(q) = \hU(\tau)\hz\hU^{-1}(\tau)\circ\Psi(q)
     =e^{-\frac{i}{\hbar}\hH(q^0-\tau)}\hz\circ\psi_\tau(q^i),
 \ee
where
 \be
  \hU(\tau)=e^{-\frac{i}{\hbar}\hH(q^0-\tau)}
 \ee
is the unitary transformation defined in (\ref{npp:utt}). It is now manifest
from this analogy that we can identify $q^0$ with the time
even in quantum theory, and that the $\hZ(\tau)$ are the ``evolving''
Heisenberg operators.

Of course, if we
desire, we can also evaluate the action of the operators $\hZ(\tau)$ on a fixed
\Sch\ Hilbert space, say corresponding to $q^0=\tau_0$:
 \be \label{npp:cpopt}
  \hZ(\tau)\circ\psi_{\tau_0}(q^i)=\left[ e^{\frac{i}\hbar\hH (\tau-\tau_0)}
              \hz e^{-\frac{i}\hbar\hH(z) (\tau-\tau_0)}
            \right]\circ\psi_{\tau_0}(q^i).
 \ee
As expected, we have lost all reference to $q^0$, and have obtained the
complete deparametrization of the theory to the usual Heisenberg/\Sch\ picture.

In retrospect we see
that we could have worked always in the covariant picture, with the
$\tau$-dependent Heisenberg operators defined in (\ref{npp:cpop}). However, we
would
then have lost the analogy
with the classical picture, and hence, both the interpretation of $q^0$ as time
as well as the motivation for the introduction
of the ``evolving'' observables. It is in order to see the unfolding of the
``hidden'' dynamics that we have to break the covariance of the space of
solutions and introduce on this space a ``foliation'' corresponding to time
evolution and the resulting sequence of \Sch\ states.

To conclude this section, note that nowhere in the kinematical construction
to find the inner product
was it necessary to treat $q^0$ in a special manner. We found the inner product
without explicitly eliminating the ``time'' $q^0$ and without integrating over
only the true degrees of freedom $q^i$; the reality conditions on $\cV_{phy}$
suffice to give us the inner product. This is an important point, since it
illustrates that it is not necessary to isolate time in order to construct the
kinematical quantum theory and a Hilbert space of physical states. (For an
original and more complete discussion of this aspect of the issue of time in
\qg, see \cite{aa:osgood6,aa:rst6,newbook6}.) Note however, that to complete
the analysis and make physical
predictions, as in the \Sch\ picture, one may need to find
explicit solutions by diagonalizing the ``true'' Hamiltonian $H$. In addition
to
the states, one will also have to construct explicit expressions for a
complete set of interesting operators.

Another important lesson that we have learnt is the following: {\em if}\
the constraint for any system can be deparametrized
and expressed in the form of a \Sch\ equation, {\em then} the problem of
constructing a {\em kinematical} quantum theory (by which I mean the
identification of a complete set of observables, and a unitary representation
on a Hilbert space) and a dynamical interpretation is formally completely
solved!

Finally, note
that it is trivial to extend this discussion to allow for a $q_0$-dependence in
the expression of the Hamiltonian (by appropriately normal ordering the
$U(\tau)$)
or to replace the Euclidean space by a 3-manifold.

\mysection{Reduced space quantization and interpretation}

In the previous section we saw that if a dynamically constrained system could
be deparametrized and a time-like variable identified, then one could formally
construct the Hilbert space of physical states. Furthermore, there is
well-defined framework in which to tackle the physical interpretation of the
theory. However, there are two obvious disadvantages to the above framework. In
the first place it is absolutely essential that the constraint is
deparametrized and expressed as a \Sch\ equation. Secondly, even when this can
be done, the solutions (\ref{npp:qsol}) one obtains are only formal
expressions, and in general the observables (\ref{npp:cpop}) pose formidable
factor-ordering problems. In fact, as in ordinary \Sch\ quantum mechanics, the
only known way to obtain explicit expressions for (\ref{npp:cpop}) begins with
diagonalizing the Hamiltonian \cite{kk:pc6}.

However, there are many systems in which, though a time variable can be
identified, the constraint equation does not simplify to the \Sch\ equation.
For example, consider the two coupled oscillator system quantized in chapter 5.
For this model, the natural time variable is angular. The momentum conjugate to
the time is then a ``radius-squared'' variable. (Roughly, $dp_1\wedge
dx^1=d(\frac{r^2}{2})\wedge d\theta$ is the symplectic structure.) The
constraint is linear in this new momentum variable, however, there is a {\em
non-holonomic} constraint that the new momentum should be positive. Due to
this, nontrivial subtleties arise in the quantum theory.

Similarly, there are systems for which the constraint equation is in a second
order, Klein-Gordon form. For example, consider (super-)stationary Bianchi
cosmologies, i.e.\ ones which admit a causal KVF on minisuperspace. In such
cases a positive/negative-frequency decomposition of the constraint has to be
carried out to identify a ``true'' Hamiltonian \cite{atu:I6}. Due to the
``square-root'' nature of the Hamiltonian, for such systems the approach of
section 2 may be too cumbersome or difficult to implement. If on the other hand
such systems are classically exactly solvable, then we can construct the
reduced space quantum theory. In this section I will discuss how a physical
interpretation of such systems can be obtained, without a \Sch\ like constraint
equation. Admittedly, the examples I have given above {\it can} be treated in
the Dirac framework. However, there are some difficulties, as I have mentioned,
and therefore it seems worthwhile to elucidate the classical, reduced space
approach. In this approach I have taken after Rovelli \cite{cr:I6}.

Consider a $2n+2$ dimensional \ps\ with a single constraint $C$, with a
specific choice of lapse. As in
spatially compact \gr, the constraint is also the generator of dynamics. Let
$z^\mu,\>\mu=1...2n+1$ be a set of independent functions on the constraint
surface%
\footnote{For the sake of simplicity, I will assume that the set is complete
 but not overcomplete, so that there are no algebraic identities amongst them.
 In the more general case, the range of $\mu$ will be greater, and there will
 be the appropriate number of algebraic identities, so that there are again
 only $2n+1$ independent functions, even though one may not be able to solve
 for them explicitly. In the following discussion, the counting will then
 change in only a trivial manner.}.
Our aim is find the observables of the theory, which correspond to functions on
the reduced \ps. Denote these by $ Z^i, i=1...2n$.

The constraint generates \ct s whose orbits lie in the constraint surface
itself. (The reduced \ps, as we know, is the space of orbits.)
The infinitesimal \ct\ is given by the Hamiltonian vector field $X_C$
of the constraint. Let $\tau$ be an affine parameter along these orbits. (Note
that $\tau$ is really the same as the $\lambda$ used in section 1; however, we
use $\tau$ here to emphasize that this is a dynamically constrained system.)
Now,
the observables $ Z^i$ pull-back from the reduced \ps\ to functions on the
constrained surface which are {\it constant} on the gauge orbits. At this
point, a very useful choice can be made: choose $\tau$
such that $( Z^i,\tau)$ is a new set of coordinates on the constraint
surface. (Note that nothing we have said so far prevents us from making this
choice, in fact this can always be done.) This choice
fixes the direction of the
coordinate vector field of $\tau$: it
has to be parallel to the Hamiltonian vector field of the constraint. Further,
since by definition $\tau$ is an affine parameter along the gauge orbits, we
know in fact that $(\frac{\d}{\d\tau})=X_C$!

So, we have {\it two} sets of coordinates on the constraint surface $\Gbar$:
the ``old'' coordinates $(z^\mu)$ induced from $\Gamma$, and the new
coordinates $( Z^i,\tau)$ ``induced'' from the reduced \ps\ $\hat\Gamma$.
Our problem is to relate these two sets of coordinates: To do this we have to
integrate
the gauge orbits.

Let us return therefore to the transformation
generated by the constraint.
This \ct\ induces a 1-parameter family of
transformations on the algebra of functions. The infinitesimal transformation
on a function $f$ is the Lie derivative, $\Lie{X_C} f$. {\it Now}, since
$(\frac\d{\d\tau})=X_C$, we can identify the infinitesimal \ct\ with the $\tau$
derivative of $f$: $\dot{f}:=\frac{\d f}{\d\tau}=\{f,C\}$! This identification
helps us to determine the relation between the two sets of coordinates.

On the constraint
surface,
the \ct\ is completely specified by
 \be\label{gic:heom}
  \dot{z}^\mu=\frac{d}{d\tau} z^\mu=\{z^\mu,C\}.
 \ee
Note that this is a set of coupled, but {\it ordinary} differential equations
for the $z^\mu$ as functions of $\tau$.
The algebra of observables is generated by the initial data for the above
equations, which we can identify with the $ Z^i$. Since the space of orbits is
$2n$ dimensional, the initial values of
$2n$ of the $2n+1$ $z^\mu$ specify the orbit. The initial value of the
remaining one, say $z^{(2n+1)}$, serves to fix the initial slice ($\tau=0$)
itself; assume for simplicity that $z^{(2n+1)}(0)=0$.
For later use, note that the formal solution to the above Hamiltonian equation
of
motion is given by the Taylor series
 \be\label{Taylor}
   z ^\mu(\tau)=\sum_N^\infty \frac{\tau^N}{N!} \left. \left(\frac{d^N z ^\mu}
    {d\tau^N} \right)\right|_{\tau=0}
   \equiv\sum_N^\infty \frac{\tau^N}{N!} \left. \left(\Lie{X_C}^N z^\mu
               \right)\right|_{\tau=0} ,
 \ee
where the coefficients of $\tau^N$ on the RHS are functions of $ Z^i$. If
it can be found, the solution of (\ref{gic:heom}) is in the form
 \be \label{gic:sol}
   z ^\mu= z ^\mu( Z^i,\tau),
 \ee
which is the coordinate transformation we were looking for on the constraint
surface between the set of coordinates ($z^\mu$) on the one hand and the set
($ Z^i,\tau$) on the other. This coordinate transformation can be inverted,
to yield the time and the constants of motion explicitly as functions on the
constraint surface
 \bea\label{gic:solinv}
  \tau&=&\tau(z^\mu) \cr
   Z^i&=& Z^i(z^\mu).
 \eea
The $ Z^i$ are classical Dirac observables. They parametrize the reduced
\ps\ $\hat\Gamma$, and the Poisson bracket on $\hat\Gamma$ can be derived from
(\ref{gic:solinv}) (equivalently, the symplectic structure on $\hat\Gamma$ can
be obtained from (\ref{gic:sol}). Now, for each value of $\tau$, one obtains a
cross-section of the orbits. Evaluating (\ref{gic:sol}) at a fixed value of
$\tau$, the RHS is a function on the reduced \ps, and hence an observable.
Thus, (\ref{gic:sol}) defines a 1-parameter family of observables corresponding
to each $z^\mu$. These are the Heisenberg observables. Note that as Heisenberg
observables, the Poisson brackets between $z^\mu$ are to be evaluated on
$\hat\Gamma$, and can be evaluated at different times $\tau$.

To construct the reduced space quantum theory, one has to represent the
classical Dirac observables $ Z^i$ as (Hermitian) operators $\hZ^i$ on
some Hilbert space of physical states. Physical interpretations and dynamical
information about any function on $\bar\Gamma$ can be obtained from the
operator analog of (\ref{gic:sol}):
 \be \label{gic:qsol}
   \hat{z}{}^\mu(\tau)= \>{\bf :}z ^\mu(\hZ{}^i,\tau){\bf :}\>,
 \ee
where ${\bf :}z^\mu{\bf :}$ indicates that difficult factor ordering problems
may have to be resolved in order to make the RHS a well-defined operator. For
each value of the parameter $\tau$, (\ref{gic:qsol}) defines an operator on
physical states corresponding to the classical function $z^\mu$. We can easily
generalize this to arbitrary functions $f(z^\mu)$ on the constraint surface.
Corresponding to $f$ we have the 1-parameter family of physical operators
 \be
  \hat{f}(\tau)=\>{\bf :} f(z^\mu(\hZ^i,\tau)){\bf :}\> .
 \ee

As always with the reduced space approach to quantum theory, the complete
classical solution is necessary before one can even begin to quantize. In the
absence of the classical solution, one can not even ``get off the ground''.

On another note, an important conceptual difference between this approach and
that of the previous section is that here the full constraint is thought of as
the generator of dynamical evolution, whereas in the \Sch\ approach, a piece of
the constraint is considered as the true Hamiltonian. Correspondingly, while
the time variable one obtains in the \Sch\ approach is an ``internal'' time,
obtained explicitly right from the start as a specific function of the
elementary operators, the time variable in the reduced space approach is an
``external'' time, only implicitly defined by the solutions (\ref{gic:sol}).
Since the time is an external variable (it belongs to a different set of
coordinates on $\Gbar$) the $z^\mu$ are treated democratically in the sense
that not one of them is singled out beforehand as a time variable.

\mysection{Bianchi I model}

In this section I will illustrate the above approach to the physical
interpretation of constrained dynamical systems by studying Bianchi I model in
some detail. As a practical application I will analyze the issue of
singularities in quantum cosmology. As we saw in section 3.4 (see also
\cite{atu:II6}), the reduced space and Dirac quantum theories of the solvable
models are kinematically equivalent. Since the issues will be conceptually
easier to grasp, I will take the reduced space approach to the canonical
quantization of Bianchi I.

On the spatially homogeneous (SH) slices introduce a SH co-basis, $\omega^i$
(see \cite{mah6,m:mini6,RandS6}). For Bianchi I this basis of 1-forms satisfies
$d\omega^i=0$. In this basis, the components of the 3-metric on the SH slices
are constants. Further, there is a \st\ symmetry \cite{aa:sam6} which allows
one
to eliminate the off-diagonal components of the 3-metric $q_{ij}$ and its
conjugate momentum $p^{ij}$. In this gauge, the diffeomorphism constraint of
\gr\ vanishes identically. The gravitational \ps\ $\Gamma$ is thus the
cotangent bundle over the configuration space \cC\ parametrized by the 3
diagonal components $q_i$ of the metric. Define new coordinates on \cC\ by
 \be \label{b1:newpar}
  \left( \begin{array}{c}
        \bz \cr
        \bp \cr
        \bm \cr
        \end{array} \right) =  \dfrac{1}{2}
  \left( \begin{array}{ccc}
           1/3   & 1/3          & 1/3  \cr
           1/6   & 1/6          & -1/3 \cr
     1/2\sqrt{3} & -1/2\sqrt{3} & 0    \cr
        \end{array} \right)
  \left( \begin{array}{c}
         \ln q_1 \cr
         \ln q_2 \cr
         \ln q_3 \cr
        \end{array} \right)\ .
 \ee
The canonically conjugate momenta are
 \be \label{b1:newmom}
  \left( \begin{array}{c}
        \pz \cr
        \pp \cr
        \pmm \cr
        \end{array} \right) = 2
  \left( \begin{array}{ccc}
        q_1            & q_2             & q_3   \cr
        q_1            & q_2             & -2q_3 \cr
        \sqrt{3}\, q_1 & -\sqrt{3}\, q_2 & 0     \cr
        \end{array} \right)
  \left( \begin{array}{c}
         p^1 \cr
         p^2 \cr
         p^3 \cr
        \end{array} \right)\ ,
 \ee
so that the symplectic structure is $\Omega=d\pi_A\wedge d\beta^A,\> A=0,\pm$.
In terms of the momenta, the diagonal components of the extrinsic curvature can
be obtained via $K_i=\frac{1}{V}\left(p^iq_i^2- \half(\sum_j
p^jq_j)q_i\right)$, where $V=(q_1q_2q_3)^{1/2}$ is the spatial volume element.

In cosmological models, in the gauge $N =$ const., the supertime (the affine
parameter along the dynamical trajectories in \ps) can be identified with the
proper time of an unaccelerated observer whose world line is orthogonal to the
SH slices. In this physical time gauge, the initial singularity in classical
theory occurs at a finite time in the past.

Choosing $N=4$, the scalar constraint for Bianchi type I is
 \be \label{b1:con}
  C=\tfrac16e^{-3\bz}\eta^{AB}\pi_A \pi_B =
        \tfrac16e^{-3\bz}(-\pz^2+\pp^2+\pmm^2).
 \ee
For an expanding universe, there is an additional nonholonomic constraint
$\pi_0<0$. (See subsection 4.2 for a complete discussion of this choice.)

Let $\tau$ be an affine parameter along the ``gauge'' orbits generated by the
constraint, chosen as in the previous section such that
$\dot{f}\equiv\frac{\d f}{\d\tau}$. Then the Heisenberg
equations of motion are $\dot{f}=\{f,C\}$. For Bianchi type I these can be
easily solved on the constraint surface, to yield
 \bea \label{b1:clsol}
   \pi_\pm &=& \rp^\pm\ , \cr
   -\pz    &=& \rpz := \sqrt{(\rpp)^2+(\rpm)^2}\ ,\cr
   \bz(\tau) &=& \tfrac13\ln(\rpz\tau)\ ,\cr
   \beta^\pm &=& \rb_\pm + \dfrac{\rp^\pm}{3\rpz} \ln\tau\ ,
 \eea
where $\tau\ge0$, the initial value of $\bz$ has been chosen to simplify the
solutions, and the initial data $(\rb_\pm, \rp^\pm)$ are canonical coordinates
on the reduced \ps\ $\hat\Gamma$. The reduced symplectic structure (which can
be obtained by pulling back the symplectic structure on $\Gamma$ to the
constraint surface, and then evaluating at some fixed value of $\tau$, say
$\tau=1$) is $\hat\Omega=d\rp^{\skhatA}\wedge d\rb_{\skhatA},\> \skhatA=\pm$.
Thus $\hat\Gamma$ is the cotangent bundle over the reduced configuration space
$\hat{\cal C}=(\rpp,\rpm)$.

The algebra of elementary, reduced space observables is given by the
commutation relations
 \be  \label{b1:redccr}
  [\wh\rb_\hA,\wh\rp^\hB]=i\hbar\delta_\hA{}^\hB.
 \ee
{}From the point of view of later calculations, it is convenient to choose the
($\rp$)-representation for quantum theory. The states $\psi=\psi
(\rpp,\rpm)$ are $C_0^\infty$ functions on $\hat{\cal C}$, and the operators
are represented by
 \bea \label{b1:redrep}
  \wh\rp^\pm\circ\psi(\rpp,\rpm) &=& \rp^\pm\cdot\psi(\rpp, \rpm)\cr
  \wh\rb_\pm\circ\psi(\rpp,\rpm) &=&
    i\hbar\dfrac\d{\d\rp^\pm} \psi(\rpp,\rpm).
 \eea
One can immediately see that the operators are symmetric with respect to the
inner product
 \be  \label{b1:ip}
  \IP\psi\phi=\int_{\hat{\cal C}}\> d^2\rp\>\bar\psi\,\phi.
 \ee

\subsection{Physical interpretation}

To extract a physical interpretation, we have to ``quantize'' the classical
solutions (\ref{b1:clsol}). The 1-parameter families of quantum ``Heisenberg
observables'' are
 \bea  \label{b1:hobs}
   \wh{\pi}_\pm &= \wh\rp{}^\pm\ ,\qquad
   -\wh{\pi}_0 = \wh\rp{}^0 =
               \sqrt{(\wh\rp{}^+)^2 + (\wh\rp{}^-)^2}\ ,\cr
   \wh\bz(\tau) &= \fr13\ln(\wh\rp{}^0\tau)\ ,\qquad
   \wh\beta^\pm(\tau) = \wh\rb_\pm +
                        \dfrac{\wh\rp{}^\pm}{3\wh\rp{}^0} \ln\tau\ .
 \eea
Note that while the $\beta^A$ are not by themselves Dirac observables, and are
thus not well defined operators on the space of physical states, when they are
expressed as Heisenberg operators (\ref{b1:hobs}) then (up to factor-ordering)
they do have a well-defined action on the physical states. We can represent the
Heisenberg observables on physical states by
 \bea \label{b1:hrep}
   \wh{\pi}_\pm\circ\psi &=& \rp^\pm\psi\ , \cr
   \wh{\pi}_0\circ\psi &=& \rpz\psi\ ,\cr
   \wh\bz(\tau)\circ\psi &=& \tfrac13\ln(\rpz\tau)\psi\ ,\cr
   \wh\beta^\pm(\tau)\circ\psi &=& i\hbar\dfrac\d{\d\rp^\pm}\psi +
                          \dfrac{\rp^\pm}{3\rpz} (\ln\tau)\psi\ .
 \eea
For Bianchi I we have the good fortune that there is a representation in which
there are no (insurmountable) factor-ordering problems, at least for the
elementary Heisenberg observables.

Note that the above procedure allows one to represent {\it any} \ps\ function
as a 1-parameter family of Heisenberg operators on physical states.
Corresponding to any classical \ps\ function $f(\beta^A,\pi_A)$, by
substituting (\ref{b1:hrep}) we have the Heisenberg operator
 \be  \label{b1:hop}
  \wh{f}(\tau)=\>
   {\bf :}f(\beta^A(\wh\rb_\hA,\wh\rp{}^\hA,\tau),\pi_A(\wh\rb_\hA,\wh\rp
  {}^\hA,\tau)){\bf :}
 \ee
where ${\bf :}f{\bf :}$ indicates an appropriate factor-ordering.

Some observables of particular interest are the 3-volume $V=\exp(3\bz)$ and the
trace of the extrinsic curvature ${\rm tr}K=-\fr14 \pz \exp(-3\bz)$. The
Heisenberg operators are
 \be  \label{b1:vol}
   \wh{V}(\tau)\circ\psi = \wh\rp{}^0\tau\circ\psi=\tau\rpz\psi\ ,\qquad
   \wh{{\rm tr} K}(\tau)\circ\psi = \frac1{\tau}\psi\ .
 \ee
We see from the above expression that the quantum operator corresponding to the
spatial volume vanishes as $\tau\rightarrow0$; and that the operator
corresponding to the trace of the extrinsic curvature becomes unbounded. This
gives us a hint of things to come in the next section.

Ideally, one would like to eliminate the $\tau$ entirely from the above
equations, since in \gr\ we do not have an external time. (For a a recent
review of the issue of time in \qg, see \cite{kk:time6}.) Rovelli's suggestion
for the general case \cite{cr:I6} is that this be can be done by substituting
the quantum version of the solution $\tau=-\frac{4}\pz \exp(3\beta^0)$ back
into the rest of the Heisenberg observables (\ref{b1:hrep}). However, in
general, this is an ill-defined procedure, since it is the quantum analog of
substituting part of a coordinate transformation back into itself.

In this framework, there is an alternate viewpoint one can take on the issue of
time, which is to {\em eliminate the time $\tau$ interpretationally}, by asking
only {\em relational} questions. Consider Bianchi I. As an Heisenberg operator,
$\wh{{\rm tr}K}$ commutes with all other operators. Conceivably, we can measure
${\rm tr} K$, and this provides us with a ``time''. Now we can
simultaneously measure any other \ps\ function of interest, say the volume.
As a result of such an experiment (which involves 2 measurements) one would
make a physically meaningful statement of the form: ``At the time tr$K=3$, the
universe was found to be half full.''.

\subsection{ The initial singularity in quantum cosmology}

Classically, for generic solutions, the initial singularity is a curvature
singularity in Bianchi type I: the Weyl curvature squared scalar
$W^2:=C_{abcd}C^{abcd}$ blows up at a finite time in the past. What happens to
this behaviour in the quantum theory?

The operator we would like to analyse is the Heisenberg observable
corresponding to $W^2$. Thus we want to express $W^2$ in terms of the Cauchy
data: either the 3-metric and extrinsic curvature or, equivalently, the pair
$(\beta^A,\pi_A)$. By a $P2C2E$%
\footnote{``Process Too Complicated To Explain'' \cite{sr:sea6}.}
it follows that \cite{jp:cu6}
 \be  \label{b1:ws}
   W^2=\fr1{216}e^{-12\bz}\pz^4(1+\cos3\theta)\ ,
 \ee
where $\theta=\tan^{-1}(\pmm/\pp)$ is an angular coordinate on the $\pi_\pm$
plane. (Note that here as in full \gr, the Weyl scalar is independent of the
lapse (and shift) \cite{jp:cu6,ct:pc6}.)

Substituting the classical solutions (\ref{b1:clsol}), the classical Heisenberg
observable, in terms of the reduced space observables is
 \be\label{b1:hws}
  W^2(\tau)=\frac{32}{27\tau^4}(1 + \cos3\Theta)
 \ee
where $\Theta=\tan^{-1}(\rpm/\rpp)$ is an angular coordinate on the reduced
configuration space. Fortunately, in the ($\rp$)-representation the factor
ordering of (\ref{b1:hws}) is trivial. The action of the Heisenberg operator
corresponding to the Weyl scalar on physical states is
 \be\label{b1:wsop}
   \wh{W}^2\circ\psi(\rpp,\rpm)=\frac{32}{27\tau^4}(1 + \cos3\Theta)
   \cdot\psi(\rpp,\rpm)\ .
 \ee
As $\tau\rightarrow0$, the whole spectrum of $W^2$ blows up. Clearly, the
classical \st\ singularity persists in quantum theory. Similar results have
been obtained previously by Gotay and Demaret \cite{mjg:jd6}; and for Gowdy
models by Husain \cite{vh6}. Such results for Bianchi models have also been
hinted at in \cite[pp. 198]{RandS6}.

\subsubsection*{Remarks:}
\remarks
\remark
\noindent Since $\tau$ is thought of as a fiducial parameter, one is perhaps
more
interested in relating $W^2$ to some other physical quantity, say $\frac1{{\rm
tr}K}$. (As Gotay and Demaret have pointed out \cite{mjg:jd6}, since this is a
relativistic model, one is free to rescale the parameter time so that the
singularity does not occur at a finite parameter time. When one does this in
the classical theory, though, certainly no one claims that the singularity has
been eliminated.) Then, the relational
statement is that $W^2$ blows up when $\frac1{{\rm tr}K}$ vanishes. However,
there is no guarantee that $\frac1{{\rm tr}K}$ itself vanishes at some finite
time as measured by a physical clock. I have chosen a time parameter
$\tau$ for the quantum theory which appears to be closest to a classical time
in which the Universe has existed for a finite time only, and corresponds to
the proper time for an unaccelerated observer.
\remark
\noindent Functions with support only on the regions
${\Theta}=\pi,\pm\fr{\pi}3$
are indeed annihilated by the Weyl scalar operator (\ref{b1:wsop}), for all
times $\tau$. However, these functions are not normalizable states in the
Hilbert space. Furthermore, the classical spacetime constructed from initial
data which lies in this region is just locally flat space%
\footnote{Louko \cite{jl:pc6} has pointed out that while the origin
 (in the $(\rpp,\rpm)$ plane) corresponds
 to Minkowski space, the other points on ${\Theta}=\pi,\pm\fr{\pi}3$ correspond
 to \st s with conical singularities. To study the behaviour in quantum theory
 of these solutions, one could, for example, express the deficit angle as a
 function on the \ps\ and then use (\ref{b1:hop}) to construct the
 corresponding quantum operator on physical states.}.
So, for this region of the reduced \ps, there is no curvature singularity
classically either.
\remark
\noindent Note that the same result can be obtained even if one quantized the
theory using the operator constraint method of Dirac, since in the $\beta^A$
representation the observables (\ref{b1:hobs}) and the Weyl scalar
(\ref{b1:hws}) are at most linear in momenta and do not pose any
factor-ordering problem. Alternately, since the constraint (\ref{b1:con}) can
be written in the form of a nonrelativistic free particle (\ref{npp:con})
 \be
  C\equiv p_0+H=0, \quad\hbox{where}\quad H=\sqrt{(p_+)^2+(p_-)^2},
 \ee
we can carry out the analysis using the formalism of section 2. Though this
corresponds to a different time parametrization, the qualitative results will
be the same. The persistence of the singularity in quantum theory is
independent of whether or not one ``quantizes'' the constraint, and the
singularity can not be avoided by exploiting any factor ordering ambiguity.
\remark
\noindent At least some of the potential ambiguities and inconsistencies in a
quantum theory of gravity arise from a choice of a time variable which
corresponds to {\em 3-dimensional, spatial} scalars (like the volume $V$ or
${\rm tr}K$). In order to eliminate some of these inconsistencies, and possibly
even for aesthetic reasons, one may require the time variable to be {\em
4-dimensional, spacetime} scalar. One such choice would be to use the Weyl
scalar itself (or some other curvature scalar) as a time variable. One obvious
disadvantage,
indicated by (\ref{b1:wsop}), is
that with this requirement we may have to give up a state-independent notion of
time. However, this appears
to be a promising avenue for exploration.
\remark
\noindent This way of formulating the problem relies heavily on the solvability
of
the model. Over and above the solvability, though, it was essential that the
final expression for the Weyl scalar as a Heisenberg observable was relatively
simple. Consider what happens in the next simplest homogeneous cosmology.
Bianchi type II, as we saw in section 3.4, is kinematically completely solved.
One can even find an explicit expression for the classical Heisenberg
observable corresponding to the Weyl scalar. However, this expression is
horrendous, involving products of exponentials and quartics in the elementary
reduced space operators, and I have been unable so far to construct a good
quantum operator.

In spite of the factor-ordering difficulties one can make some qualitative
claims: There is again a pre-factor $1/\tau^4$ in the classical expression of
the Heisenberg observable corresponding to the Weyl scalar. Thus if there is
{\it any} factor-ordering which makes the operator well-defined and non-trivial
at some non-zero time, it is sure to blow up at $\tau\rightarrow0$. This is
likely to be a fairly generic feature of all the solvable cosmological models
(see \cite{atu:II6}).

\subsection{(\dag) Expanding or contracting Universe, or both?}

I will now discuss the choice of sign of $\pz$. When solving the constraint, I
assumed a specific choice, $\pz<0$, corresponding to an expanding universe.
Suppose that one does not make this choice. Then, the projection of the
constraint surface into the configuration space is the entire null cone,
$-\pz^2+\pp^2+\pmm^2=0$. However, the future (contracting Universe) and past
(expanding Universe) null cones are ``dynamically disconnected'' from each
other (and in fact from the origin): in the sense that there are no
(continuous) observables which will map a point in the future null cone to a
point in the past null cone, since all vector fields on the cone have to vanish
at the origin. There is of course a discrete transformation, $T:\pz\mapsto
-\pz$, corresponding to time reversal. The corresponding operator is
superselected, it commutes with all the rest of the elementary Dirac
observables $(\rb_\pm,\rp^\pm)$. Thus the reduced \ps\ consists of two
disconnected sectors, namely the cotangent bundles over the past and future
null cones. The Hilbert space of states on any one half carries a faithful
unitary representation of the algebra of (continuous) Dirac observables
$(\rb_\pm,\rp^\pm)$. The Dirac operators $\wh\rb_{\skhatA},
\wh\rp{}^{\hat{B}}$ are block diagonal and are represented on each half of the
Hilbert space by (\ref{b1:redrep}). The two halves are distinguished by the
sign of $\pz$, $\pz=\pm\rp^0=\pm\sqrt{(\rp^+)^2+(\rp^-)^2}$ on the future
and past halves respectively.

However, the Heisenberg observables {\it do} have different representations on
each half. (\ref{b1:hrep}) represents the Heisenberg observables
(\ref{b1:hobs}) on the past half of the Hilbert space. On the future half, the
representation can be obtained simply by replacing $\rp^0\mapsto -\rp^0$.
Hence we see that on the future (contracting) half, due to our choice of the
initial value of $\beta^0$, $\tau\le0$, and the singularity occurs at a finite
time in the future.

Since there is no tunneling between states in either half of the Hilbert space,
if the universe is in an expanding state, it is consistent to do quantum theory
entirely in this half. Else, one has to contend with the following scenario:
one can find appropriate choices of  the initial values for $\beta^0$ such that
the ranges of $\tau$ in each half overlap. Now consider ``mixed'' states, say
eigenstates of $\hat{T}$. In such a state, the Weyl scalar operator blows up at
both a finite time in the past and a finite time in the future. Clearly, it is
only sensible to restrict oneself to one half of the Hilbert space, say the
expanding half.

\mysection{Discussion}

Both the \Sch\ and the reduced space approach have noteworthy features:
\begin{enumerate}
\item In the \Sch\ approach of section 2, it is absolutely essential that the
constraint be deparametrized and written in the form $C=p_0+H$, where the true
Hamiltonian $H$ is its own $\star$-adjoint, so that it is represented on
physical
states by a Hermitian operator. Furthermore, as I have emphasized
before, the solutions are only formal, and a substantial amount of work is
still
required in order to obtain physical predictions. However, there is at least a
kinematical framework within which various issues can be addressed.
\item On the other hand, the reduced space approach works for any form of the
constraint, as long as it is classically solvable. When the complete classical
solution is known, one can again construct the kinematical quantum theory. The
difficulties have to do with factor-ordering {\it known} classical expressions
for the
Heisenberg observables.
\item Time plays a slightly different role in the two approaches. In the \Sch\
approach, it is essential to single out a time variable before one can proceed
further. This time is an {\em internal time} in the sense that this variable
and its conjugate momentum are part of the set of elementary variables.
Dynamics, or evolution in this internal time, is generated not by the
constraint, but by the true Hamiltonian, via
$\dot{\hat{f}}=\frac1{i\hbar}[\hat{f},\hat{H}]$.

This is in constrast to the reduced space approach (which is more closely
related to ordinary gauge theories), in which to begin with time is just an
affine parameter along the gauge orbits, {\em undetermined} as a function on
$\Gbar$ until {\it after} the classical equations of motion have been solved.
Then,
it is part of a new set of coordinates on $\Gbar$, related to the elementary
variables by a \ct\ generated by the constraint. In this picture, time
evolution is generated by the constraint itself, via $\dot{f}=\{f,C\}$. This
approach is more democratic, since {\it a priori} all the elementary variables
are treated on an equal footing. Further, since only relational questions are
asked, this is a more relativistic approach. If at all, a particular \ps\
function is identified with time only after the complete solutions are
obtained, and then, this is done only interpretationally.
\item Of course, in the \Sch\ approach one can attempt standard perturbation
techniques. In addition, both approaches lend themselves to another
approximation technique. The idea here is to truncate the equations of motion
in ``time''. In the \Sch\ approach, this amounts to expanding either
(\ref{npp:cpop}) or (\ref{npp:cpopt}) using the Baker-Hausdorff lemma and
retaining only a finite number of terms. On \Sch\ states, the approximate
Heisenberg observables are then given by:
 \be
  \hZ_{(k)}(\tau)
   =\hz+ \frac\tau{i\hbar}\left. [\hz,\hat{H}]\right|_{\tau=0} +
   ... +\left(\frac\tau{i\hbar}\right)^k \left.
         [..[[\hz,\hat{H}],\hat{H}]..,\hat{H}]^k\right|_{\tau=0}
 \ee
where it is understood that the terms on the RHS, since they are evaluated at
$\tau=0$, are functions of $\hZ$. The Hilbert space structure can be used to
make various truncated operators well defined. The constraint is obviously
exactly satisfied. Such an approach can be used on certain Bianchi models
\cite{atu:I6} which are not exactly soluble, but where the existence of a
causal
supersymmetry permits one to do a $\pm$ frequency decomposition of the
constraint, identify the true Hamiltonian and construct a physical Hilbert
space.

In the reduced space approach an approximation is needed before one can even
construct a Hilbert space, but the approximation itself is done classically, so
perhaps it has some advantages. The idea here is to truncate the Taylor series
for the observables (\ref{Taylor}) at finite order in $\tau$:
 \be\label{Taylor2}
   z ^\mu_{(k)}(\tau)=\sum_N^k \frac{\tau^N}{N!} \left. \left(\frac{\d^N z
^\mu}
    {\d\tau^N} \right)\right|_{\tau=0} ,
 \ee
where $\frac{\d f}{\d\tau}=\{f,C\}$ and the RHS is in terms of $ Z$.
Pictorially, one identifies a particular cross-section of the gauge orbits with
$\hat\Gamma$ and then, to extract dynamical information one approximates the
gauge orbits by higher and higher order polynomials in $\tau$. The constraint
can either be solved exactly, or to a particular order only.

Clearly, such approximation techniques will {\em not} provide answers about the
existence or structure of singularities, the issue of chaos in Bianchi IX etc.
It is not immediately obvious which interesting issues in \qg\ can be
addressed by applying {\em such} approximations.
\item Yet another possibility is a hybrid approach. The major flaw of the \Sch\
approach is that the constraint has to be deparametrized, while in the reduced
space approach one has to find the full classical solutions. In a hybrid
approach, one might try to repeat the construction of section 3, but quantum
mechanically. One might look for quantum versions of the solutions
(\ref{Taylor}), or undeparametrized versions of (\ref{npp:cpop}):
 \be
  \hZ(\tau)=e^{\frac{i}\hbar \hat{C}\tau}\hz e^{-\frac{i}\hbar \hat{C}\tau}.
 \ee
Unfortunately, this naive extension of the approach of section 3 to Dirac
quantization is inconsistent, for a multitude of reasons. However, some
possibilities appear worth investigating.
\end{enumerate}

I do not want to give the impression that the approaches described in sections
2 and 3 are the only way to obtain physical interpretations of a theory.
Quantum gravity will certainly have many inter-related ``phases'': isolated,
asymptotically flat systems in which one may be interested in quantum effects
on the spectrum of gravitational radiation from binary stars; strong
curvature/black hole regimes with particle production; microstructure of \st,
wherein will lie the justification for and understanding of quantum field
theory; cosmology and large scale structure. These phases may or may not all be
described by the ``same'' quantum theory, and certainly the questions one asks
will be very different. The above approach to physical interpretations is
clearly not applicable in all the regimes. Also, many classically important,
Kepler-like questions, e.g. the deflection of light, the Shapiro time delay,
the precession of the perihelion, decay of the orbital frequency of binaries,
etc. may not be relevant questions in quantum gravity.

However, one question we do know will be important is whether a quantum theory
of gravity is free of curvature and other singularities (which we know are
generic in classical theory) or renders them irrelevant, or leads to a better
picture of their ``internal structure''. The above framework for physical
interpretation is applicable to this question, at least in the context of
quantum cosmology. The result, that the classical singularity persists in
quantum cosmology, leads one to the conclusion that homogeneous cosmologies as
models of full \gr\ are far too simplistic and fail to share some of its
essential features. For example, due to the imposition of spatial homogeneity,
the behaviour of the geometry of \st\ on the micro-scale is frozen out. The
possible discrete nature of the spatial geometry on the small scale, which
would be one
avenue to avoid singularities and infinities in quantum theory obviously never
plays a role in the quantum theory. In this light the above result is not
surprising.

On the other hand, the result agrees with the ``rule of unanimity''. Wheeler's
conjecture \cite{jw:rule6} was that in cosmological models, if {\it generic}
initial data evolved classically to a singularity in finite time, then generic
physical states would have support on such data and the singularity would still
be present in the quantum theory. Where classically there are
powerful theorems that fairly generic initial data evolve in finite
time to a singularity, is it too speculative to wonder if such a
rule of unanimity applies to full \qg\ as well?

%
\newpage
\mbox{}
\chapter{DIALOGUE}
\pagestyle{myheadings}
\markboth{{\sf Chapter 7}}{{\sf Dialogue}}

\newcommand{\ach}{\item[{\it Achilles:}]}
\newcommand{\tor}{\item[{\it Tortoise:}]}
\def\npp{non-relativistic parametrized particle}
\def\dcs{dynamically constrained system}

\begin{quotation}
{\sl (This concluding chapter is in the form of a conversation, mostly
between two characters: Achilles and the Tortoise.
I have borrowed this format from Hofstadter \cite{geb}.
Other characters put in cameo appearances, the
resemblance of any of these characters to real persons is coincidental.)}
\end{quotation}

\begin{description}

\ach {\large\bf D}oubtless, Mr.\ T, this thesis has presented a reformulation,
in
an algebraic framework, of the Dirac approach to the quantization of
constrained
systems. Some rather simple illustrative examples have also been worked out.
But what are really the new elements in algebraic quantization? After all,
consider the issue of finding the inner product using the Hermiticity
conditions on certain operators. Isn't this exactly what we do in ordinary
quantum mechanics?

\tor {\large I} would object that this not quite so. In ordinary quantum
mechanics, Achilles, one has a metric on the configuration space and its volume
element defines an inner product on quantum states. One then {\it checks} that
various interesting operators are actually Hermitian in this inner product; if
they are not, one symmetrizes them. This leads e.g.\ to the addition of a
$\frac1{r}$ term in the representation of the radial momentum operator. Here on
the other hand, one stands this procedure on its head. We work with a {\it
fixed} representation of the operators on a complex vector space, and {\it use}
the Hermiticity conditions on them to {\it find} an inner product. As we saw in
various examples ($q-z$ variables for the oscillator and the Bianchi II model),
in the configuration representation this process yields differential equations
for the measure, which can usually be solved. In some cases of course, no
solution may exist. Then one has to start all over again by choosing another
linear representation of the algebra of operators.

\ach {\large R}adical! But what about constrained systems?

\tor {\large A}s discussed in chapter 1, for constrained systems there really
weren't any criteria established for introducing the physical inner product.
Typically one could use available background structure (e.g.\ the metric)  to
construct the inner product. Or, as in quantum cosmology, this issue was
considered of secondary importance and often left unresolved. Consider, e.g.,
the Bianchi II models. Previously, in the $\bar\beta$-representation a large
number of solutions to the scalar constraint were known. However, it was not at
all clear in what sense this set of solutions is complete, or what inner
product one could use on these states.

\indent The algebraic framework provides criteria which resolve both these
issues: The space of physical states should be large enough to carry a faithful
representation of the physical operator algebra; and, the physical inner
product on these states should be such that physical operators corresponding to
real functions on \ps\ are Hermitian. Obviously, in order to implement the
criteria we need a complete set of physical operators. However, once these
criteria were established, we were able to apply them, e.g.\ to the Bianchi II
model, and construct the complete kinematical quantum theory. In this case a
clever choice of elementary variables gave us a complete set of physical
observables and we then used the Hermiticity relations amongst these to
construct an inner product on physical states.

\ach {\large C}onsider that in Bianchi models --and indeed in full \gr-- the
scalar constraint contains a term quadratic in momenta whose coefficient
defines a (super)metric on the configuration space. Couldn't we have used the
volume element of this metric to introduce the inner product?

\tor I don't think so. This volume element does define an inner product on
$\cV$, the vector space which carries a representation of the elementary
operators themselves. One can certainly use this inner product as a technical
device for various intermediate calculations. However, in general physical
states are not normalizable with respect to it (as happens in Bianchi II
models), and a new inner product has to be introduced. There is another
potential problem, illustrated by the coupled oscillators and the A-H model. In
these cases, one {\it can} introduce an inner product on the representation
space \cV, make the constraint a Hermitian operator on the resulting Hilbert
space and then solve the constraint equation. There are normalizable solutions.
However,  if one does this, the space of physical states may be just too small
to capture all the physics.

\indent Consider also the quantum theory of chaotic systems. Perturbation
calculations for some systems indicate that the spectrum of the (Hermitian)
Hamiltonian becomes non-degenerate as one enters the chaotic regime. Now,
certain relativistic cosmologies may well be classically chaotic. For these
models, the Hamiltonian is constrained to vanish. If one introduces an inner
product on ${\cal V}$ itself, and if the spectrum of the Hamiltonian constraint
is non-degenerate, we would find  at most {\it one} physical state. Thus this
quantum description would again be highly incomplete.

\ach I see now that one should first solve the quantum constraints for the
physical states, find physical observables and only then construct an inner
product, on the space of physical states, that makes the observables Hermitian.
But, in previous versions of Dirac quantization most of the emphasis was placed
on the solutions to the quantum constraint. Why do we need an inner product on
these states? Don't they already contain all physical information?

\tor To extract any physical interpretation one needs a {\it Hilbert} space of
physical states, and a set of operators corresponding to first class functions
on the \ps. One needs an inner product on physical states for the same reason
as in ordinary quantum mechanics. One is not interested in just the
wavefunctions and their support, but in probability {\it densities}. A
particular physical state might be found to have large support in some
interesting area of the configuration space, but if the measure itself is very
small in that region, then the state will not contribute much in quantum
theory.

\ach Aha! the A-H model of chapter 4 provides an extreme example of this. There
we found {\em solutions} to the quantum constraint equations
which actually had support {\em entirely} in
the classically forbidden, negative energy regions. Since they were
normalizable in the fiducial (Euclidean) metric inner product, one was tempted
to conclude all sorts of interesting things in the quantum theory. However,
since the {\it physical} inner product {\it vanishes} in that region, as
elements of the Hilbert space, these solutions are identified with the zero
state.

\tor Glory be, Achilles! What a good example of the pitfalls associated with
jumping to conclusions about the quantum solutions themselves.

\ach Another issue that has me confused is the following: In the language of
chapter 6, \gr\ is a ``\dcs'', in which the generator of dynamics is
constrained
to vanish. Time is considered to reside in one of the canonical variables. For
the sake of discussion, choose a polarization in which the time parameter is a
configuration variable. In the Dirac theory, in the configuration
representation, wavefunctions --and in particular the physical states-- will
depend on this time variable. So, any inner product would appear to necessarily
involve an integral over the time variable too. Now, if one repeats this in
Schr\"odinger quantum mechanics (where too the states depend on time) --i.e.,
if one integrates the usual inner product over time to define a {\it new} inner
product-- the states will not be normalizable in the new inner product. Doesn't
this indicate that in \dcs s too, one should first isolate time from the
``true'' degrees of freedom, and {\it then} look for an inner product,
involving, say, an integral over only the true degrees of freedom? We did this,
e.g., in the reduced space quantization of the Bianchi type I and type II
models.

\tor True enough, but one is not {\it forced} to take this approach. Let us
consider what happens in the \npp\ model, which is a paradigm for some \dcs s.
There, as far as the inner product is concerned, we did {\it not} isolate a
time variable. In fact, at the level of the kinematical quantum theory, we did
not even {\it explicitly identify} a time variable. Now, returning to the inner
product, this is a 4-dimensional integral over {\it all} the configuration
variables $(q^0,q^i)$. We followed the algebraic program ``blindly'', and
imposed the Hermiticity conditions on the observables. Lo and behold, we {\it
did} find the inner product on physical states.

\indent Similarly, in the Dirac quantum theory of the Bianchi II models, we did
not
identify any time variable. Yet, there too, the Hermiticity conditions on
observables yield the physical inner product.

\indent In sum, to find the physical inner product, it is {\it not} necessary
to
isolate or even identify a time variable. If we have a sufficient number of
observables, we can follow the prescription of the algebraic program, and
proceed to use the Hermiticity conditions on observables to find an inner
product. Time itself could be an approximate or interpretational notion,
identified {\it after} the kinematical quantum theory has been completed.

\ach I think I understand the separation of the interpretational or
semi-classical role of time from its role in finding an inner product. But, if
I recall correctly, something unexpected happened in the \npp, as we proceeded
to construct the inner product. Even after we imposed the Hermiticity
conditions on all the observables, the dependence of the measure on one of the
variables, $q^0$, was left undetermined. Using the properties of physical
states, this allowed us to choose the measure so that the integral over $q_0$
was finite; and thus reduce the inner product to a 3-dimensional integral over
only the variables $q^i$.

\tor If you think about it, we do not encounter this unexpected behaviour only
in \dcs s. A similar thing happened in the A-H model, but we didn't pay much
attention to it, since it was so incidental at the time. Recall that in the A-H
model, the $\theta$ dependence of the measure on physical states was
undetermined by the Hermiticity conditions on observables. Although one could
have done so, {\it we} certainly did not consider $\theta$ to be a time-like
variable.

\indent What is common to both models is the following: Recall, that in each
model, one of the momentum operators --$p_0$ and $p_\theta$, respectively-- is
represented by a {\it multiplication} operator on {\it physical} states. Thus,
we do not have to {\it require} that $i(\d/\d q^0)$ (or $i(\d/\d\theta)$ is a
Hermitian operator, and the corresponding derivative of the measure is left
undetermined.

\indent So you see, this ``unexpected'' behaviour is quite common, and
perfectly
reasonable.

\ach If we do consider $\theta$ to be a timelike variable in the A-H model
--i.e.\ if we consider the constraint itself as the generator of dynamics--
then it is a {\it Euclidean} time, in the sense that if the overall sign for
the constraint is chosen so that the potential is negative --and thus
corresponds to bound states-- then the kinetic energy is positive. In the
Bianchi models, the
natural time variable is Lorentzian, and in the \npp, the time is Newtonian.
Yet in all of these models, we were able to find an inner product.

\tor Hmm.., yes. It certainly appears as if this issue --whether in a
particular model the  time is Euclidean, Newtonian or Lorentzian-- does not
play a terribly significant role as far as finding the inner product is
concerned.

\ach Let me turn now to the second ``new'' feature: the algebraic relations and
their role in quantum theory. Why are these rarely encountered, even in
constrained systems like gauge theories?

\tor Typically the configuration spaces one deals with are linear spaces; even
for gauge field theories, the configuration space --the space of
connections-- is an affine space. Hence one has the luxury of introducing
Cartesian coordinates and doing quantum theory. Or, as in problems with
spherical symmetry, one is sloppy and uses spherical ``coordinates''. As
elementary variables for the quantum theory, such coordinates are complete but
not overcomplete and there are no algebraic relations between them. In simple
cases, one can get by because one has experience in handling pathologies that
arise at points (such as the poles on the 2-sphere) where the ``coordinate
systems'' fail. In general, however, one can simply get incorrect results.

\item[{\it Beaver}:] But Mr. T, why do you call this feature ``new''? I think
one does encounter algebraic relations in ordinary quantum mechanics. Consider
something like $\wh{xp}=\half(\hat{x}\hat{p} + \hat{p}\hat{x})$, isn't it an
algebraic relation?

\tor In the usual treatment of quantum mechanics, $x$, $p$ and $1$ are the
elementary variables; $xp$ is not. The relation you mentioned is actually
a definition of the {\it new} operator $\wh{xp}$ in terms of the known
expressions for the operators $\hat{x}$ and $\hat{p}$.

\ach I am still somewhat unfamiliar with this concept. Are there situations in
which algebraic relations are unavoidable?

\tor Yes. One situation occurs when one cannot find Cartesian coordinates.
Consider the case when the configuration space is a non-trivial manifold (e.g.
$S^1$), and does not admit a global coordinatization. Unfortunately, one does
not know how to do quantum mechanics on (collections) of patches. One has two
options: the first is to imbed the configuration space into some $\real^n$,...

\ach Which is usually how the configuration space is given to us in the first
place.

\tor ... then the finite number of imbedding functions are the algebraic
relations on the elementary variables. For example, for $T^\star S^1$, we used
the set $(1,\, q_1=\sin\theta,\, q_2=\cos\theta, \,p)$, and the imbedding
function is $(q_1)^2+(q_2)^2-1=0$.

\indent The second option is to consider the space of a large class of --or
indeed,
all-- functions on configuration space and a similar class of functions linear
in momenta, as we did for  the Bianchi II model. There are then infinitely many
algebraic relations, which have to be represented in quantum theory.

\ach Could it happen that the \ps\ itself is a well-defined manifold, but for
which there is no configuration space?

\tor Yes, such situations arise when one constructs the reduced \ps\ for a
classical constrained system. For example, for the model in section 5.7 (two
oscillators with the energy sum as a constraint), the reduced \ps\ is $S^2$.

\indent There is another situation in which algebraic relations are
unavoidable, in
spite of the trivial topology of the \ps. This situation can arise because we
are interested in the manifold as a {\it symplectic} space, not just as a
topological or metric space. The Cartesian coordinates on the \ps\ may not be
canonical, and in fact may not generate a vector space closed under Poisson
brackets. Then one is forced to enlarge the set of operators, even though it is
complete, and this leads to algebraic relations. For example, when the energy
difference is imposed as a constraint on a system of two oscillators (chapter
5) the reduced \ps\ is topologically $\real^2$, and Cartesian coordinates are
$J_\pm$. However, $\{J_+,J_-\}=(J_+J_-+\delta^2)^{1/2}$; as a symplectic
manifold, it is natural to consider the \ps\ to be the future mass shell in
3-dimensional Minkowski space. In order to do quantum theory, one has to
introduce the additional function $J_z=(J_+J_-+\delta^2 )^{1/2}$, and the
appropriate algebraic relation.

\indent So, algebraic relations on \cS\ arise due to the twin requirements that
\cS\
should be (over)complete {\it and} closed under Poisson brackets. When there
are algebraic relations, as we learned from the example of the particle on
$S^1$, the anticommutation relations have to imposed in quantum theory to
ensure that one stays on the correct, physical sector of the theory.

\ach Hmm... ({\it long pause as he digests all this}) now I am confused about
something else. General relativity in the new variables was mentioned as part
of the motivation for formulating an approach which encompasses such algebraic
relations. But it is a theory of the dynamics of a connection and so the
configuration space is an affine space. Why are there algebraic relations in
the connection-dynamical formulation?

\tor True, the algebraic relations are neither unavoidable nor necessary in the
classical theory. However, it currently appears that {\it loops} play a
fundamental role in \qg. For example, the loop variables seem indispensable to
express and regulate physically interesting operators without reference to
background fields such as a metric or a connection.  It is, in particular,
natural to consider the holonomies of the connection around loops as the
elementary configuration variables for quantum theory. These holonomies are
genuinely {\it over}complete, and hence it is perhaps not surprising that there
are algebraic identities between them. However,...

\ach However,...

\tor ...since the loop variables are nonlocal quantities...

\ach ...since \gr\ is a field theory...

\tor ...the counting of the number of algebraic identities is quite
non-trivial.

\ach ...the counting of the number of algebraic identities is quite
non-trivial.

\indent I see now that the algebraic approach provides a framework for {\em
canonical} quantization which is general enough to apply to a large class of
constrained systems, including \gr. Also, I think I understand the role of the
two new ingredients in this approach, the use of the Hermiticity conditions on
observables to fix the physical inner product, and the presence and
incorporation of algebraic relations. However, I still have some questions
about the general framework. In the first place, it is not really a
constructive {\it procedure} for quantization. For example, one is instructed
``to find a representation of \cA'', but {\it how} one should do this is left
unspecified. On the other hand, in Schr\"odinger mechanics one is explicitly
given a representation. Even in path integral quantization, fairly detailed
instructions are issued, it is another matter that there are immense technical
difficulties in carrying them out.

\item[{\it Giselle}:] In this sense perhaps geometric quantization is also
``not constructive''; one is told to ``find a polarization'' on \ps, but one is
not told how to explicitly construct it. However, I get your general drift,
Achilles.

\tor Yes, there is certainly much more freedom in the algebraic framework.
There are specific steps at which crucial choices have to be made. The first is
in the selection of the set of elementary functions and a linear representation
of the resulting operator algebra. In general there will exist many
overcomplete sets, and representations of the corresponding algebras may be
equally easy (or  difficult) to find. However, these choices will have
ramifications later, in terms of the ability to complete the quantization. A
poor choice for \cS\ may make it difficult to solve the constraints or find
physical operators.

\item[{\it Owl}:] {\large\bf P}ardon me for interrupting, but a dramatic
example is provided by the Bianchi II model. In the $\bar\beta$ variables, even
though the constraint had been solved ``explicitly'' in terms of power series
solutions, a complete set of physical operators was not known. In retrospect we
can see why: in terms of the $\bar\beta, \bar\pi$ variables the classical
expressions for the {\it physical observables} are ridiculously complicated.
While apparently suitable in the sense that they satisfy the requirements of
completeness and closure, the $\bar\beta,\bar\pi$ variables are not adapted to
the symmetries of the constraint and the algebra of observables. In contrast,
when we introduce the symmetry adapted variables $(\tilde\beta,\tilde\pi)$, the
construction of the complete quantum theory is almost trivial.

\tor {\large A}s another example of a poor choice of \cS, consider the coupled
oscillators, in real Cartesian coordinates and the Schr\"odinger
representation. A few minutes of toying around will convince anyone that this
choice of variables and representation is obviously ill-suited to construct
solutions to the constraint equation, particularly for the generic case when
the energy difference is non-integer. Solutions are products of generalized
Hermite polynomials with non-integer indices, and things are a total mess.

\item[{\it van Gogh}:] {\large R}ight, but considerable improvement is achieved
by using a $z_1,z_2$ representation. However, a naive power series ansatz or
the holomorphic representation does not yield all physical states, since
solutions to the constraint contain real --rather than integral--  powers of
$z_1,z_2$. In a more sophisticated representation, physical states are Bessel
functions, and while it is quite unobvious and somewhat technically difficult,
a physical inner product can be found.

\tor {\large I} have no objection in principle to technical difficulties.
However in this case it seems that the technical difficulties may thwart one
from considerable physical insight. Consider on the other hand the angular
momentum like representation for the coupled oscillators. This is the ideal
situation: the representation is well adapted to the algebra of
observables and chosen such that the constraint is actually diagonal, solving
it is then a piece of cake.

\ach {\large T}hat brings up another question. Is there freedom in the choice
of the physical observables too?

\tor {\large Y}es. A trivial source of freedom is just the re-coordinatization
of the reduced \ps. Consider the coupled oscillators again. Since the reduced
\ps\ is topologically $\real^2$, one could introduce operators corresponding to
radial/angular coordinates.

\ach {\large\bf E}xcept that in quantum theory, there are obvious difficulties
with this, since the radius (squared) and the angle are then
conjugate to each other.

\tor {\large I} agree. To continue my point, a choice with non-trivial
implications presents itself in the quantization of the A-H model. There was an
obvious choice of a set of physical operators which is locally overcomplete.
However, as we saw, this set failed to capture some of the {\em global}
structure of the reduced \ps, and led to results that are qualitatively
different from the correct ones. In most other examples (recall e.g., the
Bianchi I model) the choices were rather obvious, and so it is difficult to
construct good examples of bad choices.

\ach {\large N}ow, in the next step, we have to implement the $\star$-relations
on \cAp\ to find the physical inner product. If I recall correctly, in both the
Bianchi II model and the coupled oscillator model, even after all the
Hermiticity conditions are imposed there is freedom to choose the inner
product. For, as we have seen, the $\star$-relations fix only the {\em
relation} between the representation and the inner product. We are free to
choose an inner product, as long as the representation is fixed appropriately,
or vice versa.

\tor {\large S}ure. But in the examples you mention, this freedom is rather
trivial. It is easy to see that these choices yield the same, i.e. unitarily
equivalent, quantum theories. It is still useful though, since we can use this
freedom to simplify the representations of interesting observables.

\ach {\large T}o summarize then, the power of this approach arises from a
tension between two contrasting aspects: the tight set of criteria that various
choices should satisfy, and the looseness in prescribing what those choices
should be.

\tor {\large E}xactly. When taking the algebraic approach to the quantization
of a problem, one has to use one's physical intuition for the problem in making
these choices. One has to anticipate the structures (mainly \cAp) that arise in
later steps, and use this knowledge to guide the decision-making in the earlier
steps, just as in a conversation one sometimes has to anticipate the other
person's thoughts in order to decide what to say. For example, in many of the
models we considered, we used detailed insight into the algebra of observables
before finally selecting a representation of the elementary algebra \cA\
itself.

\ach {\large I} was thinking ... ({\it Appears lost in thought.}) ...  about
the
A-H model again. First, it clearly indicates that one cannot always {\it
sequentially} follow the steps outlined in section 2.3. In this model, after
step 5, we have a large set of solutions to the quantum constraint equation.
However, the operators that leave this space invariant do not form a
$\star$-algebra, and we seemed to be at an impass\`e. Inspite of this we tried
to implement step 7 partially: the $\star$-relations on part of \cAp\ then
forced us to go back to step 5 and discard many of the solutions from the space
of physical states. The operators that leave this newly defined, smaller ${\cal
V}_{phy}$ invariant do form a $\star$-algebra and we are finally able to
complete the quantization of this model.

\indent Second, this model emphasizes the distinction between solutions to the
constraint and physical states. All physical states are solutions to the
constraints, but not vice versa.

\indent I am sorry, I missed something you said a short while ago.

\tor {\large N}ot much. I was just saying that in carrying out the quantization
program one has to go back and forth, correcting poor decisions taken in the
early steps, as ones intuition for the problem develops. As was emphasized in
the discussion of the A-H model, we can insist only that the final
quantum theory is consistent and complete. We cannot quibble about the
intermediate steps. Of course, a complete quantum theory is one in which we
have obtained a (faithful) $\star$-representation of a complete algebra of
physical observables. The step-by-step approach, and the conditions spelt out
in chapter 2 are just guidelines to achieve this final result.

\ach {\bf D}o we obtain an inner product
or a quantum theory which is {\em unique} --up to unitary equivalence-- if we
satisfy all these conditions?

\item[{\it St. Joseph}:] Oh no, Achilles! Even for quantum mechanics on a
finite dimensional {\it manifold} there are no such uniqueness theorems. The
only
known statement, the Stone-von Neumann theorem, states only that there is a
unique, weakly continuous representation of the Weyl operator algebra
constructed on $T^\star\real^n$.

\tor Right! In fact, as you know, there are inequivalent unitary
representations of
the Weyl algebra for the particle on the real line, in which the spectrum of
the momentum operator includes all real numbers, but this spectrum is {\em
discrete}! The only problem is that these representations are not weakly
continuous. Since we did not introduce topological considerations, such
discrete representations may well arise in the algebraic approach.

\indent Furthermore, there are counter-examples, in this thesis itself, to a
naive ``uniqueness  conjecture''. Recall the fractional spin representations
for the particle on $S^1$, and the interval's worth of ambiguity in the
representation of the coupled oscillator model when the energy difference is
less than $\half$.

\ach I would have intuitively expected to obtain a unique inner product by
imposing the Hermiticity conditions on a complete set of generators of
\cAp.

\tor Now you are jumping to conclusions again. The test to check that the set
of physical observables is (over)complete on the reduced \ps\ guarantees only
{\em local} completeness. The observables may not ``know'' enough about the
global structure. We have already seen the importance of the role of
additional, discrete physical operators (which arise from discrete symmetries
of the constraint) in reducing the ambiguities in the inner product. Another
source of ambiguity in the inner product arises if the physical operators are
complete only {\it almost} everywhere on the constraint surface, i.e.\ if
there are lower dimensional submanifolds on which the physical observables are
not complete.

\ach Even so, this algebraic approach to the quantization of constrained
systems is
certainly very powerful. We have been able to use it to quantize numerous
models of \gr, some of which were previously only incompletely solved. However,
all the difficulties and subtleties in the
quantization of these relatively simple {\em finite dimensional} models, which
you have so kindly commented on, make me wonder about the prospects for a
quantum theory of gravity, which after all, is a {\it nonlinear field theory}.
While
a large (but certainly incomplete) set of solutions to the quantum scalar
constraint is known in the new variables, there is still not a single known
local observable. Perhaps you have sold the quantization program to me too
strongly, but currently, it {\it appears} to me that physical observables are
crucial even in \qg; not just to construct the kinematical quantum theory, but
perhaps also, as we saw in chapter 6, to understand dynamics and obtain a
certain type of physical interpretation. Given all this, I wonder if I can ask
you one last question. Do you expect to see a complete theory
of \qg\ in your lifetime, Mr. T?

\tor ...possibly, Achilles, but then you know how long-lived tortoises are!

\end{description}

\newpage\mbox{}
\appendix
\chapter{QUANTIZATION OF SECOND CLASS CONSTRAINTS}
\pagestyle{myheadings}
\markboth{{\sf Appendix A}}{{\sf Quantization of Second Class Constraints}}


\def\hzo{\hat{z}_1}
\def\hzbo{\hat{\bar{z}}_1}


In the main part of my thesis, I have dealt only with first class constraints,
since, as is well known, second class constraints have to be solved
classically. If we were to represent the constraint functions as operators,
then because they are second class, i.e. because they do not commute with each
other weakly, there is no common kernel \cite{pamdA}. (Consider the action of
the commutator on states annihilated by both constraint operators, clearly this
vanishes. On the other hand, the commutator is proportional to the identity
operator since the constraints are second class, and so we have an
inconsistency.) More intuitively, if there is a Hermitian inner product on the
representation space (so we can calculate expectation values), then we see that
a state annihilated by both second class constraint operators violates the
uncertainty principle.

In this appendix I will show that there is a mathematically well-defined sense
in which one can ``solve'' second class constraints in quantum theory. Of
course, since the argument in the preceding paragraph is a strict mathematical
theorem, one of its hypotheses must be violated by our solution. We will
construct quantum states which are ``solutions'' to the pair of second class
constraints. These are not solutions in the standard sense: they are not
annihilated by both constraint operators.  The idea is to first write the
second class pair as complex conjugate functions of each other. Then, solve
{\em one} of the constraints using a {\em holomorphic} $\delta$ function, which
I will describe in section A.1. Call these solutions the physical states.  Now
one can introduce an inner product such that the action of the second
constraint on the above physical states results in states {\it orthogonal} to
the space of physical states, thus, it sends physical states to the zero
element of the physical Hilbert space. I will clarify this issue after I have
presented the details of the solution in section A.2.

There are not yet any applications of this approach, in which we quantize and
then solve second class constraints. However, there are situations in which
this approach is potentially useful, and I will comment briefly on these at the
end of the appendix.

\mysection{Holomorphic $\delta$ function}

Let us define a {\em holomorphic distribution} (or a generalized function)
$\delta(z)$ as follows \cite{newbookA}: it is a complex linear mapping from the
space of functions of the type $\sum f_i(z)g_i(\ovr{z})$, where $f_i(z)$ are
entire holomorphic functions and $g_{i}(\ovr{z})$ are entire anti-holomorphic
functions, to the space of entire anti-holomorphic functions:
 \be\label{holod}
  \delta(z)\circ \sum_i f_i(z)g_i(\ovr{z}) = \sum_i f_i(0)g_i(\ovr{z}).
 \ee
We can also define the anti-holomorphic distribution $\delta(\ovr{z})$ simply
by taking the complex conjugate of $\delta(z)$ and this new distribution has
the action:
 $$\delta(\ovr{z})\circ \sum_i f_i(z)g_i(\ovr{z}) = \sum_i f_i(z)g_i(0)
  \eqno(\arabic{section}.\ovr{\arabeq})$$
We will also need the product of a polynomial $a(z,\ovr z)$ with a
distribution ${\cal F}(z)$. This is a distribution defined by:
 \be
  [a(z,\ovr z){\cal F}(z)]\circ\sum_if_i(z)g_i(\ovr z):={\cal F}(z)\circ
  \>a(z,\ovr{z})\sum_if_i(z)g_i(\ovr z)
 \ee
Using, as usual, the Leibnitz rule as a motivation, one can define the
derivative of a distribution ${\cal F}(z)$, as
 \bea \label{leib}
  \left[\dfrac{d}{dz}{\cal F}(z)\right]&\circ&
      \sum_i f_i(z)g_i(\ovr z)\cr
   &:&= \dfrac{d}{dz}\left({\cal F}(z)\circ\sum_if_i(z)g_i(\ovr z)\right)
      - {\cal F}(z)\circ\dfrac{d}{dz}\>\sum_if_i(z)g_i(\ovr z),
 \eea
and similarly for the derivative w.r.t.\ $\ovr{z}$. Applying this to
(\ref{holod}), for example, we find
 \be
  \frac{d}{d\ovr{z}}\delta(z) = 0, \quad\hbox{and}\quad
   \left[\frac{d}{dz} \delta(z)\right] \circ \sum_i f_i(z)g_i(\ovr{z})
  = \left. -\sum_i \frac{df_i(z)}{dz}\right|_{z=0}g_i(\ovr{z}).
 \ee
Thus, {\em the $\delta$-distribution is {\em holomorphic} and its derivative
with respect to $z$ is a distribution with the expected property.} Finally, we
notice that the product of the two distributions (\ref{holod}) and
(\ref{holod}$'$) is well-defined:
 \be
  [\delta(z)\delta(\ovr{z})]\circ\sum_i f_i(z)g_i(\ovr{z}) =\sum_i
f_i(0)g_i(0).
 \ee
Note that this is just the two dimensional $\delta$-distribution and therefore
admits the standard integral representation (which is useful as a mnemonic for
the definitions (\ref{holod}-\ref{leib})):
 \bea
  \lint dq\wedge dp\> \delta^2(q,p;0,0)\>\sum_i
                f_i(z)g_i(\ovr{z})&& \cr
   := [\delta(z)\delta(\ovr{z})]&\circ&\sum_i f_i(z)g_i(\ovr{z})
      =\sum_i f_i(0)g_i(0)
 \eea
where, we have used $z=q-ip$. Thus, one can regard $\delta(z)$ as the
``holomorphic square-root'' of the standard 2-dimensional $\delta$-distribution
on the 2-plane, (picked out by the complex structure).

\mysection{Second class constraints in quantum theory}

Consider the $n$ dimensional system with real configuration coordinates $q^i,\,
i=1,...n$ and the conjugate momenta $p_i$. The symplectic structure is $\Omega=
dp_i \wedge d q^i$. The system is constrained by the two {\em second class}
constraints $q^1=0,\> p_1=0$. The reduced \ps\ is coordinatized by the true
degrees of freedom $(q^{i'},p_{i'}),\> i'=2,...n$.

We want to solve the pair of second class  constraints in quantum theory. Let
us first introduce complex coordinates on (a portion of) the \ps, $z_1:=
(p_1-iq_1)/ \sqrt2$, and $\bar{z}_1:= (p_1+iq_1)/ \sqrt2$. Hence, the algebra
\cA\ of elementary operators contains the pair $\hzo,\hzbo$, with the
commutation relation $[\hzo,\hzbo]=-1$. The $\star$-relation between these
operators is $\hzo{}^\star=\hzbo$. The pair of second class constraints can now
be expressed as: $\hzo=0,\>\hzbo=0$.

Classically, we know that $z_1=0\Rightarrow\bar{z}_1=0$. In the quantum theory,
if there is an inner product that implements the $\star$-relations, then
$\hzo=0\Rightarrow\hzo{}^\star=0\Rightarrow\hzbo=0$. Therefore the strategy is
to first find a representation of \cA\ and impose the $\star$-relations to find
an inner product such that $\hzo^\dagger=\hzbo$, and then solve only the
constraint $\hzo=0$. The inner product will automatically implement $\hzbo=0$.

Let us choose as a representation space the vector space of holomorphic
distributions $\Psi(z_i)$ of the kind introduced in section 1. The elementary
operators are represented by:
 \bea
                   \hzo\circ\Psi &=& z_1\Psi  \cr
  \hbox{and}\quad \hzbo\circ\Psi &=& \dfrac{\d}{\d z_1}\Psi.
 \eea
Let
 \be
  \IP\Psi\Phi:=\bar{\Psi}\Phi\circ\mu(z_1,\bar{z}_1)
 \ee
be an ansatz for the inner product. A straightforward calculation --using the
definitions from section A.1 and discarding the surface term, as usual--
shows that the inner product is given by
 \be
  \mu=e^{-z_1\bar{z}_1}.
 \ee

The constraint equation $\hzo\circ\Psi=0$ is solved by distributions of the
form
 \be\label{dsol}
  \Psi(z_i)=\delta(z_1)\psi(z_{i'}).
 \ee
Call these the physical states, and denote the vector space by \cVp.
Note that $\hzo\circ\Psi$ vanishes as a {\em distribution}, as can be confirmed
by
using the definition (\ref{holod}) of the delta function; and also that one has
no need of the inner product to do this calculation. However, physical states
are normalizable \wrt\ this inner product.

Now, on solutions (\ref{dsol}), it is not true that $\hzbo\circ\Psi=0$, in
fact, as a distribution,
 \be\label{dzbar}
  \hzbo\Psi\circ f=-\Psi\circ\frac{\d f}{\d z_1} = - \left. \frac{\d f}{\d z_1}
   \right|_{z_1=0}\cdot\psi \not= 0,
 \ee
where $f$ is an arbitrary test function of the type used in section 1. However,
$\hzbo\circ\Psi$ is orthogonal to the physical states. Namely, for
$\Psi,\Phi\in\cVp$,
 \bea
  \bra\Psi\hzbo\ket\Phi&=&\bar{\Psi}(\bar{z}_i)\left(\dfrac{\d}{\d z_1}
  \Phi(z_i)\right) \circ\mu\cr
  &=& -\bar{\Psi}\Phi\circ\dfrac{\d\mu}{\d z_1}=\bar{\Psi}\Phi\circ
     \bar{z}_1\mu\cr
  &=& \bar{\psi}\delta(\bar{z}_1)\phi\delta(z_1)\circ \bar{z}_1\mu=0.
 \eea
Hence, we have a representation in which both $\hzo=0$ and $\hzbo=0$. In this
sense, the second class constraints have, apparently, been imposed as if they
are first class.

How have we circumvented the argument, presented in the introduction, against
solving second class constraints in the quantum theory? This question can be
asked and analyzed at various levels. First, note that at the vector space
level --i.e.\ prior to the introduction of an inner product-- there is no
contradiction since we have {\em not} solved for $\hzbo=0$: the space of
physical states is defined simply by $\hat{z}_1\circ\Psi=0$. Second, when we do
introduce an inner product, on physical states there is again no contradiction,
since $\hzbo$ is not a physical operator. The inner product is defined not only
on physical states, but on all distributions. Now, the action (\ref{dzbar}) of
$\hzbo$ takes physical states {\em out} of \cVp. The resulting state is still a
distribution, and we can compute its inner product with a physical state. It is
this {\em projection} that vanishes. Finally, using the inner product, we can
define a projection to the physical states, and consequently a new operator
$\hzbo|_{phy}$. On physical states, $\hzbo|_{phy}\circ\Psi=0$. However, this
new operator is (by construction) a physical operator: it is easy to see that
$[\hzo,\hzbo|_{phy}] =0$ on physical states. However, now $\hzo$ and
$\hzbo|_{phy}$ do {\it not} satisfy the CCR. Thus, at this level, there is no
contradiction with the theorem since the CCR is itself changed.

This seems like an awful lot of heavy artillery brought to bear on a simple
problem that one knows how to deal with, and there are no obvious applications
of this approach. However, if solving the second class constraints complicates
the remaining first class constraints, or results in complicated algebraic
relations on the variables, then ``quantizing'' second class constraints may be
helpful. Perhaps a more useful application could be to the following situation:
Suppose that factor ordering and/or regularizing constraints which are {\it
classically first class} leads inevitably to second  class constraints in the
quantum theory (and assume that this can be done while keeping the true degrees
of freedom the same). This presents us with a dilemma, since the second class
constraints are {\it derived} and {\it defined} in the quantum theory itself.
Our approach to solving second class constraints in quantum theory might
provide a way out of this dilemma.

%
\newpage\mbox{}
\chapter*{Bibliography}
\addcontentsline{toc}{chapter}{Bibliography}
\pagestyle{myheadings}
\markboth{}{Bibliography}
\newcounter{listnumber}

\setcounter{listnumber}{1}
\begin{itemize}
\item[[\thelistnumber]] \stepcounter{listnumber}
Ashtekar A and Geroch R 1974 \RPP\ {\bf37} 1211-56
\item[[\thelistnumber]] \stepcounter{listnumber}
Ashtekar A and Magnon A 1975 {\it Proc. R. Soc. (Lond.)} {\bf A346} 375
\item[[\thelistnumber]] \stepcounter{listnumber}
Ashtekar A 1980 \CMP\ {\bf 71} 59
\item[[\thelistnumber]] \stepcounter{listnumber}
Ashtekar A and Horowitz G H 1982 \PR\ {\bf D26} 3342-53
\item[[\thelistnumber]] \stepcounter{listnumber}
Ashtekar A and Magnon A 1982 \CMP\ {\bf86} 55-68
\item[[\thelistnumber]] \stepcounter{listnumber}
Ashtekar A and Stillerman M 1986 \JMP\ {\bf 27} 1319
\item[[\thelistnumber]] \stepcounter{listnumber}
Ashtekar A 1986 \PRL\ {\bf57} 2244
\item[[\thelistnumber]] \stepcounter{listnumber}
Ashtekar A 1987 \PR\ {\bf D36} 1587
\item[[\thelistnumber]] \stepcounter{listnumber}
Ashtekar A 1989 in {\it Conceptual Problems in Quantum Gravity} Ashtekar A and
Stachel J (eds) (Birkh\"auser Boston)
\item[[\thelistnumber]] \stepcounter{listnumber}
\newbook
\item[[\thelistnumber]] \stepcounter{listnumber}
Ashtekar A and Samuel J 1991 \CQG\ {\bf8} 2191
\item[[\thelistnumber]] \stepcounter{listnumber}
Ashtekar A, Tate R S and Uggla C 1992 \SUpp\ SU-GP-92/2-5
\item[[\thelistnumber]] \stepcounter{listnumber}
Ashtekar A, Tate R S and Uggla C 1992 \SUpp\ SU-GP-92/2-6
\item[[\thelistnumber]] \stepcounter{listnumber}
Ashtekar A and Tate R S 1992 \SUpp\ in preparation
\item[[\thelistnumber]] \stepcounter{listnumber}
Arnowitt R, Deser S and Misner C W 1962 {\it Gravitation: An Introduction to
Current Research} Witten L (ed) (Wiley New York)
\item[[\thelistnumber]] \stepcounter{listnumber}
Br\"ugman B, Gambini R and Pullin J 1992 \PRL\ {\bf68} 431-4
\item[[\thelistnumber]] \stepcounter{listnumber}
Boulware D G 1983 \PR\ {\bf D28} 414-16
\item[[\thelistnumber]] \stepcounter{listnumber}
Dirac P A M 1964 {\it Lectures on Quantum mechanics, Belfer Graduate School
Monograph Series No. 2} (Yeshiva University New York)
\item[[\thelistnumber]] \stepcounter{listnumber}
Feynmann R P and Hibbs A R 1965 {\it Path Integrals and Quantum Mechanics}
(McGraw Hill, New York)
\item[[\thelistnumber]] \stepcounter{listnumber}
Gelfand I M and Naimark M A 1943 {\it Mat. Sobernik} {\bf12} 197
\item[[\thelistnumber]] \stepcounter{listnumber}
Geroch R 1974 {\it University of Chicago Lecture Notes on Quantum Mechanics}
unpublished
\item[[\thelistnumber]] \stepcounter{listnumber}
Gotay M J and Demaret J 1983 \PR\ {\bf D28} 2402
\item[[\thelistnumber]] \stepcounter{listnumber}
Gotay M J 1986 \CQG\ {\bf3} 487-91
\item[[\thelistnumber]] \stepcounter{listnumber}
Hajicek P 1990 \CQG\ {\bf 7} 871
\item[[\thelistnumber]] \stepcounter{listnumber}
Husain V 1987 \CQG\ {\bf4} 1587
\item[[\thelistnumber]] \stepcounter{listnumber}
Isham C J, Penrose R and Sciama D W (eds) 1981 {\it Quantum
Gravity 2: A Second Oxford Symposium} (Clarendon Press Oxford); see in
particular the overview by Isham C J
\item[[\thelistnumber]] \stepcounter{listnumber}
Isham C J 1984 in {\it Relativity, Groups and Topology II, Les Houches 1983}
DeWitt B S and Stora R (eds) (North-Holland Amsterdam)
\item[[\thelistnumber]] \stepcounter{listnumber}
Kucha\v r K 1981 in {\it Quantum Gravity 2: A Second Oxford Symposium}
Isham C J, Penrose R and Sciama D W (eds) (Clarendon Press Oxford)
\item[[\thelistnumber]] \stepcounter{listnumber}
Kucha\v r K 1981 \JMP\ {\bf 22} 2640-54
\item[[\thelistnumber]] \stepcounter{listnumber}
Kucha\v r K 1982 \JMP\ {\bf 25} 1647-61
\item[[\thelistnumber]] \stepcounter{listnumber}
Kucha\v r K 1992 {\it Time and interpretations of Quantum Gravity} in {\it
Proceedings of the 4th Canadian Conference on General Relativity and
Relativistic Astrophysics} Kunstaatter G, Vincent D and Williams J
(eds) (World Scientific Singapore)
\item[[\thelistnumber]] \stepcounter{listnumber}
Loll R 1990 \PR\ {\bf D41} 3785
\item[[\thelistnumber]] \stepcounter{listnumber}
MacCallum M A H 1979 in {\it Physics of the Expanding Universe} Demianski M
(ed) (\SV\ Berlin)
\item[[\thelistnumber]] \stepcounter{listnumber}
Misner C W 1972 in {\it Magic Without Magic} Klauder J (ed) (Freeman
San Francisco)
\item[[\thelistnumber]] \stepcounter{listnumber}
Romano J D and Tate R S 1989 \CQG\ {\bf 6} 1487-500
\item[[\thelistnumber]] \stepcounter{listnumber}
Rosquist K, Uggla C and Jantzen R T 1990
 \CQG\ {\bf 7} 611
\item[[\thelistnumber]] \stepcounter{listnumber}
Rovelli C 1990 \PR\ {\bf D42} 2638
\item[[\thelistnumber]] \stepcounter{listnumber}
Rovelli C 1991 \PR\ {\bf D43} 442-56
\item[[\thelistnumber]] \stepcounter{listnumber}
Rovelli C 1991 \CQG\ {\bf8} 1613-75
\item[[\thelistnumber]] \stepcounter{listnumber}
Ryan M P 1972 {\it Hamiltonian Cosmology: Lecture notes in Physics} {\bf 13}
Ehlers J {\it et al} (eds) (\SV\ New York)
\item[[\thelistnumber]] \stepcounter{listnumber}
Ryan M P and Shepley L C 1975 {\it Homogeneous Relativistic Cosmology}
(Princeton \UP Princeton)
\item[[\thelistnumber]] \stepcounter{listnumber}
Segal I E 1947 {\it Bull. Amer. Math. Soc.} {\bf53} 73
\item[[\thelistnumber]] \stepcounter{listnumber}
Smolin L 1992 in {\it Proc. XXII Gift Intl. Seminar on Theo. Phys., Quantum
Gravity and Cosmology, June 1991 Catalonia Spain} (World Scientific,
Singapore)
\item[[\thelistnumber]] \stepcounter{listnumber}
Sniatycki J 1980 {\it Geometric Quantization and Quantum mechanics} (\SV, New
York)
\item[[\thelistnumber]] \stepcounter{listnumber}
Stillerman M 1985 {\it Ph.D. thesis} Syracuse University
\item[[\thelistnumber]] \stepcounter{listnumber}
Tate R S 1992 {\it Constrained systems and Quantization, Lectures at the
Advanced
Institute for Gravitation Theory, Session: December 91 Cochin} \SUpp\ No
SU-GP-92/1-4
\item[[\thelistnumber]] \stepcounter{listnumber}
Taub A H 1951
{\it Ann.\ Math.\/} {\bf 53} 472
\item[[\thelistnumber]] \stepcounter{listnumber}
Wald R M 1984 {\it General Relativity} (University of Chicago Press, Chicago)
\item[[\thelistnumber]] \stepcounter{listnumber}
Woodhouse N J M 1981 {\it Geometric Quantization} (\OUP, Oxford)
\item[[\thelistnumber]] \stepcounter{listnumber}
Wheeler J 1977 \GRG\ {\bf 8} 713

\end{itemize}

\end{document}